\documentclass[preprint,12pt]{elsarticle}

\usepackage{setspace} \singlespace % \doublespacing

\usepackage[T1]{fontenc}
\usepackage[utf8]{inputenc}
\usepackage[version=4]{mhchem}

\usepackage[pdftex,colorlinks]{hyperref}
\usepackage{booktabs}
\usepackage{float}
\usepackage{caption}

\usepackage{cleveref}
\crefname{algocf}{pseudocode}{pseudocodes}
\Crefname{algocf}{Pseudocode}{Pseudocodes}

\newcommand\etal{\textsl{et al.}}
\newcommand\ie{, \textsl{i.e.},}

\newcommand\Eh{\ensuremath{\textrm{E}_\textrm{h}}}

\usepackage{amsthm} % Theorem Formatting
\usepackage{upgreek}
\usepackage{url}
\usepackage{longtable}
\newcommand{\xtwodhf}{\texttt{x2dhf}}
\newcommand{\helfem}{\texttt{HelFEM}}
\newcommand{\opmks}{\texttt{OPMKS}}
\newcommand{\gaussian}{\texttt{GAUSSIAN}}

\newcommand{\citeref}[1]{ref. \citenum{#1}}

\renewcommand{\v}[1]{\ensuremath{\mathbf{#1}}} % for vectors
\newcommand{\gv}[1]{\ensuremath{\mbox{\boldmath$ #1 $}}}
\newcommand{\avg}[1]{\left< #1 \right>} % for average
\newcommand{\pdd}[2]{\frac{\partial^2 #1}{\partial #2^2}}
\let\baraccent=\= % rename builtin command \= to \baraccent
\renewcommand{\=}[1]{\stackrel{#1}{=}} % for putting numbers above =

\renewcommand{\=}[1]{\stackrel{#1}{=}} % for putting numbers above =

\theoremstyle{definition}

\theoremstyle{remark}

\newcommand{\fbf}{\textbf{f}}
\newcommand{\Rbf}{\textbf{R}}

\newcommand{\sbf}{\textbf{s}}
\newcommand{\ph}{\phantom}

\usepackage{xcolor}
\usepackage{lmodern}
\usepackage{listings}

\definecolor{mGreen}{rgb}{0,0.6,0}
\definecolor{mGray}{rgb}{0.5,0.5,0.5}
\definecolor{mPurple}{rgb}{0.58,0,0.82}
\definecolor{backgroundColour}{rgb}{0.95,0.95,0.92}

\lstdefinestyle{CStyle}{
    backgroundcolor=\color{backgroundColour},
    commentstyle=\color{mGreen},
    keywordstyle=\color{magenta},
    numberstyle=\tiny\color{mGray},
    stringstyle=\color{mPurple},
    basicstyle=\footnotesize,
    breakatwhitespace=false,
    breaklines=true,
    captionpos=b,
    keepspaces=true,
    numbers=left,
    numbersep=5pt,
    showspaces=false,
    showstringspaces=false,
    showtabs=false,
    tabsize=2,
    language=C
}

\biboptions{comma,square,compress}

\newcounter{bla}

\journal{Computer Physics Communications}

\usepackage[linesnumbered,ruled,vlined,boxed]{algorithm2e}
\usepackage{algpseudocode}

\begin{document}
\newcommand{\hsss}{\hspace*{2em}}
\newcommand{\var}[1]{$#1$}
\newcommand{\extraskip}{3pt}
\SetAlgorithmName{Pseudocode}{List of pseudocode names}

\begin{frontmatter}

\title{Review of the finite difference Hartree--Fock method for atoms and diatomic molecules, and its implementation in the \xtwodhf{} program}

\author[a]{Jacek Kobus\corref{author}}
\author[b]{Susi Lehtola}

\cortext[author] {Corresponding author.\\\textit{E-mail address:} jacek.kobus@umk.pl}
\address[a]{Instytut Fizyki, Uniwersytet Mikołaja Kopernika, Grudziądzka 5, 87-100 Toruń, Poland}
\address[b]{Department of Chemistry, University of Helsinki, P.O. Box 55 (A. I. Virtasen aukio 1),
FI-00014 University of Helsinki, Finland}

\begin{abstract}
  We present an extensive review of the two-dimensional finite
  difference Hartree--Fock (FD HF) method, and present its implementation in the newest
  version of \xtwodhf{}, the FD HF program for atoms and diatomic
  molecules.  The program was originally published in this journal in
  1996, and was last revised in 2013. \xtwodhf{} can be used to obtain
  HF limit values of total energies and multipole moments for a wide
  range of diatomic molecules and their ions, using either point
  nuclei or a finite nuclear model.  Polarizabilities ($\alpha_{zz}$)
  and hyperpolarizabilities ($\beta_{zzz}$, $\gamma_{zzzz}$,
  ${A}_{z,zz}$, ${B}_{zz,zz}$) can also be computed by the program
  with the finite-field method.  \xtwodhf{} has been extensively used
  in the literature to assess the accuracy of existing atomic basis
  sets and to help in developing new ones. As a new feature since the
  last revision, the program can now also perform Kohn--Sham density
  functional calculations with local and generalized gradient
  ex\-change-cor\-re\-la\-tion functionals with the Libxc library of
  density functionals, enabling new types of studies. Furthermore, the
  initialization of calculations has been greatly simplified. As
  before, \xtwodhf{} can also perform one-particle calculations with
  (smooth) Cou\-lomb, Green--Sellin--Zachor and Krammers--Henneberger
  potentials, while calculations with a superposition of atomic
  potentials have been added as a new feature.  The program is easy to
  install from the GitHub repository and build via CMake using the
  \texttt{x2dhfctl} script that facilitates creating its single- and
  multiple-threaded versions, as well as building in Libxc support.
  Calculations can be carried out with \xtwodhf{} in double- or
  quadruple-precision arithmetic.
\end{abstract}

\end{frontmatter}


% All CPiP articles must contain the following
% PROGRAM SUMMARY.

{\bf NEW VERSION PROGRAM SUMMARY}

\begin{small}
  \noindent
{\em Program Title: \xtwodhf{}}                                          \\
{\em CPC Library link to program files:} (to be added by Technical Editor) \\
{\em Developer's repository link:} \url{https://github.com/x2dhf/x2dhf} \\
{\em Code Ocean capsule:} (to be added by Technical Editor)\\
{\em Licensing provisions(please choose one):} GPLv3  \\
{\em Programming language: Fortran 95, C}                                   \\
{\em Supplementary material:}                                 \\
  % Fill in if necessary, otherwise leave out.
{\em Journal reference of previous version:}  \cite{Kobus:2013} \\
{\em Does the new version supersede the previous version?: Yes}   \\
%Only required for a New Version summary, otherwise leave out.
{\em Reasons for the new version:}\\
Code modularisation with Fortran 95, parallelisation via OpenMP and
Portable Operating System Interface threads (pthreads), support for density
functional theory using the Libxc library \cite{Lehtola2018_S_1-head},
simplified initialization of calculations, build process facilitated by
CMake and \texttt{x2dhfctl} (a Bash script), testing facilitated
by \texttt{testctl} (a Bash script) and a host of test suites.\\
{\em Summary of revisions:}\\
Code overhauled, modularised and streamlined with Fortran 95 standard,
parallelisation of the self-consistent field (SCF) process and the
successive overrelaxation (SOR) algorithm, corrected implementation of GGA
functionals and support for the Libxc library of density functionals
\cite{Lehtola2018_S_1-head}, improved initialisation of the SCF process via
HF or LDA atomic orbitals and the superposition of atomic potentials
\cite{Lehtola2019_JCTC_1593-head}, an enlarged test suite of input data and
the corresponding outputs (235 in all) and \texttt{xhf} and
\texttt{testctl} scripts to run and examine the tests. Script
\texttt{x2dhfctl} added to control the build process via CMake,
\texttt{pecctl} to automate calculations of potential energy curves and
\texttt{elpropctl} to calculate
(hyper)\-po\-lar\-is\-abil\-ities. \\
{\em Nature of problem:}\\
The program finds numerically exact solutions of the HF or Kohn--Sham
density functional equations for atoms, diatomic molecules, and their ions
by determining the lowest energy eigenstates of a given irreducible
representation and spin.  Density functional calculations can be carried
out using various exchange and correlation functionals provided by the
Libxc library \cite{Lehtola2018_S_1-head}. The program can also be used to
perform independent particle calculations with the (smooth) Coulomb,
Green--Sellin--Zachor, Krammers--Henneberger, and superposition of atomic
potentials \cite{Lehtola2019_JCTC_1593-head}, as well as two-particle HF
calculations for the harmonium atom.
\\
% Describe the nature of the problem here. \\
{\em Solution method:}\\
%Describe the method solution here.
Factoring out the analytical angular solution around the bond axis,
two-di\-men\-sion\-al numerical single-particle functions (orbitals) are
used to construct an antisymmetric many-electron wave function according to
the restricted open-shell HF or density functional theory (DFT) model. The
HF/DFT equations are written as coupled two-di\-men\-sion\-al second-order
(elliptic) partial differential equations (PDEs), which are discretized by
an eighth order central difference stencil on a two-di\-mensional grid,
whereas quadrature is performed with a Newton--Cotes rule.  The resulting
large and sparse system of linear equations are solved by the (multicolour)
successive overrelaxation ((MC)SOR) method, and the orbitals and potentials
are solved by simultaneous SOR iterations on the corresponding Poisson
equations.  The convergence of the SCF procedure is monitored with that of
the orbital energies and normalisation factors.  The precision of the
obtained solutions depends on the grid and the system under consideration,
and one can typically obtain orbitals that yield total and orbital energies
with up to 12 significant figures using double precision arithmetic. If
more precise results are needed, \xtwodhf{} can also be compiled in
quadruple precison floating-point arithmetic.
\\
{\em Additional comments including restrictions and unusual features:}\\
CMake (ver. 3) and gfortran/ifort compiler are required to compile and
build the program. The incomplete gamma function is needed to evaluate
hydrogenic orbitals and its values are calculated by means of the
\texttt{dgamit.F} subroutine written by Fullerton
\cite{fullerton:1977-head} which uses FORTRAN 90 versions of
\texttt{d1mach} and \texttt{i1mach} functions.

\end{small}

\makeatletter
\renewcommand{\l@section}{\@dottedtocline{1}{1.5em}{2.6em}}
\renewcommand{\l@subsection}{\@dottedtocline{2}{4.0em}{3.6em}}
\renewcommand{\l@subsubsection}{\@dottedtocline{3}{7.4em}{4.5em}}
\makeatother

\tableofcontents

\section{Introduction \label{sec:introduction}}

A great deal of effort has been spent over the last 50 years on the
development of computational methods to model the electronic structure of
atoms and molecules. Thanks to the resulting improvements in the ease of
use and accuracy of these models, electronic structure calculations are
nowadays routine, and a significant part of the computing power available
to the scientific community is used to extract atomistic understanding of
the physical and chemical behaviour of molecular and solid state systems
with a range of methods.

Practically all ab initio electronic structure calculations start out by
solving the Hartree--Fock (HF) or Kohn--Sham \cite{Kohn1965_PR_1133}
equations for the molecular orbitals.  Naturally, these unknown functions
have to be discretized in some way to allow for a computational solution.
In mainstream computational chemistry, the molecular orbitals are typically
expressed as linear combinations of atomic basis functions, since this
makes modeling systems of any composition and geometry straightforward, and
the resulting method easily scales to calculations on large systems.

However, atomic basis sets are usually far from complete.  Accordingly, any
computed properties suffer from basis set truncation errors, which can be
difficult to assess and control.  Atomic-orbital basis sets do have a major
benefit here, in that they tend to benefit from systematic error
cancellation for many types of observables, such as reaction or excitation
energies and other types of molecular properties. In specific, as errors
made in the energetically important core region will be similar across
geometries and electronic states, they cancel out when computing such
observables.

To work around the limitations posed by basis set truncation error, atomic
basis sets typically come in families that span various sizes, commonly
ranging from split-valence or double-$\zeta$ quality to polarized
triple-$\zeta$ or polarized quadruple-$\zeta$ quality, some families going
even further to quintuple-$\zeta$, hextuple-$\zeta$, or beyond. The access
to different sized basis sets allows users to find the sweet spot in cost
and accuracy for their application.  Several families of atomic orbital
basis sets have been developed over many decades in order to make
calculations of various systems and properties feasible and credible
\cite{Huzinaga1985_CPR_281, Davidson1986_CR_681, Hill2013_IJQC_21,
  Jensen2013_WIRCMS_273}.

Typical basis set families also offer further variants that have been
augmented with diffuse functions for modeling extended electronic
states, such as those found in many anionic species. Although these
functions typically have little effect on total energies of neutral
species in their ground state, they are also often important for the
reliable modeling of electronic excited states, which may even require
the addition of several sets of diffuse functions to obtain a
converged result. Similarly, reliable modeling of the response of the
ground state to an external electric field also usually requires the
inclusion of diffuse functions, and multiple augmentation may again be
necessary to reach a numerically converged result.

As the quality of a computational solution critically depends on two
things---the discretization error and the error inherent in the
computational model itself---it is important to separate these two when
examining and developing novel discretizations of the electronic structure
problem for a given level of theory. Knowing the right result---the
complete basis set (CBS) limit---for the studied quantum chemical model is
key to the design of new discretizations, or families of basis sets that
approach the CBS limit in a systematic and error-balanced sequence.

The design of atomic orbital basis sets therefore often starts from
establishing fully numerical reference values at the CBS limit.  Comparison
to fully numerical reference values enables one to assess the accuracy of
the designed approximate atomic basis sets, and this knowledge is useful
for designing cost-balanced basis sets.

As we will shortly review, fully numerical calculations on atoms and
diatomic molecules have been possible for a long time, and an extensive
review of the topic has been recently published
\cite{Lehtola2019_IJQC_25968}. In the following, we will go over the key
studies leading to the approach used in the \xtwodhf{} program for fully
numerical calculations on atoms and diatomic molecules; for further
references and other approaches, we invite the reader to read the
discussion in \citeref{Lehtola2019_IJQC_25968}. We will also undertake a
review of further related literature later on in this work (see
\cref{sec:review}).

The proper description of the nuclear Coulomb cusp is key to the
numerical accuracy of any electronic structure method, as most of the
total energy of an atom can be found close to the nucleus.  Atoms
feature a high degree of symmetry, and the polar spherical coordinate
system allows for an efficient handling of atomic problems, as the
one-electron solutions can be written as a product of a radial
function with an analytic angular solution in terms of spherical
harmonics. The radial problem is straightforward to solve by numerical
methods, since the singular Coulomb potential is regularized by the
$r^2$ factor in the volume element of the coordinate system, and the
resulting one-dimensional problem is facile to solve to high accuracy.
As a result, fully numerical calculations on atoms were feasible
already in the 1950s \cite{Froese1957_MNRAS_615}.

Riding on the success of the atomic approach, there were attempts in
the early 1960s to solve the HF problem for molecules with one-center
expansions \cite{AlbasinyC:1961, AlbasinyC:1963, AlbasinyC:1965,
  AlbasinyC:1966, Keefer:1969}.  However, as polyatomic calculations
lack the symmetry of the atomic problem, the off-center nuclear
Coulomb singularities are not killed off by the $r^2$ Jacobian, and as a
result, the one-center expansions converge extremely slowly with
respect to the angular expansion, which is where the off-center
nuclear Coulomb cusp problem lies.  Alternative avenues for polyatomic
molecules are thereby needed.

Like all linear molecules, diatomic molecules have cylindrical
symmetry that allows one to factor out the ``angular'' part of the
problem and solve it analytically
\cite{Lehtola2019_IJQC_25968}. However, what makes diatomic molecules
special is that they can be fully described within the prolate
spheroidal coordinate system, where there are no issues with nuclear
cusps: the volume factor in this coordinate system turns out to be
proportional to ${\rm d}V \propto r_A r_B$, which again regularizes
the singularities in the Coulomb nuclear attraction terms $-Z_A /
r_A^{-1}$ and $-Z_B / r_B^{-1}$ that are the bane of general
three-dimensional approaches for atoms and molecules
\cite{Lehtola2019_IJQC_25968}. This elimination of the nuclear cusps
in the prolate spheroidal coordinates enables facile numerical
approaches to the diatomic problem, as well, leading to a
two-dimensional problem instead of the one-dimensional radial problem
found in atoms.

Let us now delve a bit deeper. Let us place nuclei A and B in
Cartesian coordinates along the $z$ axis at points ${\bf R}_A =
(0,0,-R/2)$ and ${\bf R}_B = (0,0,+R/2)$, $R$ being the internuclear
distance. The prolate spheroidal coordinates are then given by the
``radial'' coordinate
\begin{equation}
  \xi=(r_{A}+r_{B})/R, \label{eq:xi}
\end{equation}
the ``relative'' coordinate
\begin{equation}
  \eta =(r_{A}-r_{B})/R, \label{eq:eta}
\end{equation}
and the azimuthal angle $\theta$ ($ 0\leq \theta \leq 2\pi$) measured
around the $z$ axis, where $r_A=|{\bf r}-{\bf R}_A|$ and $r_B = |{\bf
  r} - {\bf R}_B|$ are the distances of a given point ${\bf r}$ from
the two nuclei.

McCullough made the first successful numerical attempt to solve the
(multi-configuration) HF equations for diatomic molecules by the
so-called partial-wave self-consistent-field method (PWSCF)
\cite{McCullough:1974, McCullough:1986} in this coordinate system.  As
a result of the cylindrical symmetry, the molecular orbitals and the
corresponding Coulomb and exchange potentials can be expressed in the
form $f(\eta,\xi)e^{im\theta}$, where $m$ is an integer, and the
three-dimensional HF equations for diatomic molecules reduce to a set
of two-di\-mensio\-nal problems defined by the functions $f(\eta,\xi)$
for each $m$.

In the PWSCF method, the function $f$ is expanded in $\eta$ in
terms of an orthonormal polynomial basis set,
\begin{equation}
  f (\eta,\xi) = \sum_{l=m}^{l_{\rm max}} X^{ml}(\xi) P_l^{m}(\eta) \label{eq:pwscfexp}
\end{equation}
where $P_l^{m}(\eta)$ are associated Legendre polynomials.  McCullough
then solved the unknown functions $X^{ml}(\xi)$ with finite
differences on a grid.

Since the expansion in \cref{eq:pwscfexp} is in terms of Legendre
polynomials while $X^{ml}(\xi)$ was solved on a grid, McCullough
referred to this method as \textit{semi-numerical}. However, the
expansion in $P_l^{m}$ is in principle a completely valid fully
numerical technique, which typically rely on the use of orthonormal
polynomials: calculations can be straightforwardly converged to the
CBS limit by studying larger and larger values for the truncation
parameter $l_{\rm max}$. Of course, also the finite difference grid
for $\xi$ needs to be converged to reach the CBS limit.

Following McCullough, Becke developed a numerical approach for solving
density functional equations for $f$ in a basis of cubic polynomial
splines on a two-di\-mensional grid in the early 1980s \cite{Becke:1982, Becke:1983,
  Becke:1985, Becke:1986}.  Importantly, Becke proposed using a
transformed coordinate system \cite{Becke:1982}: employing the
$(\nu,\mu,\theta)$ system of coordinates given by
\begin{equation}
  \nu=\cos^{-1} \eta \label{eq:nu}
\end{equation}
and
\begin{equation}
  \mu=\cosh^{-1}(\xi) \label{eq:mu}
\end{equation}
instead of the $(\eta, \xi, \theta)$ coordinates originally employed by
McCullough, exponential functions centered at the nuclei become Gaussians,
and molecular orbitals have a definite parity in $\nu$.  A combination of
this coordinate system, the idea of the original PWSCF approach, and
high-order finite elements for the $X^{ml}(\mu)$ expansion has been
recently discussed by Lehtola \cite{Lehtola2019_IJQC_25944}, and is
available in the \helfem{} program \cite{helfem}.

Laaksonen, Pyykk\"o, and Sundholm joined the effort on fully numerical
solutions of diatomic molecules in the early 1980s.  Following
proof-of-concept work \cite{Laaksonen1983_IJQC_309,
  Laaksonen1983_IJQC_319}, Laaksonen, Pyykk\"o, and Sundholm quickly
found the coordinate system proposed by Becke to be useful in a number
of studies at the HF and multiconfiguration self-consistent field
levels of theory \cite{Laaksonen1983_CPL_1, Laaksonen1984_CPL_573,
  Sundholm1984_CPL_1} (see \citeref{Lehtola2019_IJQC_25968} for
further references).  Laaksonen, Pyykk\"o, and Sundholm chose to
employ two-dimensional finite differences for the solution of
$f(\nu,\mu)$.  The second-order partial differential equations for the
orbitals and potentials were discretized by means of the sixth-order
cross-like stencil, and the ensuing large and sparse systems of linear
equations were solved by the iterative successive
overrelaxation (SOR) method \cite{LaaksonenPS:1986}. This approach
will be referred to as the finite difference HF (FD HF) method in the
remainder of this work.

The first author got interested in the FD HF method in the late 1980s
and has been involved in its development and applications ever since
\cite{Kobus:1993, Kobus:1994, Kobus:1997}, taking over the work begun
by Laaksonen, Pyykk\"o, and Sundholm.  An improved version of the FD
HF program was announced in 1996 \cite{KobusLS:1996}, and it was again
revised in 2013 \cite{Kobus:2013}. In this work, we present the
current version (3.0) of the \xtwodhf{} program; this work is thus an
update to the two earlier papers describing previous versions of the
program \cite{Kobus:2013, KobusLS:1996}.

In short, the content of the paper is as follows. We
begin with a thorough discussion of the theory behind \xtwodhf{}, and
present the latest revisions to the method: various improvements have
resulted in a considerable increase in its efficiency.  We then
present the internal organization of the program; the refactorings
performed in the code allow for further development to take place
wherever necessary. Next, we review the literature applications of the
program, and then demonstrate its capabilities with some example
calculations: the method can now be routinely and confidently applied
to medium-size diatomic molecules\ie{} systems with 35--45 electrons.

In specific, the layout is the following.
The introduction is followed by a general description of
the restricted open-shell HF method for diatomic molecules in
\cref{sec:theory}, which can also be used to solve the Kohn--Sham
equations. Also model potentials, such as the Green--Sellin--Zachor
and superposition of atomic potentials, as well as the
Kramers--Henneberger potential can be modeled within similar
approaches by omitting the dielectronic terms. The harmonium atom can
be approached with minor modifications to the equations, as well.

Having written down the Fock and Kohn--Sham equations in \cref{sec:theory},
the solution of such elliptic partial differential equations (PDE) by the
SOR method is discussed in \cref{sec:solvingPDEs}.  The accuracy of the
discretization of the PDEs is analysed on calculations on a model Poisson
equation. The handling of the proper boundary conditions is described, and
the numerical stencil used to relax the solution at every inner grid point
is presented.  The SOR method and its multi-colour variant (MCSOR) are
discussed within the context of the solution of the self-consistent field
(SCF) equations in terms of macro- and micro-iterations, and the update
procedure of the boundary values is also discussed in this context.  The
crucial problem of choosing the overrelaxation parameter for orbitals and
potentials is analysed in detail, as it is critical for the efficiency of
the resulting FD HF method.

The subsequent \cref{sec:program-description} describes the \xtwodhf{}
program itself.  Pseudocode is used to illustrate the most important
routines. The evaluation of first and second derivatives over the $\nu$ and
$\mu$ variables is cast into a matrix-times-vector form which can be
performed efficiently on modern hardware. The script used to build the
program with OpenMP and Portable Operating System Interface thread
(pthread) support is also presented, and the efficiency of the
parallelisation of the SCF process is discussed.

Having described the theoretical foundations and the program's
structure, related literature complementing the recent review
\cite{Lehtola2019_IJQC_25968}, such as literature applications of the
\xtwodhf{} program, are discussed in \cref{sec:review}.
\Cref{sec:example-results} contains some example results obtained
with \xtwodhf{} for the harmonium atom and Ar-C in the highly repulsive
region.  In addition, \cref{sec:example-results} also contains tests
of the Libxc interface for various exchange and correlation
functionals on closed-shell atoms, as well as plots of the kinetic
potential and its ingredients for the FH molecule at the HF and local
spin density (LDA) levels of approximation.  The article concludes in
\cref{sec:conclusions}.  Hartree atomic units are used throughout the
text, unless specified otherwise.  Further details are discussed in
the appendices (\crefrange{sec:nuclear-models}{sec:appendix-initArrays}).

\section{Problem formulation \label{sec:theory}}
\subsection{The restricted open-shell HF method \label{sec:rohf}}

\subsubsection{General formulation \label{sec:rohf-formulation}}
Let us study HF calculations for an open-shell diatomic
molecular system. We thus approximate the solution of the
Schr\"odinger equation with a single Slater determinant ansatz
\begin{equation}
  \Phi=\frac{1}{\sqrt{M!}}\det|\phi_1(1),\phi_2(2),\ldots,\phi_M(M)|, \label{eq:slaterdet}
\end{equation}
which corresponds to having the electrons occupy the $M$ lowest-lying
spin-orbitals $\phi_{a}=\phi_{a}({x,y,z},\sigma)$.  The HF
wave function is found by minimizing the energy functional obtained
as the expectation value of the Hamiltonian
\begin{equation}
E[\Phi] = \langle \Phi |H| \Phi \rangle= \left \langle \Phi \left| -\sum_{a} \frac{1}{2}\nabla_{a}^2
- \frac{Z_A}{r_{aA}} - \frac{Z_B}{r_{aB}} + \sum_{a<b} \frac{1}{r_{ab}} +
\frac{Z_{A} Z_{B}}{R} \right| \Phi \right \rangle \label{eq:totenergy}
\end{equation}
with respect to the orbitals, with the added constraint that the
orbitals be orthonormal. This procedure leads to the HF equations (also known as Fock equations) for
the occupied orbitals: they are a set of $M$ coupled one-particle
integro-differential equations of the form
\cite{Kobus:2013, KobusLS:1996, Kobus:2000}
\begin{equation}
{F}_{a} \phi_{a} = \sum_{b=1}^M \varepsilon_{ab} \phi_{b} ,\,\,\;\; {a}=1,\ldots,{M}, \label{eq:fock1a}
\end{equation}
where $\varepsilon_{ab}$ are the Lagrangian multipliers that were
introduced to enforce orbital orthonormality in the variation.
The Fock equations for the orbitals are written in full as
\begin{equation}
-\frac{1}{2}\nabla^{2}\phi_{a} =
-\left(- \frac{{Z}_{A}}{{r}_{aA}} - \frac{{Z}_{B}}{{r}_{aB}}
+\sum_{b}\left({V}_{C}^{b}-{V}_{x}^{ab}\right)
-\varepsilon_{a}\right)\phi_{a}
+ \sum_{b \neq a}\varepsilon_{ab}\phi_{b}, \label{eq:fock1}
\end{equation}
where the electron-electron Coulomb $V_C$ and
exchange $V_x$ potentials can be determined from orbital densities and orbital products,
respectively, by solving the Poisson equations
\begin{align}
\nabla^{2} {V}_{C}^{b} = -4\pi \phi_{b}^{*}\phi_{b}^{}, \label{eq:coulomb} \\
\nabla^{2} {V}_{x}^{ab} = -4\pi \phi_{a}^{*}\phi_{b}^{}. \label{eq:exchange}
\end{align}

\subsection{Solution by relaxation of Poisson equations \label{sec:poisson}}
The Fock equations for the orbitals (\cref{eq:fock1}) are solved by the
iterative SCF procedure, which is started from a suitable initial
guess. The right-hand side of \cref{eq:fock1} is known in each SCF
iteration, as it is determined by the current best estimate for the
orbitals; the Coulomb and exchange potentials are also determined by the
same orbitals as solutions to the Poisson equations of
\cref{eq:coulomb,eq:exchange}, respectively.

Note that \cref{eq:fock1} is a Poisson equation, as well.  This means that
one only needs to iteratively solve sets of Poisson equations to solve SCF.
Let us furthermore note that when the density changes considerably between
SCF iterations, exact solutions of \cref{eq:fock1,eq:coulomb,eq:exchange}
are not necessary, as approximate solutions to these equations are
sufficient to approach the SCF solution at a similar rate as if numerically
exact solutions of \cref{eq:fock1,eq:coulomb,eq:exchange} were used,
instead: after all, the right-hand side of these equations is not
accurately known at this stage anyway, since it needs to be found by the
SCF procedure.

Thus, when the SCF process is far from convergence, approximate solutions
to the Poisson equations for the orbitals and potentials are
sufficient. Then, as the SCF approaches convergence, the orbitals and
potentials undergo smaller and smaller changes. And in this case, an
iterative solution of the Poisson equations will also converge relatively
quickly onto the exact solution.

The special feature of the FD HF method is that the SCF iterations are
interwoven with a number of SOR relaxation sweeps to improve the solutions
to the Poisson equations for the potentials and the orbitals (see
\cref{sec:interweave} for in-depth discussion).
An analysis of numerical complexity showed the FD method to be about
ten-fold more efficient than analogous partial-wave self-consistent-field and
finite-element methods due to this interweaving \cite{Kobus:2000}.

\subsubsection{Angular and radial discretization \label{sec:angrad-discretization}}
As was discussed in the Introduction, the prolate spheroidal coordinate system defined by \cref{eq:nu,eq:mu}
is attractive for discretizing the electronic structure of diatomic molecules:
not only does the nuclear Coulomb cusp not pose a problem in this coordinate system,
but the choice of the $(\nu,\mu)$ coordinates also guarantees
that $\phi_a$ is a quadratic function of $\mu$
and $\nu$ in the vicinity of the $z$ axis.\footnote{$\mu =0$ corresponds to
  the Cartesian coordinates $(0,0,-R/2 \le z \le R/2)$, $\nu=0$ to
  $(0,0,z \ge R/2)$ and $\nu=\pi$ to $(0,0,z \le -R/2))$; see
\cref{sec:symmprop} for details.} 

As was also mentioned in the Introduction, the cylindrical symmetry of
diatomic systems allows for handling the $\theta$ part of the orbitals
and the potentials analytically by expressing them in the factorized form
\begin{equation}
\left.
\begin{tabular}{c}
$\phi_{{a}}$ \\[0.2cm]
${V_{C}^a}$ \\[0.2cm]
${V_x^{ab}}$
\end{tabular} \right\} = {f(\nu,\mu)e^{im\theta}}. \label{eq:ffunct}
\end{equation}
In this equation, $m$ defines the rotation symmetry of the orbital or the
potential: for example, $\sigma$, $\pi$, $\delta$ and $\varphi$ orbitals
correspond to $m=0$, $m=\pm 1$, $m=\pm2$, and $m=\pm3$, respectively.
Orbitals of higher symmetry than $\varphi$ are not relevant for the ground
states of ordinary diatomic molecules at the HF level of theory.

Since the angle $\theta$ does not appear in the Hamiltonian, it is known
that the orbitals with $m=\pm|m|$ have the same radial part in exact
theory.  Although this symmetry can be broken in HF as well as in DFT, in
\xtwodhf{} the orbitals with $m={\pm} |m|$ are handled as same-shell
orbitals, that is, the same $f(\nu,\mu)$ expansion is used for both.

The orbitals are also spin-unpolarized, which is why the resulting method
is referred to as restricted open-shell HF (ROHF), which reduces to
restricted HF (RHF) for a closed-shell configuration.  As a result, while
$\sigma$ orbitals can fit up to 2 electrons, other types of orbitals can
fit up to 4 electrons, since they consist of two separate spatial orbitals
for ${\pm} |m|$.

We note that spin-polarized as well as broken-symmetry solutions can be
targeted with \helfem{} \cite{Lehtola2019_IJQC_25944}, which supports
calculations with unrestricted HF as well as DFT calculations with LDA,
generalized gradient approximation (GGA), and meta-GGA functionals.

\subsubsection{Working equations \label{sec:working}}
Now, the total energy expression for the ROHF
method reads
\begin{equation}
\label{eq:eqds1}
E = \sum_{a} q_a \left\langle \phi_a \left| -\frac{1}{2}\nabla^2+V_n \right| \phi_a \right\rangle  +
\sum_{a,b} U_{ab} \left\langle \phi_a \left| V_C^b \right| \phi_a \right\rangle  -
\sum_{a,b} W_{ab} \left\langle \phi_a \left| V_x^{ab} \right| \phi_b \right\rangle
\end{equation}
where $V_n=-Z_A/r_{A}-Z_B/r_{B}$ is the nuclear potential energy operator,
$q_a$ is the occupation number for orbital $a$, and $U_{ab}$ and $W_{ab}$
are the corresponding occupation-number-dependent factors for the Coulomb
and exchange energy contributions. Their values are configuration dependent,
and are determined by the Slater--Condon rules for evaluating the expectation
values of the two-particle operators with the single Slater determinant wave function.

In the transformed prolate spheroidal coordinates
$(\nu,\mu,\theta)$, the ``radial'' part of the Laplacian reads
\begin{equation}
{4 \over R^2(\xi^2-\eta^2)}
\left\{{\partial^2 \over \partial\mu^2} +{\xi \over \sqrt{\xi^2-1}} {\partial \over \partial \mu}
+{\partial^2 \over \partial\nu^2}
+{\eta \over \sqrt{1-\eta^2}} {\partial \over \partial \nu}
-m^2_a u(\eta,\xi) \right\} \label{eq:laplacian}
\end{equation}
where we have introduced the helper function
\begin{equation}
  u(\eta,\xi)=\frac {1} {\xi^2-1} + \frac {1} {1-\eta^2}. \label{eq:u}
\end{equation}
Therefore, by multiplying the Fock equation (\cref{eq:fock1}) by $- R^2(\xi^2-\eta^2)/2 = - r_A r_B / 2$, we arrive
at the working equation for the spatial part of orbital $a$, $f_a(\nu,\mu)$,
in the transformed prolate spheroidal coordinates
\begin{align}
\Bigg\{ {\partial^2 \over \partial\mu^2}
+{\xi \over \sqrt{\xi^2-1}} {\partial \over \partial \mu}
& +{\partial^2 \over \partial\nu^2}
+{\eta \over \sqrt{1-\eta^2}} {\partial \over \partial \nu}
-m^2_a u(\eta,\xi) + v(\eta,\xi) \nonumber \\
& \left. - \frac{R}{\xi} (\xi^2-\eta^2)\tilde{V}_C
+ \frac{R^2}{2}(\xi^2-\eta^2)\varepsilon_a \right\} f_a(\nu,\mu) \nonumber\\
 & +\frac{R}{\xi}(\xi^2-\eta^2)\left(\tilde{V_x}^a
+ \frac{R\xi}{2} \sum_{b\ne a} \varepsilon_{ab} f_b(\nu,\mu)\right) = 0
\label{eq:eq14}
\end{align}
where
\begin{equation}
  v(\eta,\xi)=R[(Z_A+Z_B)\xi+(Z_B-Z_A)\eta] \label{eq:v}
\end{equation}
is the nuclear potential term whose singularities at both nuclei have
been cancelled out by the choice of the coordinate system, and the
modified Coulomb (${\tilde{V}}_C$) and exchange potentials
(${\tilde{V}}_x^a$) are defined as
\begin{align}
\tilde{V}_C &= \sum_a \tilde{V}_C^a =\sum_a \frac{R\xi}{2} V^a_C, \label{eq:eq13}\\
\tilde{V}_x^a &= \sum_{b\ne a} \tilde{V}_x^{ab}f_b(\nu,\mu)=
\sum_{b\ne a} \frac{R\xi}{2} V_x^{ab}f_b(\nu,\mu). \label{eq:eq15}
\end{align}
The diagonal orbital energy parameters $\varepsilon_{a}$ in \cref{eq:eq14} are calculated as
\begin{equation}
\varepsilon_a = \langle \phi_a | h_a | \phi_a \rangle = \left\langle \phi_a \left| -\frac{1}{2}\nabla^2+V_n
+\frac{2}{R\xi}(\tilde{V}_C-\tilde{V}_x^a) \right| \phi_a \right\rangle, \label{eq:eq10}
\end{equation}
while the off-diagonal parameters $\varepsilon_{ab}$ are obtained as
\begin{equation}
\varepsilon_{ab} = \frac{q_b}{q_b + q_a}
\left( \langle \phi_b | h_a | \phi_a \rangle + \langle \phi_a | h_b | \phi_b \rangle \right), \label{eq:eq10a}
\end{equation}
where $q_a$ and $q_b$ are again the occupation numbers of orbitals $a$
and $b$.  The formulae for evaluating the kinetic and nuclear
potential energy terms as well as the Coulomb and exchange energy
contributions in \cref{eq:eq10} are given in
\cref{sec:appendix-integrals}.

We note here that \xtwodhf{} program can also perform calculations
with finite nuclear models, wherein the nuclear charge distribution is
described by either a Gaussian or Fermi distribution instead of a
point nucleus. As the potentials of such spherically symmetric
nuclear charge distributions can be written in the form
\begin{equation}
V(r)=-\frac {Z(r)} {r}, \quad r>0, \label{eq:vpot}
\end{equation}
the finite-nucleus implementation is as simple
as replacing $Z_A \to Z_A(r_A)$ and $Z_B \to Z_B(r_B)$ in
\cref{eq:v}. The parameters of the finite nuclear models have been
taken from the table of atomic masses of Wapstra and Audi
\cite{WapstraA:1983, AudiW:1993} (see \cref{sec:nuclear-models} for
details on the implementation).

The working equation for the exchange potentials is obtained in an
analogous manner to the Fock equation. The Poisson equation
(\cref{eq:exchange}) reads in the transformed prolate spheroidal
coordinates as
\begin{align}
\Bigg\{ \frac{\partial^2}{\partial\mu^2}
+\left(\frac{1}{\sqrt{\xi^2-1}}
-\frac{2\sqrt{\xi^2-1}}{\xi}\right) \frac{\partial}{\partial \mu}
& +\frac{\partial^2}{\partial\nu^2}
+\frac{\eta}{\sqrt{1-\eta^2}}\frac{\partial}{\partial \nu} -\frac{2}{\xi^2} \nonumber \\
\left. \phantom{-\frac{2}{\xi^2}} -(m_a-m_b)^2u(\eta,\xi)\right\}\tilde{V}^{ab}_x
= & -\frac{\pi R^3}{2}\xi(\xi^2-\eta^2)
f_a(\nu,\mu)f_b(\nu,\mu) \label{eq:eq16}
\end{align}
and is solved by overrelaxation, like the Poisson equation for the orbitals
(\cref{eq:eq14}).

It can be seen from \cref{eq:coulomb,eq:exchange} that the expression for
the Coulomb potential can be obtained from that for the exchange by setting
$a=b$. The expression is thus obtained from \cref{eq:eq16} by setting $a=b$
and removing the vanishing term, obtaining
\begin{align}
\Bigg\{ \frac{\partial^2}{\partial\mu^2}
+\left(\frac{1}{\sqrt{\xi^2-1}}
-\frac{2\sqrt{\xi^2-1}}{\xi}\right)\frac{\partial}{\partial \mu}
& +\frac{\partial^2}{\partial\nu^2}
+\frac{\eta}{\sqrt{1-\eta^2}}\frac{\partial}{\partial \nu} -\frac{2}{\xi^2}\Bigg\}\tilde{V}^{a}_C \nonumber \\
& = -\frac{\pi R^3}{2}\xi(\xi^2-\eta^2)
f_a^2(\nu,\mu). \label{eq:eq17}
\end{align}
We note here in passing that the variational solution of
Poisson's equation in the prolate spheroidal coordinate system has
been discussed by Csavinszky \cite{Csavinszky1986_IJQC_305}.

\subsection{DFT method \label{sec:dft}}

The method described in the previous subsection can be tailored to
solve a few other problems of interest. In particular, \cref{eq:fock1}
can adapted for solving the Kohn--Sham \cite{Kohn1965_PR_1133}
equations of density functional theory (DFT) \cite{Kohn1965_PR_1133,
  Hohenberg1964_PR_864}
\begin{equation}
-\frac{1}{2}\nabla^{2}\phi_{a} =
-\left(- \frac{{Z}_{A}}{{r}_{aA}} - \frac{{Z}_{B}}{{r}_{aB}}
+\sum_{b}{V}_{C}^{b}-v_{xc}^{\sigma}[n_{\alpha},n_{\beta}]
-\varepsilon_{a}\right)\phi_{a}
+ \sum_{b \neq a}\varepsilon_{ab}\phi_{b} \label{eq:ks1}
\end{equation}
where $v_{xc}^{\sigma}[n_{\alpha},n_{\beta}]$ is the exchange–correlation
(xc) potential.

\xtwodhf{} has long supported \cite{KobusLS:1996} calculations within the
so-called Hartree--Fock--Slater (HFS) model \cite{Slater1951_PR_385}, but
the details of implementation have not been described in the previous
publications \cite{Kobus:2013, KobusLS:1996}. Since the HFS model may not
be familiar to most readers, we point out that it is obtained from
Hartree--Fock theory by replacing the exact exchange term with Slater's
$X \alpha$ functional \cite{Slater1951_PR_385}. However, as the $X \alpha$
functional is just scaled LDA exchange \cite{Bloch1929_ZfuP_545,
  Dirac1930_MPCPS_376}, the HFS model can be thought of as a variant of
exchange-only LDA.

Importantly, the HFS implementation in \xtwodhf{} employs an
atom-specific weighting of the $X \alpha$ functional that is optimized
to lead to the best agreement with Hartree--Fock total energies
\cite{Schwarz1972_PRB_2466}. The parameter corresponding to the
heavier atom in the molecule is chosen for the calculation in
\xtwodhf{}. The original aim of such a system dependent functional was
to provide a faster alternative to Hartree--Fock, since there is no
need to relax exchange potentials in this method. However, this
system-dependent HFS model is obviously not a good starting point for
computing binding energies $\Delta E_\text{XY} = E(\text{X}) +
E(\text{Y}) - E(\text{XY})$, for example, as there will be no
systematic error cancellation if two different functionals are used to
evaluate the terms.

Although the LDA exchange functional itself \cite{Bloch1929_ZfuP_545,
  Dirac1930_MPCPS_376}---whose definition is not system dependent---is
also available in \xtwodhf{}, many density functional approximations
whose definition is not system dependent like the HFS model and that
are also more accurate than the LDA have been developed in recent
decades.  As a new feature in \xtwodhf{}, the xc potential can be
evaluated with Libxc \cite{Lehtola2018_S_1} from the total densities
for the spin-up ($\alpha$) and spin-down ($\beta$) electrons
\begin{equation}
n_\alpha =\sum_{b} q_b \phi_b^{*\alpha}\phi_b^{\alpha}, \quad \quad \quad
n_\beta  = \sum_{b} q_b \phi_b^{*\beta} \phi_b^{\beta},
\end{equation}
respectively, for self-consistent calculations with LDA or GGA
functionals. While LDAs require no special consideration, GGA
functionals introduce dependence on the density gradients $\nabla
n_\alpha$ and $\nabla n_\beta$, which needs to be taken into account
when solving \cref{eq:ks1}; the relevant details are discussed in
  \cref{sec:appendix-libxc}.
Note that although some GGA functionals were included in the previous
version of the program \cite{Kobus:2013}, the gradient dependence of
the potential was not considered, and as a result, GGA results
obtained with old versions of the program do not reproduce the correct
ground state.

Finally, although \cref{eq:ks1} has been written in the form for
semi-local functionals, global hybrid functionals such as B3LYP
\cite{Stephens1994_JPC_11623}---which also include a fraction of exact
exchange---are also supported in \xtwodhf{}, following the methodology
already presented above in \cref{sec:working}.

We will verify and exemplify the DFT implementation and Libxc
interface below in \cref{sec:libxc} for calculations with various
semi-local and hybrid LDA and GGA functionals.

\subsection{Atomic model potentials \label{sec:model-potentials}}

Independent particle model potentials are often useful for
understanding the arrangement of energy levels in diatomic
molecules. Calculations with such potentials avoid the need to solve
SCF equations, instead obtaining orbitals and orbital energies from
simple model potentials that reproduce the atoms' electronic
shell structure. While such models do not yield estimates for total
energies due to the lack of interatomic Pauli repulsion effects, they do tend to
yield quite reasonable orbitals and orbital energies.

In the model potential approach, the Hamiltonian of
\cref{eq:totenergy} is replaced by the following one-electron Hamiltonian
\begin{equation}
E[\Phi] = \langle \Phi |H| \Phi \rangle= \left \langle \Phi \left| -\sum_{a} \frac{1}{2}\nabla_{a}^2
- \frac{Z^{\rm eff}_A(r_{aA})}{r_{aA}} - \frac{Z^{\rm eff}_B(r_{aB})}{r_{aB}} \right| \Phi \right \rangle \label{eq:modelenergy}
\end{equation}
where the effective nuclear charges now depend on the radial distance
from the nuclei (cf. \cref{eq:vpot}).
The implementation of this model within the FD HF
approach is extremely straightforward, as there is no need to solve
for Coulomb or exchange potentials. Instead, the value of the one-particle
potential can straightaway be tabulated on the $(\nu,\mu)$ grid.

In the late 1960s, Green, Sellin, and Zachor (GSZ) \cite{Green1969_PR_1,
  Green1973_AQC_221} proposed a simple radial potential for
phenomenological studies
\begin{equation}
  Z_A^{\rm eff}(r) = (Z_A-1)\Omega(r) +1 \label{eq:gsz}
\end{equation}
that was found to yield good agreement with experiment in early
studies on diatomic molecules \cite{Whalen1972_AJP_1484,
  Miller1974_JCP_2617, Sawada1974_PRA_1130}.
The function $\Omega$ in \cref{eq:gsz} is given by
\begin{equation}
  \Omega(r) = \frac 1 {H(\exp(r/d)-1) +1} \label{eq:gsz-omega}
\end{equation}
and fitting to atomic HFS calculations led to
\begin{equation}
  H = d \alpha (Z-1)^{\nu} \label{eq:gsz-H}
\end{equation}
with $\nu=0.4$ and $\alpha \approx 1.05$ \cite{Green1969_PR_1}. The $d$
parameter in \cref{eq:gsz-H} is atom-specific, and suitable values have
been determined by optimizing the HF energy of the resulting wave function
\cite{Green1969_PR_1}. Support for the GSZ potential in \xtwodhf{} was
already briefly mentioned in the previous publication \cite{Kobus:2013}. As
setting $\alpha=1$ for simplicity causes no significant loss in accuracy
\cite{Green1969_PR_1}, this choice has been made in the \xtwodhf{}
implementation.

As a new feature, \xtwodhf{} can now also be used to solve orbitals arising
from the superposition of atomic potentials \cite{Lehtola2019_JCTC_1593,
  Lehtola2020_JCP_144105}. The idea of this scheme is similar to that of
the GSZ approach: instead of employing a potential with fixed analytic form
as in \cref{eq:gsz,eq:gsz-omega,eq:gsz-H}, self-consistent radial
potentials $V(r)$ that contain classical Coulomb, exchange, as well as
correlation effects are instead determined in atomic calculations at the
CBS limit, and then
\begin{equation}
  Z^{\rm eff}(r) = - rV(r) \label{eq:sap}
\end{equation}
is computed and tabulated numerically for future use. The tabulated
atomic potentials in terms of $Z^{\rm eff}(r)$ come from exchange-only
LDA \cite{Bloch1929_ZfuP_545, Dirac1930_MPCPS_376} calculations in
\helfem{} \cite{Lehtola2019_IJQC_25945, Lehtola2020_PRA_12516}, as
this level of theory is simple while yielding excellent agreement with
the optimized effective potential method \cite{Lehtola2020_JCP_144105}.

\subsection{Kramers--Henneberger atom \label{sec:kramers}}

Continuing with model potentials, like the previous version of the
program \cite{Kobus:2013}, also the current version supports
calculations \cite{DziubakM:2009} with the Kramers--Henneberger
potential \cite{Henneberger:1968} defined in the cylindrical
coordinates $z$ and $s$ as
\begin{equation}
V_{KH}(z,s)= \frac{1}{2}\left(\frac{m}{s}\right)^2\frac{1}{T}\int_{0}^{T}
V(z,\beta(t),a) {\rm d}t \label{eq:kh1}
\end{equation}
where $m$ is the magnetic quantum number of the state being calculated,
\begin{equation}
V(z,\beta,a) = \frac{-V_0}{\displaystyle \frac{2\pi}{\omega} \sqrt{a^2+\beta^2+z^2}} \label{eq:kh-v} \\
\end{equation}
where $a$ and $V_0$ are the width and depth of the soft-core
potential, respectively; $\epsilon$ is the intensity and $\omega$ the
cycle frequency of the laser field; and
\begin{equation}
\beta(t) = s+\alpha_0+\alpha_0 (\cos(\omega t)-1), \quad \alpha_0=\epsilon/\omega^2
\label{eq:kh2}
\end{equation}
is the electron's classical trajectory of oscillatory motion.  The
numerical integration in \cref{eq:kh1} is carried out in \xtwodhf{}
using composite Simpson quadrature.

\subsection{Harmonium atom \label{sec:harmonium}}

The method for solving the HF equations for diatomic molecules can
also be applied to the harmonium atom, which is also known as Hooke's
(law) atom. If the Hamiltonian in \cref{eq:totenergy} is replaced by
that of harmonium
\begin{align}
H &= -\frac{1}{2}\nabla_a^2 +\frac{1}{2}\omega^2 \gv{r}_a^2
-\frac{1}{2}\nabla_b^2 +\frac{1}{2}\omega^2 \gv{r}_b^2 + \frac{1}{|\gv{r}_a-\gv{r}_b|} \nonumber\\
&= -\frac{1}{2}\nabla_a^2 -\frac{1}{2}\nabla_b^2
+ \frac{1}{2}\omega^2(r_{a}^2 + r_{b}^2) + \frac{1}{r_{ab}} \label{eq:harmonium}
\end{align}
we see that instead of the nuclear potential, $V_n$, we now have to
deal with the harmonic potential $\frac{1}{2}\omega^2(r_{a}^2 +
r_{b}^2)$, where $r_{i}$ is the distance from the geometric centre,
$r_i=\xi_i\eta_i/\sqrt{\xi_i^2+\eta_i^2-1}$, and $r_{ab}$ is the
distance between the two electrons.

The Schr\"odinger equation for harmonium can be solved exactly, as the
Hamiltonian of \cref{eq:harmonium} can be decoupled in the center of
mass coordinate system given by the centroid
$\gv{R}=\frac{1}{2}(\gv{r}_1+\gv{r}_2)$ and the relative coordinate
$\gv{r}=\gv{r}_1-\gv{r}_2$, yielding
\begin{equation}
 H = -\nabla_r^2 +\frac{1}{4}\omega^2 {r}^2 + \frac{1}{r}
-\frac{1}{4}\nabla_R^2 + \omega^2 {R}^2. \label{eq:harmonium-decoupled}
\end{equation}
The centroid and relative coordinates are sometimes also referred to
as extra- and intracular coordinates, respectively. In the new
coordinates, the eigenvalue problem can be formulated as
\begin{align}
\left( -\frac{1}{2}\nabla_r^2 +\frac{1}{2}\omega_r^2 {r}^2 + \frac{1}{2r}\right)\psi_r(r)
&= \frac{1}{2}E_r \psi_r(r), \label{eq:intracule} \\
\left(-\frac{1}{2}\nabla_R^2 + \frac{1}{2} \omega_R^2 {R}^2\right)\Psi_R(r)
&= 2 E_R \Psi_R(r), \label{eq:extracule}
\end{align}
where $\omega_r=\omega/2$, and $\omega_R=2\omega$ and the eigenvalue of the
original Hamiltonian is obtained as the sum of the eigenvalues $E_r$ and
$E_R$ of the decoupled equations. The decoupling has thus split the
hard-to-solve correlated two-particle problem into two independent
one-dimensional one-particle Schr\"odinger equations, which allow either
closed-form analytical or facile numerical solution to arbitrary precision
with established methodologies, meaning that the full correlated problem in
Hooke's atom is indeed (numerically) exactly solvable.

It has been shown that the intracular problem (\cref{eq:intracule})
has analytical solutions for some values of $\omega_r$
\cite{Kais:1993}: in particular, when $\omega_r=1/2$ (corresponding to
$\omega=\sqrt{2}/2$), then $E_r=5/4$.  However, a numerical solver is
necessary to find solutions for other values of $\omega_r$.
Fortunately, \cref{eq:intracule} can be viewed as a simplified variant
of the Fock equation (\cref{eq:fock1}), and the FD HF technique can be
used to solve it. The new version of \xtwodhf{} allows calculations on
the harmonium atom as a novel feature; examples will be discussed
below in \cref{sec:harmonium-results}.

\section{Solving elliptic PDEs \label{sec:solvingPDEs}}

In the previous section, we saw that the problem of solving the Fock and
Kohn--Sham equations (\cref{eq:fock1a,eq:ks1}), the orbitals for the
Green--Sellin--Zachor potential (\cref{eq:gsz,eq:gsz-omega,eq:gsz-H}), the
superposition of atomic potentials (\cref{eq:sap}), as well as the
Kramers--Henneberger potential (\cref{eq:kh1}), and the intracule equation
for the harmonium atom (\cref{eq:intracule}) involve solving
second-order partial differential equations (PDEs) of the form
\begin{align}
\Bigg\{A(\nu,\mu) {\partial^{2}\over \partial \nu^{2}}
+ B(\nu,\mu) {\partial \over \partial \nu}
+ C(\nu,\mu) {\partial^{2} \over \partial \mu^{2}}
& + D(\nu,\mu) {\partial \over \partial \mu} + E(\nu,\mu) \Bigg\} f(\nu,\mu) \nonumber\\
&  = F(\nu,\mu), \label{eq:gelliptic}
\end{align}
where the functions are defined on a rectangular domain
$ (\nu,\mu) \in [0,\pi]\times[0,\mu_{\infty}]$ that corresponds to
$(\eta,\xi) = [-1,1]\times[1, \xi_{\infty}]$.  Within the approach used in
\xtwodhf{}, one chooses a suitable grid (\cref{sec:grid-specification}),
and approximates the first and second derivatives by finite differences of
a given order (\cref{sec:discretization-order,sec:fd-xtwodhf}).  The
resulting system of linear equations is solved by the SOR or MCSOR methods,
which are discussed below in \cref{sec:sor}.

\subsection{Grid specification \label{sec:grid-specification}}

To obtain accurate\ie{} HF limit solutions to the FD HF equations, one
has to guarantee that all orbitals and all the corresponding Coulomb
and exchange potentials have been calculated to high accuracy. In
specific, this requires converging the calculations with respect to
the employed grid size; we will exemplify this below in
\cref{sec:discretization-order} on a model Poisson equation.

Importantly, the domain definition also depends on the specification
of a $\mu_{\infty}$ (or $\xi_{\infty}$) parameter, which means that
the calculation also has to be converged with respect to the employed
$\mu_{\infty}$ parameter. In the \helfem{} program
\cite{Lehtola2019_IJQC_25944, Lehtola2019_IJQC_25945,
  Lehtola2020_PRA_12516, Lehtola2023_JCTC_2502}, this parameter only
controls how far orbitals are non-zero, because \helfem{} employs a
variational approach where the orbitals are found by diagonalization
and the potential is evaluated analytically. In \xtwodhf{}, however,
the $\mu_{\infty}$ parameter has a double role: in addition to the
above role pertaining to the finite support of the orbitals,
\xtwodhf{} also relies on the use of asymptotic expansions of the
orbitals and potentials in the region close to $\mu_\infty$ to set the
boundary conditions for the relaxation procedure.

\begin{figure*}
\caption{Distribution of grid points in the $(z,x)$ plane
  corresponding to the uniform distribution in $(\nu,\mu)$ variables
  on a [$50\times50 / 50 a_0$] grid.  The upper plot shows the whole
  grid, while the lower plot is a close-up of the region around the A
  and B nuclei at $(-1,0)$ and $(1,0)$, respectively.}
\label{fig:2dgrids}
\begin{center}
\begin{tabular}{c}
  \includegraphics[width=0.9\textwidth]{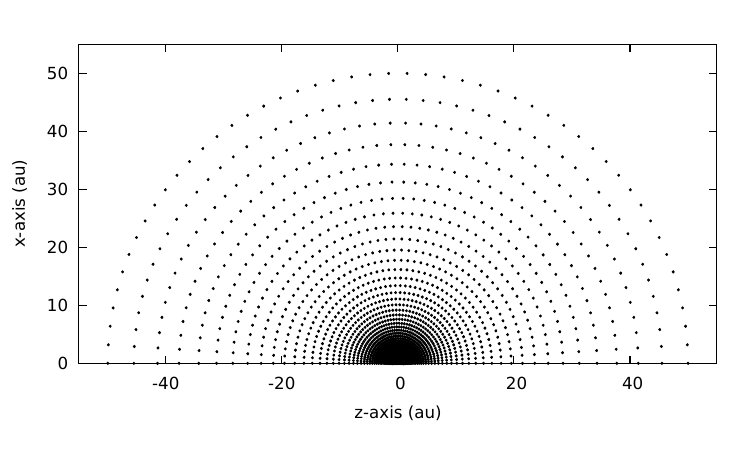} \\

  \includegraphics[width=0.9\textwidth]{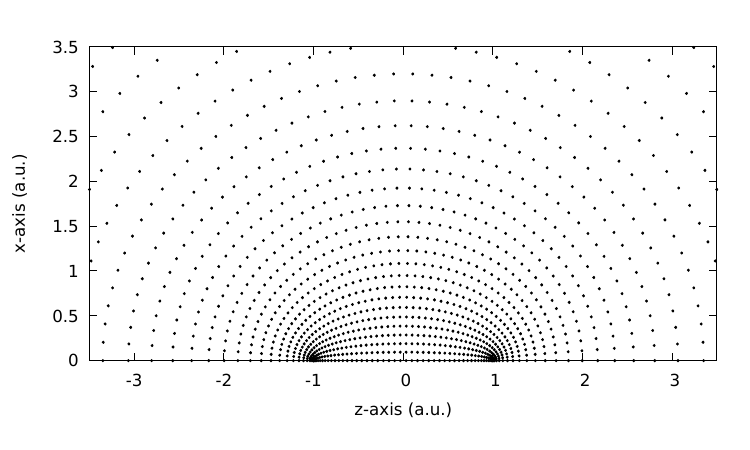}
\end{tabular}
\end{center}
\end{figure*}

Since the isosurfaces of $\mu$ approach spheres for large $\mu$ (see
\cref{fig:2dgrids}), the radial boundary value is typically chosen
with the radius of such a sphere as
\begin{equation}
  \mu_{\infty} = \cosh^{-1} \xi_{\infty}= \cosh^{-1} \frac {2r_{\infty}} {R}, \label{eq:xinfty}
\end{equation}
where $r_{\infty}$ defines the employed value for the ``practical
infinity'' \cite{Lehtola2019_IJQC_25968}.
While orbitals usually decay quite quickly, the potentials can sometimes
deviate from their asymptotic limit in a large region. The value of
$r_\infty$ must thus be chosen large enough in \xtwodhf{} to guarantee that
both the orbitals and potentials have reached their asymptotic behaviour,
as this is used to estimate their values in points in the boundary region
(see \cref{sec:orbital-boundary,sec:potential-boundary} for
details). Calculations with \xtwodhf{} thereby sometimes require the use of
extremely large values of $r_\infty$ to obtain results converged to the HF
limit \cite{Jensen2005_TCA_187}.

The grid points in the $\nu$ and $\mu$ coordinates are distributed uniformly as
\begin{align}
\nu_{i}=(i-1) h_\nu, \  h_\nu =\pi/(N_\nu-1), \phantom{ai}i=1,2,...,N_\nu\nonumber\\
\mu_{i}=(i-1) h_\mu, \  h_\mu =\mu_{\infty}/(N_\mu-1), j=1,2,...,N_\mu
\end{align}
where $N_\nu$ and $N_\mu $ are the number of grid points in the $\nu$
and $\mu$ variables, respectively, $\mu_\infty$ can be computed from
$r_\infty$ by \cref{eq:xinfty}. Such a uniform distribution of the
grid points in the $(\nu,\mu)$ plane corresponds to a non-uniform
distribution in the $(z,x)$ plane with more grid points in the
vicinity of the nuclei, as shown in \cref{fig:2dgrids}, and such a grid
will be denoted as $[N_{\nu}\times N_{\mu}/r_{\infty}]$; a similar
input is used to specify the grid in \xtwodhf{}.

According to our experience with the FD HF method, we find that the
best accuracy is typically achieved when $h_\nu \approx h_\mu$.  Since
the domain of the $\nu$ variable is always $\nu \in [0,\pi]$, the grid
can also be specified in \xtwodhf{} more simply as $N_\nu/r_\infty$,
as the value of $N_\mu$ that yields $h_\mu \approx h_\nu$ is easy to
determine automatically within the employed
discretization,\footnote{Note the limitations on the supported grid
sizes discussed below in \cref{sec:fd-xtwodhf}.} which will be discussed
next.

\subsection{Various discretizations of a model problem \label{sec:discretization-order}}

Our generalized elliptic PDE of \cref{eq:gelliptic} can be discretized on a
given numerical grid by the $r+1$ point stencil, which is based on the
$r$-th order Stirling interpolation formula \cite{Bickley:1941}. But, what
order $r$ of the discretization should one apply to \cref{eq:gelliptic}? To
try to answer this question, let us examine the following model Poisson
equation
\begin{equation}
\pdd{f}{x} + \pdd{f}{y} = -(n+1)^2 \left[ \sin((n+1)x) + \sin((n+1)y) \right] =: G(x,y)
\label{eq:modelProblem}
\end{equation}
where we have implicitly defined the source term $G(x,y)$.

We will now examine solutions of \cref{eq:modelProblem} in the square
 $(x,y) = \left[0,\pi\right]\times \left[0,\pi\right]$, when we
 pose zero boundary conditions at the four edges: $f(0,y)=0$, $f(\pi,y)=0$,
$f(x,0)=0$,  and $f(x,\pi) = 0$.  It is easy to see that
the exact solution of \cref{eq:modelProblem} in this region with these
boundary conditions is just
\begin{equation}
f(x,y) = \sin[(n+1)x]+\sin[(n+1)y],
\label{eq:modelSolution}
\end{equation}
so the parameter $n$ in \cref{eq:modelProblem} gives the
number of inner nodes ($f=0$) in each variable in the exact solution.

Let us now discretize \cref{eq:modelProblem} on a grid.
\Cref{eq:modelProblem} is an infinite family of problems; in the
following, we will study the case $n=3$, whose solutions therefore
should have three inner nodes.  We will study various numerical
stencils with $N_x=N_y=100$, $200$, $400$, and $800$ points on both
axes.

The accuracy of the resulting SOR method can be assessed by checking
how well the obtained numerical solution $f(x_i,y_j)$ agrees with the
exact solution. The question now arises on how to check this agreement
in a reasonable manner?  It turns out there is an elegant solution to
this question: combining \cref{eq:modelSolution,eq:modelProblem}, we
see that the exact solution satisfies
\begin{equation}
G(x,y) = -(n+1)^2 \sin[(n+1)x]+\sin[(n+1)y] = -(n+1)^2 f(x,y),
\label{eq:modelProperty}
\end{equation}
which we can use to assess the quality of numerical solutions of
\cref{eq:modelProblem}: our error metric is given by the maximal
deviation from the expected result
\begin{equation}
  \Delta f = \max_{1 \leq i \leq N_x, 1 \leq j \leq N_y}\left|f(x_i,y_j)-\frac{G(x_i,y_j)}{(n+1)^2}\right|.
  \label{eq:deltau}
\end{equation}

The results collected in \cref{table:sor-accuracy} that are displayed in
\cref{fig:sor-accuracy} show that---as expected---using a larger and
larger grid typically leads to a more and more accurate solution. In
addition to the grid size, the error also strongly depends on the employed
order of the finite difference rule: a higher-order stencil affords more
accurate results with exactly the same grid. The differences in accuracy
between the rules of various orders are considerable. The use of low-order
stencils is unattractive, since the errors remain large even with the
largest $800 \times 800$ grid: $\mathcal{O}(10^{-5})$ with the second-order
and $\mathcal{O}(10^{-9})$ with the fourth-order rule. The sixth order rule
already reaches $\Delta f < 10^{-11}$ with the $800 \times 800$ grid, while
the eighth order rule reaches such an accuracy with just the
$200 \times 200$ grid, and the tenth order rule with the $100 \times 100$
grid.

When a high-order approximation is used with a large grid, the error
made by approximating the PDE by the finite difference method thus
starts to be insignificant, which leaves the use of finite floating
point precision as the dominant remaining source of error.
Contrasting the data in \cref{table:sor-accuracy} computed in double
and quadruple precision, it becomes clear that numerical round-off
effects start to show up for the sixth, eighth, and tenth-order
stencils with all of the studied grids.

It can be seen from the data that the numerical error of the procedure
in double precision saturates to around 4 orders of magnitude greater
than machine precision, $\Delta f \approx 10^{-12}$ while
$\epsilon_{\rm machine} \approx 2.2 \times 10^{-16}$. Yet, for all
practical intents and purposes, this level of precision should be
sufficient; if necessary, \xtwodhf{} can also be compiled in quadruple
precision.

\begin{table}
\caption{Accuracy ($\Delta f)$ of the numerical solution of the model
  Poisson equation (\cref{eq:modelProblem}) with $n=3$ as a function of
  the grid size for finite difference stencils of various numerical
  orders. For comparison, in addition to calculations performed in
  double precision floating-point arithmetic, results are also
  included for quadruple precision arithmetic.}
\label{table:sor-accuracy}
\begin{center}
\begin{tabular}{ccccc}
\hline
\toprule
\multicolumn{1}{c}{grid/order} &
\multicolumn{1}{c}{$100\times100$} &
\multicolumn{1}{c}{$200\times200$} &
\multicolumn{1}{c}{$400\times400$} &
\multicolumn{1}{c}{$800\times800$} \\
\hline\\[-2pt]
& \multicolumn{4}{c}{$\Delta f$ (double precision)} \\\cmidrule{2-5}
2 & $ 4.2\times10^{-3\ph{1}} $
& $ 1.0\times10^{-3\ph{1}} $
& $ 2.6\times10^{-4\ph{1}} $
& $ 6.5\times10^{-5\ph{1}} $ \\
4 & $ 1.4\times10^{-5\ph{1}} $
& $ 8.6\times10^{-7\ph{1}} $
& $ 5.3\times10^{-8\ph{1}} $
& $ 3.3\times10^{-9\ph{1}} $ \\
6 & $ 5.7\times10^{-8\ph{1}} $
& $ 8.6\times10^{-10}$
& $ 1.3\times10^{-11}$
& $ 6.7\times10^{-12}$ \\
8 & $ 2.5\times10^{-10}$
& $ 1.2\times10^{-12}$
& $ 1.7\times10^{-12}$
& $ 3.1\times10^{-12}$ \\

10 & $ 1.2\times10^{-12}$
& $ 2.8\times10^{-13}$
& $ 6.9\times10^{-12}$
& $ 1.1\times10^{-12}$ \\[5pt]

% & \multicolumn{4}{c}{k=4 (quadruple precision)} \\\cmidrule{2-5}
& \multicolumn{4}{c}{$\Delta f$ (quadruple precision)} \\\cmidrule{2-5}
2 & $ 4.2\times10^{-3\ph{1}} $
& $ 1.0\times10^{-3\ph{1}} $ %0.1038D-02
& $ 2.6\times10^{-4\ph{1}} $ %0.2586D-03
& $ 6.5\times10^{-5\ph{1}} $ \\ %0.6492D-04
4 & $ 1.4\times10^{-5\ph{1}} $
& $ 8.6\times10^{-7\ph{1}} $ % 0.8589D-06
& $ 5.3\times10^{-8\ph{1}} $ % 0.5338D-07
& $ 4.0\times10^{-9\ph{1}} $ \\ %0.5396D-08 0.4025D-08 0.3967D-08
6 & $ 5.7\times10^{-8\ph{1}} $
& $ 6.6\times10^{-10}$
& $ 1.3\times10^{-11}$ %0.1329D-10
& $ 2.1\times10^{-13}$ \\ % 0.4349D-12? 10^5 iters 0.2058D-12
8 & $ 2.5\times10^{-10}$
& $ 9.5\times10^{-13}$
& $ 3.7\times10^{-15}$ % 0.3663D-14 0.3663D-14
& $ 1.4\times10^{-17}$ \\ % 0.1830D-06 0.1213D-10 0.1414D-16

10 & $ 1.2\times10^{-12}$ % 0.1205D-11
& $ 1.1\times10^{-15}$ % 0.1121D-14
& $ 1.1\times10^{-18}$ % 0.1072D-17
& $ 1.0\times10^{-21}$ \\[5pt] %0.1034D-20 0.1035D-20

\bottomrule\\[-5pt]

\end{tabular}
\end{center}
\end{table}

\def\scalea{0.68}
\begin{figure}
  \caption{Comparison of the accuracy $\Delta f$ of the numerical
    solution of the model Poisson equation (\cref{eq:modelProblem}) in
    double- and quadruple-precision (dashed and solid lines,
    respectively). See \cref{table:sor-accuracy} for details.}
\label{fig:sor-accuracy}
{\footnotesize
\begin{center}
\begin{tabular}{c}
  \includegraphics[width=\scalea\textwidth]{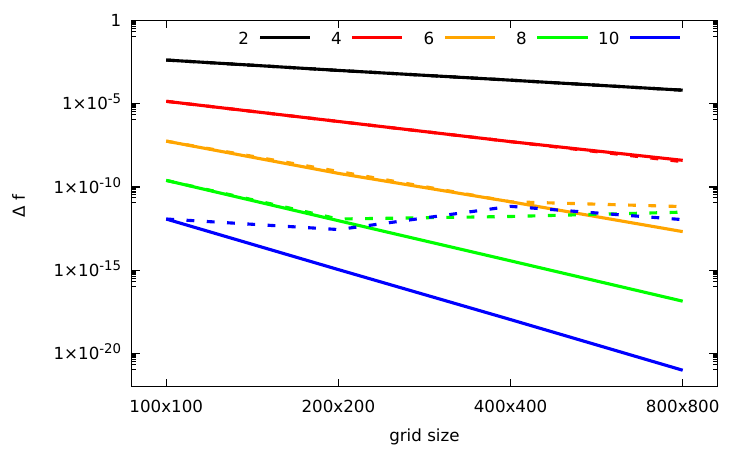}  
\end{tabular}
\end{center}
}
\end{figure}

\subsection{Discretization used in \xtwodhf{} \label{sec:fd-xtwodhf}}

A sixth-order stencil was originally used in the FD HF method
\cite{LaaksonenPS:1983a}.  In the first version of the present
program, this stencil was replaced with the eighth-order stencil,
resulting in improved accuracy and thus a substantial reduction of
computational costs \cite{KobusLS:1996}. Although the use of even
higher-order stencils would be appealing due to the greatly improved
numerical accuracy and resulting savings in grid size, the application
of higher-order rules is challenged by Runge's phenomenon
\cite{Runge1901_ZMP_224} as \xtwodhf{} employs uniformly spaced grids:
function interpolation on a uniformly spaced grid becomes numerically
ill-behaved at high order.  The same challenge also exists with finite
element approaches that employ uniformly placed nodes
\cite{Lehtola2019_IJQC_25968}.

We note here that modern mathematical approaches avoid the problem associated with
Runge's phenomenon by employing non-uniform grids. For instance,
recent works by the second author have demonstrated the feasible
stability of up to 19$^{\rm th}$ order schemes in electronic structure
calculations \cite{Lehtola2019_IJQC_25945, Lehtola2023_JCTC_4033,
  Lehtola2023_JPCA_4180}, achieving significant increases in accuracy
at similar computational cost.  It would be desirable to pursue higher
order stencils in future versions of the \xtwodhf{} program, as well.

The eighth-order stencil is thus still used in the current
version of \xtwodhf{}. This stencil is given by
\begin{align}
f'_i &= \frac{1}{840h}
\left( 3f_{i-4}-32f_{i-3}+168f_{i-2}-672f_{i-1} \right. \nonumber \\
& \left. + 672f_{i+1}-168f_{i+2}+32f_{i+3}-3f_{i+4}\right) +O(h^8)  \nonumber \\
f''_i & = \frac{1}{5040h^2}
\left( -9f_{i-4}+128f_{i-3}-1008f_{i-2}+8064f_{i-1} - 14350f_i\right. \nonumber \\
& \left.+ 8064f_{i+1}-1008f_{i+2}+128f_{i+3}-9f_{i+4}\right) +O(h^8) \label{eq:discret1}
\end{align}
where $f_i=f(x_1+ih)$ and $x$ stands for either $\nu$ or $\mu$.
However, knowing how to differentiate is not enough: a quadrature rule
is also necessary to evaluate the integrals that arise in the FD HF
method. \xtwodhf{} employs the 7-point Newton--Cotes quadrature formula
\begin{equation}
  \int_{x_1}^{x_7} f(x) {\rm d}x = \frac h {140} (41 f_1 + 216 f_2 +
  27 f_3 + 272 f_4 + 27 f_5 + 216 f_6 + 41 f_7) +
  \mathcal{O}(h^9) \label{eq:nc-quad}
\end{equation}
which is therefore asymptotically more accurate than the 8 and 9 point
expressions for estimating the first and second derivatives above in
\cref{eq:discret1}. Since the quadrature rule employs 7 consecutive
points at a time, it imposes a limitation for the total number of grid
points in the $\nu$ and $\mu$ variables. The smallest number of grids
points one could use is of course 7. The next smallest grid contains
13 points, as the last point of the first interval is the first point
of the second interval. Therefore, the grid size must be of the form
$6N+1$, where $N$ is an integer.

When the MCSOR scheme discussed below in \cref{sec:mcsor} is used, a
further restriction on the number of grid points in the $\nu$ and $\mu$
variables has to be imposed. If MCSOR sweeps are carried out with
column-major or row-major ordering (see \cref{sec:update-sweep,sec:mcsor}),
$N_{\nu}$ and $N_{\mu}$ must be of the form $(k/2+1)N'+(k/2+1)+1$, where
$k$ is the employed order of finite differences and $N'$ is another
integer. Since $k=8$ in \xtwodhf{}, we get that the number of grid points
in $\mu$ and $\nu$ must be of the form $5N'+6 = 5(N'+1) + 1$, although
different sized grids can be used for $\mu$ and $\nu$. With the aim of
enabling switching between the SOR and MCSOR relaxation methods in various
steps of any calculation, the implementation in \xtwodhf{} forces the
additional MCSOR grid size limitation in all calculations.

Taking the conditions arising from the use of the 7-point quadrature rule
of \cref{eq:nc-quad} as well as from MCSOR together, we thus see that the
number of points in $\mu$ needs to be of the form $30N+1$, where $N$ is an
integer, since $30$ is the smallest common denominator of 5 and 6. The
smallest possible grid that satisfies this limitation is therefore the
$N_{\nu}=N_{\mu}=31$ grid.

This smallest possible grid is illustrated in \cref{fig:17ps-sor}
together with the application of \cref{eq:discret1} in two dimensions,
which yields a 17-point cross-like pattern.
  
\def\scale{0.71}
\begin{center}
\begin{figure}
\hspace*{2cm}\parbox{\textwidth}{\includegraphics[width=\scale\textwidth]{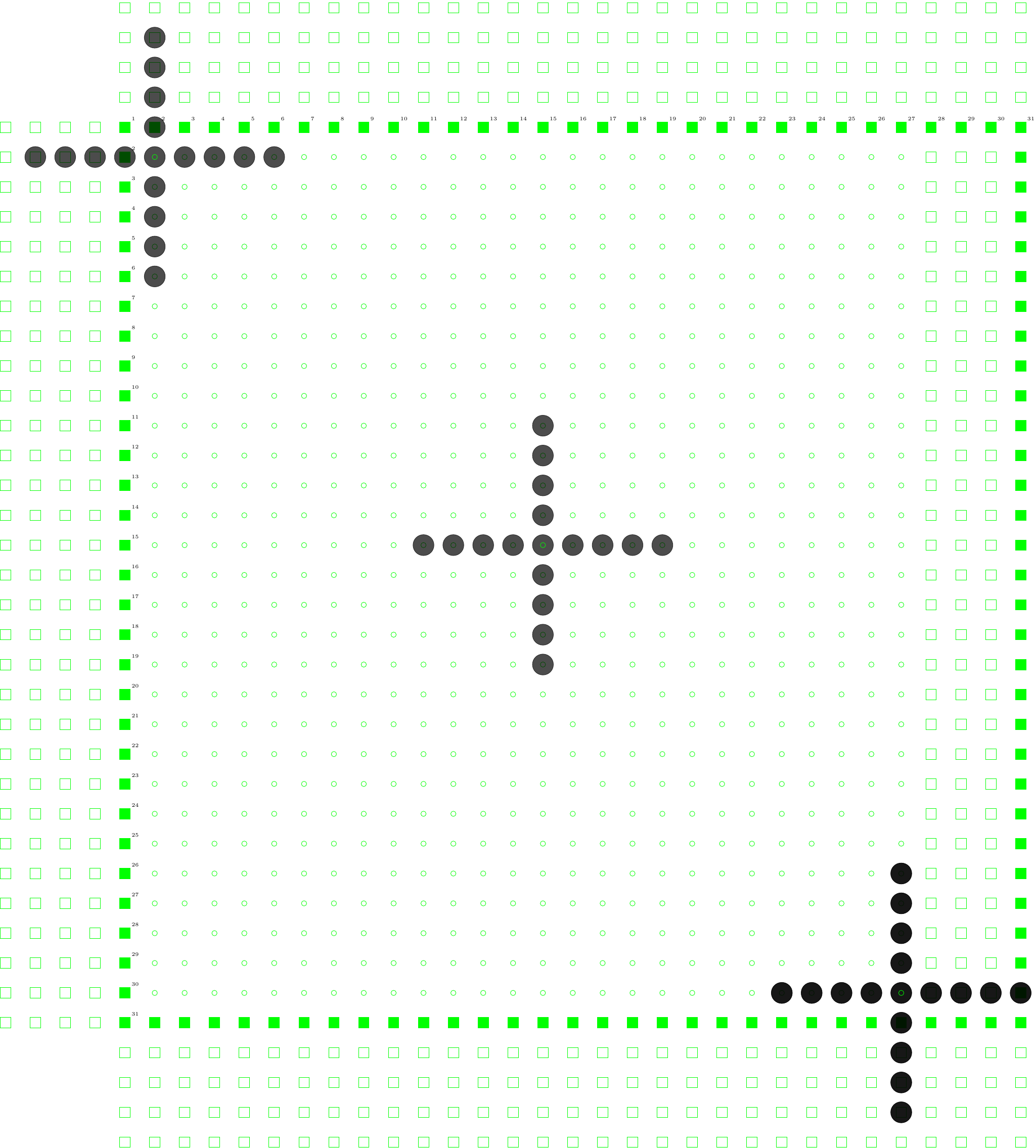}}
 \caption{A visualization of a $31\times31$ grid for $f$, and the
   17-point numerical stencil employed by \xtwodhf{}. Filled green
   squares denote the limits of the $31\times31$ grid, and points
   where the solution are found from the Poisson equation are shown as
   small empty green circles. Extra boundary values shown as empty
   green squares must be provided to be able to employ the 17-point
   stencil to relax the solution at every inner point.
   The points on the right-hand side of the grid are beyond the
   practical infinity $\mu_{\infty}$, where we assume that the
   orbitals and potentials vanish identically. The known behaviour of
   the orbitals and potentials in the large-$r$ region close to
   $\mu_\infty$ is employed to determine their values at points close
   to the boundary at $\mu_\infty$. Note that while no data exists at
   the left hand segment of the grid, as illustrated by the empty
   regions in the upper and lower left corners of the figure, the
   upper and lower right corners are required for the calculation of
   the kinetic energy in the asymptotic region. }
\label{fig:17ps-sor}
\end{figure}
\end{center}

\subsubsection{Boundary conditions \label{sec:boundaries}}

The use of \cref{eq:discret1} to evaluate derivatives at grid point
$(x_i,y_j)$ requires knowledge of the values $f(x_{i'},y_{j'})$ for
$i' \in [i-4,i+4]$ and $j' \in [j-4,j+4]$. This leads to the problem
that when \cref{eq:discret1} is used to discretize
\cref{eq:gelliptic}, one also needs to solve for grid points located
within the boundary band: if the point $(x_i,y_j)$ is close to grid
boundaries, the stencil will end up operating on points that do not
formally exist on the grid. This is a rather unwelcome complication,
because it results in a more complex algorithm.

Laaksonen, Pyykk\"o, and Sundholm used a different numerical stencil
for the boundary grid points than for the inner grid points: while
they used central finite differences for inner grid points, they used
one-sided finite differences to find the solution in the points close
to the boundary. When a high-order rule is employed, this approach
thus requires using a different set of rules for the points close to
the boundaries.

We can exemplify this complication in one dimension with the 9-point
rule of \cref{eq:discret1} at the $x=0$ boundary, employing grid
points at $x_i = (i-1) h$ for $i\in[1,N]$. Assuming $N \gg 1$, we
could then use the rule of \cref{eq:discret1} down to $i=5$, which
would use $(f_1,\dots,f_9)$ for the interpolation. The finite
difference rules for the points $i\in [1,4]$ would still use the same
set of input values $(f_1,\dots,f_9)$, but stencil would be different
for each of these points. In more dimensions, such an algorithm
requires even more complicated logic, since a point can be in or out
of the boundary region in each dimension.

However, also another approach is possible. When we dealt with the
model problem (\cref{eq:modelProblem}), we knew the exact solution,
and provided extra data by continuing the exact solution outside the
numerical grid. Thus, if we can also provide additional boundary
values for FD HF calculations, we are able to use the same stencil for
all inner grid points.

\subsubsection{Symmetry properties of coordinate system \label{sec:symmprop}}

The calculation of some of these extra boundary values relies on
knowledge of the symmetry of the coordinate system. The
Cartesian coordinates in terms of $(\nu,\mu,\theta)$ read
\begin{align}
x&={R \over 2} \sinh(\mu) \; \sin(\nu) \; \cos(\theta) \nonumber \\
y&={R \over 2} \sinh(\mu) \; \sin(\nu) \; \sin(\theta) \label{eq:cartesian2prolate} \\
z&={R \over 2} \cosh(\mu) \; \cos(\nu) \nonumber
\end{align}
where $\mu \in [0,\mu_\text{max}]$, $\nu \in [0,\pi]$ and $\theta \in
[0,2\pi]$.  It is clear from \cref{eq:cartesian2prolate} that
(fictitiously) flipping the sign of $\mu$ or $\nu$ flips the point at
$(x,y,z)$ over to $(-x,-y,z)$.
However, the same result [$(x,y,z)\to(-x,-y,z)$] is also
obtained when $\theta \to \theta +\pi$, and we know from the
analytical angular solution
\begin{equation}
  e^{im\theta}=\cos(m\theta) + i\sin(m\theta) \label{eq:angfac}
\end{equation}
that $\theta \to \theta +\pi$ leads to an additional $(-1)^m$ phase
factor. Because of this property, we can immediately identify
that flipping the sign of $\nu$ leads to
\begin{equation}
f(\nu,\mu) = (-1)^m f(-\nu,\mu) \label{eq:nu-symmetry}
\end{equation}
and also that  flipping the sign of $\mu$ leads to the
analogous property
\begin{equation}
f(\nu,\mu) = (-1)^m f(\nu,-\mu). \label{eq:mu-symmetry}
\end{equation}

We can identify one further symmetry condition for the $\nu \in [0,
  \pi]$ axis by examining \cref{eq:cartesian2prolate} around
$\nu=\pi$.  We observe that $\nu \to \pi+\nu$ flips
$(x,y,z)\to(-x,-y,-z)$ while $\nu \to \pi-\nu$ leads to $(x,y,z) \to
(x,y,-z)$. We thus observe that the two coordinates $(-x,-y,-z)$ and
$(x,y,-z)$ differ by a $(x,y)$ inversion, and we can write down the
third symmetry equation by again using the same relation obtained
above
\begin{equation}
f(\pi+\nu,\mu) = (-1)^m f(\pi-\nu,\mu). \label{eq:pi-nu-symmetry}
\end{equation}

Although the coordinate system does not formally have points at
$\mu<0$ or $\nu<0$, \xtwodhf{} employs extended grids that can have
such values to facilitate the relaxation process: as was discussed
above in \cref{sec:boundaries}, the finite difference stencil can
reach points that are formally outside the grid. Knowing the symmetry
of the orbitals and potentials around the $\mu$ and $\nu$ axes is
therefore of great value for the numerical implementation.

From the above discussion, we observe that orbitals have definite
symmetry around the $\mu$ and $\nu$ axes, and can be divided into
components of even or odd symmetry with respect to this inversion:
Orbitals of $\sigma$, $\delta$, $\ldots$ symmetry are even functions
of $(\nu,\mu)$, while orbitals of $\pi$, $\varphi$, $\ldots$ symmetry
are odd functions of $(\nu,\mu)$.

Since Coulomb and exchange potentials arise from products of orbitals,
they have analogous symmetry properties. Coulomb potentials are always
of $\sigma$ symmetry and thus even functions around the $\mu$ and
$\nu$ axes, since there is no $m$ dependent term in
\cref{eq:eq17}. Exchange potentials can be of either even or odd
symmetry, depending on the value of $m_a-m_b$ in \cref{eq:eq16}.

We observe from
\cref{eq:nu-symmetry,eq:mu-symmetry,eq:pi-nu-symmetry} that any odd
function vanishes along the $(\nu,0)$, $(0,\mu)$ and $(\pi,\mu)$
boundary lines, while even functions are finite at the boundary lines
and their values on these lines need to be determined in the SCF/SOR
procedure. Before discussing the rules for how this takes place in
\cref{sec:boundary}, we first comment on an additional symmetry found
in homoatomic molecules.

\subsubsection{Symmetry in homoatomic molecules \label{sec:homosymm}}
Homoatomic molecules have an additional symmetry with respect to
inversion $z \leftrightarrow -z$. It is easy to see that this symmetry
leads to
\begin{equation}
f(\pi-\nu,\mu) = \pm f(\nu,\mu), \ \nu \in [0,\pi] \label{eq:homo-symmetry}
\end{equation}
where the plus and minus signs are for gerade and ungerade solutions,
respectively. In principle, this symmetry could be utilised to reduce
the memory needed to store orbitals and potentials and---more
importantly---the number of grid points undergoing relaxation by a
factor of two. However, this would incur the additional hassle of
handling boundary conditions at $\nu=\pi/2$. Therefore, the
$D_{\infty\,h}$ symmetry is only used to explicitely force the
symmetry of the orbitals in post-processing in the aims to improve
SCF/SOR convergence.

\subsubsection{Determination of boundary values \label{sec:boundary}}

Since orbitals and potentials are either even or odd functions of
$\mu$ and $\nu$, the values of $f$ on the other side of the boundary
lines are known. All that remains is then to determine the values of
the orbitals and potentials on the $(\nu,0)$, $(0,\mu)$ and
$(\pi,\mu)$ boundary lines themselves.

We note again that interpolation is not needed for functions of odd
symmetry, such as orbitals and potentials of $\pi$ or $\varphi$ symmetry,
as such functions vanish on the boundary lines. The only thing that needs
to be done is therefore to derive equations for determining the values of
even functions, such as orbitals and potentials of $\sigma$ and $\delta$
symmetry, on the $(\nu,0)$, $(0,\mu)$ and $(\pi,\mu)$ boundary lines.

Assuming that $f(\nu_1,\mu_j)=f(\nu_1,-\mu_j)$,
$f(\nu_{N_{\nu}},\mu_j)=f(\nu_{N_{\nu}},-\mu_j)$, $j=2,N_\mu$, and that
$f(\nu_i,\mu_1)=f(-\nu_i,\mu_1)$, $i=2,N_\nu-1$, and using the 9-point
Lagrange interpolation formula for equally spaced abscissas, it is possible
to derive formulae for updating the function values along the boundaries
(see \cref{sec:appendix-interpolation})

% Assuming that $f(\nu_i,\mu_j)=f(\nu_i,-\mu_j)$ and that
% $f(\nu_i,\mu_j)=f(-\nu_i,\mu_j)$, it can be shown that (see
% \cref{sec:appendix-interpolation})
\begin{align}
f(\nu_1,\mu_j)&={1 \over 126}\left(210f(\nu_2,\mu_j)-120f(\nu_3,\mu_j)+45f(\nu_4,\mu_j)\right.\nonumber\\
& \phantom{{1 \over 126}}\left.-10f(\nu_5,\mu_j)+f(\nu_6,\mu_j)\right)\nonumber\\
f(\nu_{N_\nu},\mu_j)&={1 \over 126}\left(210f(\nu_{N_\nu-1},\mu_j)-120f(\nu_{N_\nu-2},\mu_j)
+45f(\nu_{N_\nu-3},\mu_j)\right.\nonumber\\
&\phantom{{1 \over 126}}\left. -10f(\nu_{N_\nu-4},\mu_j)+f(\nu_{N_\nu-5},\mu_j)\right)\nonumber\\
f(\nu_i,\mu_1)&={1 \over 126}\left( 210f(\nu_i,\mu_2)-120f(\nu_i,\mu_3)+45f(\nu_i,\mu_4)\right.\nonumber\\
&\phantom{{1 \over 126}}\left. -10f(\nu_i,\mu_5)+f(\nu_i,\mu_6)\right)\nonumber\\
& \phantom{aaaaaaaaaaaaaaaaaaaaaa} i=2,N_\nu-1,\quad j=2,N_\mu \label{eq:boundary-values}
\end{align}

So far we have
only discussed the $(\nu,0)$, $(0,\mu)$ and $(\pi,\mu)$ boundaries;
yet, the $(\nu,\mu_\infty)$ boundary line needs to be addressed, as
well. Since this boundary line corresponds to points that are far away
from the molecule, we assume that the orbitals and potentials vanish
for $\mu > \mu_\infty$. Moreover, we will estimate the values $f$ of
the orbitals and potentials that are close to the boundary,
$f(\nu_i,\mu_{N_{\mu}-4+k})$, $i=1,\dots,N_{\nu}$, $k=1,\dots,4$, with
the known asymptotic behaviour of the orbitals and potentials; see
\cref{sec:orbital-boundary,sec:potential-boundary} for in-depth discussion.

\subsubsection{Compact notation for difference formulas \label{sec:findiffform}}

One can note that the first and second derivative expressions in
\cref{eq:discret1} are odd and even functions of the grid points
around the expansion point. The expressions for the various $\mu$
derivatives needed for FD HF can then be rearranged as
\begin{align}
\left. {\partial f\over \partial \mu}\right|_{i,j}
& =  \sum_{k=1}^{4} d_{k}^{(\mu)}
\left(f(\nu_{i},\mu_{j-k}) - f(\nu_{i},\mu_{j+k})\right) \nonumber \\
& =  \sum_{k=1}^{9} \tilde{d}_{k}^{(\mu)}
f(\nu_{i},\mu_{j-5+k}) \nonumber \\
\left. {\partial^{2}f \over \partial \mu^{2}}\right|_{i,j}
& =  \sum_{k=1}^{4} d_{k}^{(\mu\mu)}
\left(f(\nu_{i},\mu_{j-k}) + f(\nu_{i},\mu_{j+k})\right)
+d_{5}^{(\mu\mu)} f(\nu_{i},\mu_{j}) \nonumber\\
& =  \sum_{k=1}^{9} \tilde{d}_{k}^{(\mu\mu)}
f(\nu_{i},\mu_{j-5+k}) \label{eq:discret2}
\end{align}
where the
coefficients $\tilde{d}_{k}^{(\mu)}$ and $\tilde{d}_{k}^{(\mu\mu)}$
are trivially identified from \cref{eq:discret1}. One can even take a
step further and collect the $\mu$-dependent part of the Laplacian as
\begin{align}
\left. \left({\partial^2f \over \partial\mu^2}
+{\cosh \mu \over \sinh \mu} {\partial f\over \partial \mu}\right)\right|_{i,j}&=
\sum_{k=1}^{9} {D}^{\mu}_k(\mu_j)
f(\nu_{i},\mu_{j-5+k}). \label{eq:discret3}
\end{align}
An analogous technique is used to evaluate the necessary derivatives
over $\nu$.

\subsection{Successive overrelaxation (SOR) method \label{sec:sor}}

Solving \cref{eq:gelliptic} by the finite difference discretization
discussed in the previous subsection leads to a matrix equation
\begin{equation}
\Rbf \fbf = \sbf \label{eq:R}
\end{equation}
where $\fbf$ and $\sbf$ are vectors of length $N_\nu N_\mu$, and
$N_{\nu}$ and $N_{\mu}$ are typically in the range $10^2$--$10^3$. The
matrix $\Rbf$ can thus be quite large, but it is also extremely
sparse. Because of this, iterative methods are extremely attractive
for the solution of \cref{eq:R}.

As was already mentioned in \cref{sec:introduction}, the orbitals and
potentials are usually solved in the FD HF method with the SOR method
\cite{Kobus:1994, Kobus:1993a, StoerB:1980}, which is a variant of the
Gauss--Seidel method for solving linear systems of equations. The
method is guaranteed to converge, if the matrix $\Rbf$ is symmetric
and positive-definite, and if the relaxation factor $\omega$ is within
the interval $0<\omega<2$ \cite{StoerB:1980,
  HagemanY:1981}.\footnote{A proof has been shown for the standard
discretization of the Poisson equation by the second-order cross-like
stencil, when the grid points are handled from the bottom to the top,
and from the left to the right; we are not aware of a proof for
higher-order discretizations.}

Every SOR sweep consists of in-place updates to the elements of the
solution array $f_p$ as
 \begin{align}
 f_{p} := R_{pp}^{-1} \left[ (1-\omega) R_{pp}f_{p} +
   \omega \Bigg( s_{p} - \sum_{q \neq p}^{N_\nu N_\mu} R_{pq}f_{q} \Bigg )\right] \label{eq:Rii}
 \end{align}
in a loop over all the elements of the array, $ p=1,\ldots,N_\nu
N_\mu$. Let us now write the SOR update of \cref{eq:Rii} for
\cref{eq:gelliptic} with the discretization of \cref{eq:discret2}. The
diagonal term $R_{pp}$, which thus yields the contribution to the grid
point $(\nu_i,\mu_j)$ arising from its old value, can be identified
from \cref{eq:gelliptic,eq:discret2} as
\begin{align}
G(\nu_{i},\mu_{j}) = &A(\nu_{i},\mu_{j}) d_{5}^{(\nu\nu)}
+ C(\nu_{i},\mu_{j}) d_{5}^{(\mu\mu)} + E(\nu_{i},\mu_{j}), \label{eq:G-diag}
\end{align}
since the first derivative operators multiplying $B$ and $D$ in
\cref{eq:gelliptic} don't contain diagonal terms (cf. the original
one-dimensional expression, \cref{eq:discret1}).

Having identified the mathematical structure of the update, we can
thus write down the SOR update arising from \cref{eq:Rii} for the
solution of \cref{eq:gelliptic} with the discretization of
\cref{eq:discret2} explicitly as
\begin{align}
f(\nu_{i},\mu_{j}):=& (1-\omega) f(\nu_{i},\mu_{j}) + \frac \omega {G(\nu_{i},\mu_{j})} \Bigg[
  F(\nu_{i},\mu_{j}) \nonumber\\
&- \sum_{k=1}^{4}
\left\{\, A(\nu_{i},\mu_{j}) d_{k}^{(\nu\nu)}
\left( f(\nu_{i-k},\mu_{j}) + f(\nu_{i+k},\mu_{j}) \right) \right. \nonumber\\
& \left.+ B(\nu_{i},\mu_{j}) d_{k}^{(\nu)\phantom{\nu}}
\left(f(\nu_{i-k},\mu_{j}) - f(\nu_{i+k},\mu_{j})\right) \right. \nonumber \\
& \left. + C(\nu_{i},\mu_{j}) d_{k}^{(\mu\mu)}
\left(f(\nu_{i},\mu_{j-k}) + f(\nu_{i},\mu_{j+k})\right) \right. \nonumber \\
& \left. + D(\nu_{i},\mu_{j}) d_{k}^{(\mu)\phantom{\mu}}
\left(f(\nu_{i},\mu_{j-k}) - f(\nu_{i},\mu_{j+k})\right) \right\} \Bigg]. \label{eq:update-rule}
\end{align}

\subsubsection{Update sweeps \label{sec:update-sweep}}

Every update sweep consists of using \cref{eq:update-rule} to update
the solution for $i=2,\ldots,N_\nu-1$ and $j=2,\ldots,N_\mu-4$, that
is, for the interior grid points. Each SOR sweep does consecutive
in-place updates to elements of the solution array ${\bf f}$, and the
update at each point thus uses the newest available information for
its neighboring points. As will be discussed below in
\cref{sec:interweave}, a number of sweeps is performed on the interior
grid points, thus propagating information from the outer boundary
region towards the centre of the molecule, as well as from the
$z$-axis back towards the outer boundary.

After the requested number of sweeps have been carried out, the
solution is updated separately at the remaining points: the line $i=1$
corresponding to the $\nu=0$ boundary line, the line $i=N_\nu$ to the
$\nu=\pi$ boundary line, the $j=1$\ie{} $\mu=0$ boundary line, and the
points $j \in [N_\mu-3,N_\mu]$ belonging to the region where the
values of the orbitals and potentials are estimated from their
asymptotic behaviour
(\cref{sec:orbital-boundary,sec:potential-boundary}).

As the update in every relaxation step happens through the employed
finite difference stencil, this means that the convergence speed
depends critically on the employed grid spacing. As the convergence
thus becomes the slower the larger a grid is used, it is recommended
to run FD HF calculations in sequences of increasing grid size, using
the solution found on the last grid size as an initial guess at each
step. The solutions are then interpolated from the previous grid to
the current one.%
\footnote{In order to facilitate this process the \xtwodhf{} program
offers a special label \textsl{interp}. See the User's Guide.} Even
when a converged solution on a smaller grid is used as an initial
guess, thousands of sweeps may be needed to meet the specified
convergence criterion.

In addition to the natural column-major (outer loop on $j$, inner loop
on $i$) update that is tailored to the memory access in Fortran,
\xtwodhf{} also supports a reverse column-major update, in which the
columns are updated in reverse order, as well as row-major (outer loop
on $i$, inner loop on $j$) updates.

In addition \xtwodhf{} implements ``middle-type'' sweeps, which are
used by default, because they lead to a faster convergence. The
ordering of the grid points in the middle-type sweeps in the SOR
method is the following. In the outer loop, the $\mu$ variable loops
first from $\mu_j=\mu_{(N_{\mu}-1)/2}$ down to $\mu_j=\mu_2$, and then
from $\mu_j=\mu_{(N_{\mu}-1)/2+1}$ up to $\mu_j=\nu_{N_{\mu}-4}$.  The
outer loop thus specifies a fixed $\mu_j$ value. In the inner loop,
the $\nu$ values loop from $\nu_i=\nu_{(N_{\nu}-1)/2}$ down to
$\nu_i=\nu_2$, and then from $\nu_i=\nu_{(N_{\nu}-1)/2+1}$ up to
$\nu_i=\nu_{N_{\nu}-1}$.

\def\vspaceb{15pt}
\def\vspacea{5pt}
\newcommand{\sepfig}{\phantom{aa}}

\subsubsection{Multicolour SOR method \label{sec:mcsor}}

The MCSOR method was originally developed for better efficiency on
vector processors \cite{Kobus:1994}.  The idea in the MCSOR
method is that the discretization through the 17-point numerical
stencil can be thought to divide the grid points into five separate
colour groups. Since the stencil only connects points by up to four
steps in either the horizontal or vertical directions, the relaxation
at, say, a ``black'' grid point depends solely on its ``red'',
``orange'', ``yellow'', and ``blue'' neighbours (see \cref{fig:mcsor}
for illustration). The relaxation of all the ``black'' grid points can
thus be performed simultaneously on a vector processor. Once the ``black''
grid points have been updated, the ``red'', ``orange'', etc. grid
points can be updated in a similar fashion, one colour at a time.

The parallel updates to the solution array are thus independent. In
contrast, the SOR method updates points consecutively, always using
the newest information for every grid point. This difference in the
way SOR vs MCSOR update the values of the solution on the grid has
implications for the resulting convergence of the overrelaxation
procedure. In our experience, the MCSOR scheme should be avoided when
the initial orbitals and potentials are of poor quality. For instance,
when hydrogenic orbitals are used to initialize calculations on
many-electron systems, it is better to use the SOR algorithm for the
first 50--100 SCF iterations.

\def\scale17{0.65}
\begin{figure}
\hspace*{2cm}\parbox{\textwidth}{\includegraphics[width=\scale17\textwidth]{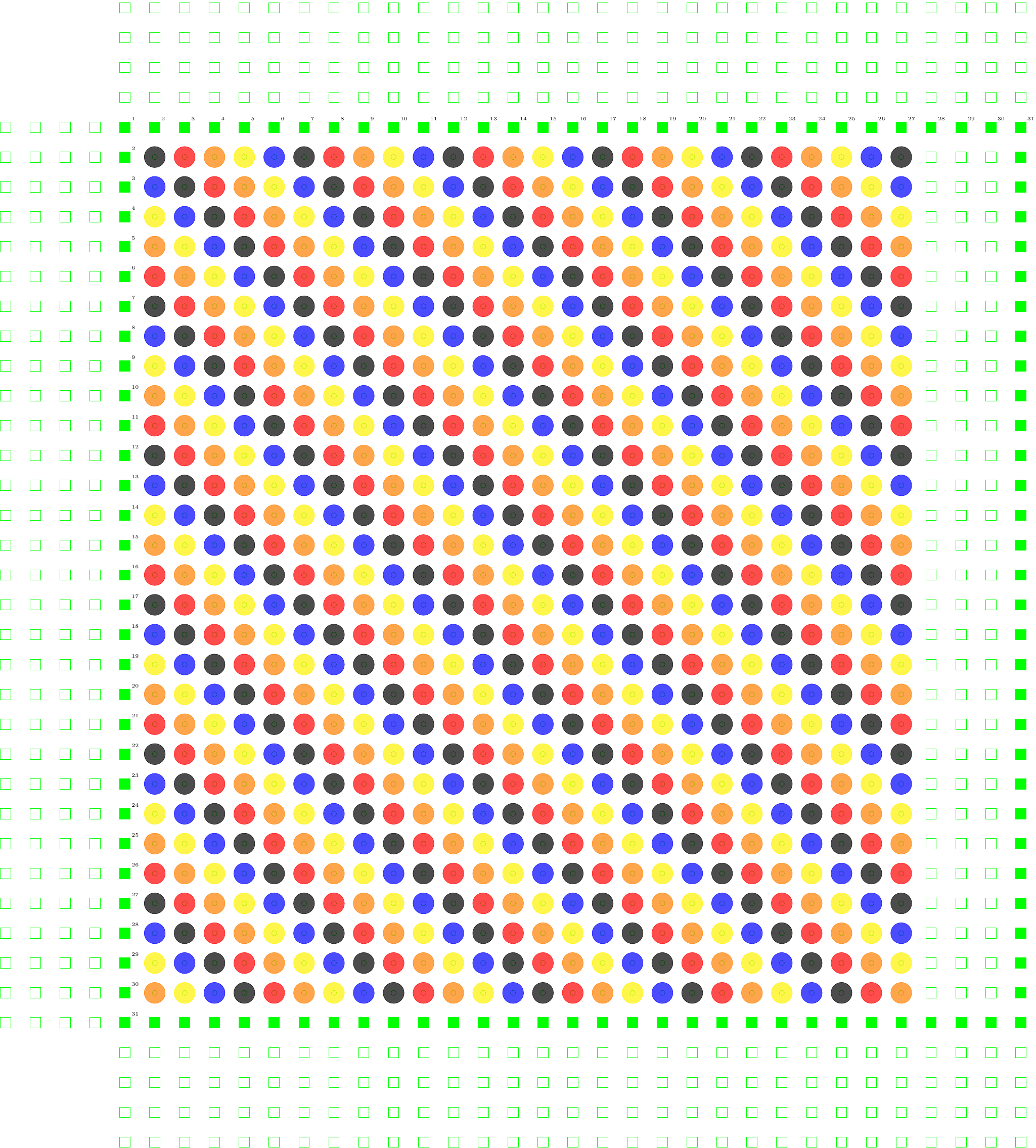}}
\caption{Discretization through the
17-point numerical stencil divides the grid points into five separate
colour groups and the relaxation at, say, a black grid point depends
solely on its red, orange, yellow, and blue neighbours.}
\label{fig:mcsor}
\end{figure}

MCSOR can be used to obtain a parallel algorithm for use on several
processors with OpenMP or Portable Operating System Interface threads
(pthreads) programming. In practice, one can expect a wall-time
speed-up factor of about 3 when using the MCSOR algorithm with OpenMP
or pthreads for orbital relaxation, at a cost of using 5 times more
processor time than an analogous calculation using a single-threaded
SOR procedure.

Importantly, the speed-up factor is grid-size dependent.  The
performance of the MCSOR method decreases when the function value
array saturates the processor's level 3 (L3) cache. Once the capacity
of the L3 cache has been exhausted, the speed-up factor drops to about
2 for larger grids, and becomes fairly independent of the number of
threads used for parallelisation (8 threads are used by default).

The ordering of the grid points in the default middle-type sweeps in
MCSOR is as follows. In the outer loop, the $\mu$ variable loops first
from $\mu_j=\mu_{(N_{\mu}-1)/2}$ down to $\mu_j=\mu_2$, as in the SOR
method. However, in the second step, the loop is handled in the
opposite order to the SOR method: we loop down from
$\mu_j=\nu_{N_{\mu}-4}$ to $\mu_j=\mu_{(N_{\mu}-1)/2+1}$, so that the
middle values are not used and modified by different MCSOR threads at
the same time, which would lead to inconsistencies.

\subsubsection{Interweaving (MC)SOR and SCF \label{sec:interweave}}

As we already discussed above in \cref{sec:rohf-formulation}, the salient
feature of the FD HF method is that the iterative SCF procedure needed to
solve the HF equations is tightly interwoven with the (MC)SOR iterations used
to compute the potentials. The main algorithm can be summarized as follows
\begin{itemize}
\item initial guess for the orbitals and potentials
\item loop over SCF iterations:
  \begin{itemize}
  \item loop over orbitals $o$ in inverse order
    \begin{itemize}
    \item relaxation step for Coulomb $V_C^{o}$ and exchange
      $V_x^{oo}$ potentials for the orbital itself%
    \item loop over orbitals $p<o$: relaxation step for exchange
      potentials $V_x^{op}$
    \item relaxation step for orbital $o$, using updated values for
      $V_C^{o}$, $V_x^{oo}$, and $V_x^{op}$ for $p<o$, and previous
      values for $V_C^{p}$ and $V_x^{op}$ for $p>o$
    \end{itemize}
  \end{itemize}
\item reorthogonalize orbitals using the Gram--Schmidt method
\item recalculate orbital energies
\item if orbital energies have changed considerably, update the
  potentials in the $\mu_\infty$ boundary region (see \cref{sec:potential-boundary})
\end{itemize}
If the maximum change in orbital energies between successive SCF
iterations has decreased by a given factor (1.15 by default; see the
\texttt{multipol} keyword in the User's Guide), the boundary values
for potentials at $\mu_{\infty}$ are re-evaluated. This approach
follows from the observation that the boundary values for the
potentials do not change considerably between consecutive SCF
iterations, which is why is no need to update them that frequently if
the orbitals are not changing much.  This approach thus saves many
processor cycles.

As we discussed above in \cref{sec:fd-xtwodhf}, each relaxation
employs the same numerical stencil for all the inner grid points; only
the boundary lines require special handling. Each (MC)SOR relaxation of an
orbital or a potential consists of $N_{\rm SOR}$ macroiterations,
which each involve $N_{\rm mSOR}$ microiterations, and the algorithm
for each relaxation step can be summarized as follows
\begin{itemize}
\item loop over $N_{\rm SOR}$ macroiterations ($N_{\rm SOR}=1$ by default):
  \begin{itemize}
  \item pad the $f(\nu_i,\mu_j)$ array with boundary values (negative $\mu$ and $\nu$ values)
  \item loop over $N_{\rm mSOR}$ microiterations ($N_{\rm mSOR}=10$ by default):
    \begin{enumerate}
    \item update $f(\nu_i,\mu_j)$ according to \cref{eq:update-rule} (various orderings possible for grid points) in a loop over inner grid points $i\in[2,N_\nu-1]$ and $j \in [2,N_\mu-4]$
    \item only for even functions: compute new function values along
      the $(\nu,0)$, $(0,\mu)$, and $(\pi,\mu)$ boundary lines from
      the relaxed solution in the interior grid points with
      \cref{eq:boundary-values} (odd functions vanish by symmetry)
    \end{enumerate}
  \item extract the solution from the padded array
  \item only if relaxing orbitals (potentials were discussed above), update boundary values in the $\mu_\infty$ boundary region (see \cref{sec:orbital-boundary})
  \end{itemize}
\end{itemize}

The above procedure thus determines the SCF cycle. When
self-consistency has been reached, the wave function matches with the
value computed from the converged potential; thereby the norm of the
orbital before reorthonormalization will already be close to unity,
and the relative change in the orbital eigenvalue will be close to
zero.  These parameters are usually defined by the user in the input
data, but the default values for these parameters are $10^{-10}$ in
\xtwodhf{}.

The distinction between the macro- and micro-iterations is a
consequence of the manner in which the boundary conditions are
applied.  The most straightforward approach would be to relax an
orbital at all inner grid points and then update the boundary values
along the internuclear axis and at infinity. In practice, there is no
need to update the boundary values at infinity in each SOR
sweep. Consequently, the micro-iterations entail updating the boundary
values at the internuclear axis in each SOR sweep, followed by an
update to the boundary values at infinity, thus forming a single
macro-iteration. The default setting is to perform a single
macro-iteration. In the case of potentials, the boundary values at
infinity are updated even less frequently, as explained above.

\def\scalea{0.68}
\begin{figure}
\caption{The compound error in orbital energies $\Delta E$
  (\cref{eq:ave-eorb-change}) as a function of the total number of SOR
  iterations taken for three choices for the number of SOR
  microiterations ($N_{\rm SOR}$) within each SCF iteration for FH
  ($3\sigma^2 1\pi^4$ configuration), \ce{KrH+} ($8\sigma^2 4\pi^4
  1\delta^4$ configuration), and TlF ($17\sigma^2 9\pi^4 4\delta^4
  1\varphi^4$ configuration). Few differences can be seen between the
  results for the three choices, confirming that the main effect comes
  just from total number of SOR iterations taken.  Note that the same
  $N_{\rm SOR}$ values are used for orbitals and potentials and that
  is why the labels read: \texttt{sor 8 8}, \texttt{sor 10 10} and
  \texttt{sor 12 12}. }
\label{fig:sorDeltae}
{\footnotesize
\begin{center}
\begin{tabular}{c}
\includegraphics[width=\scalea\textwidth]{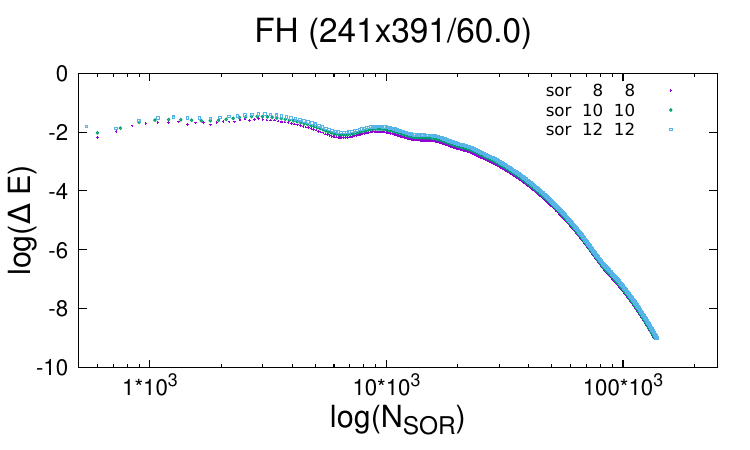}\\
\includegraphics[width=\scalea\textwidth]{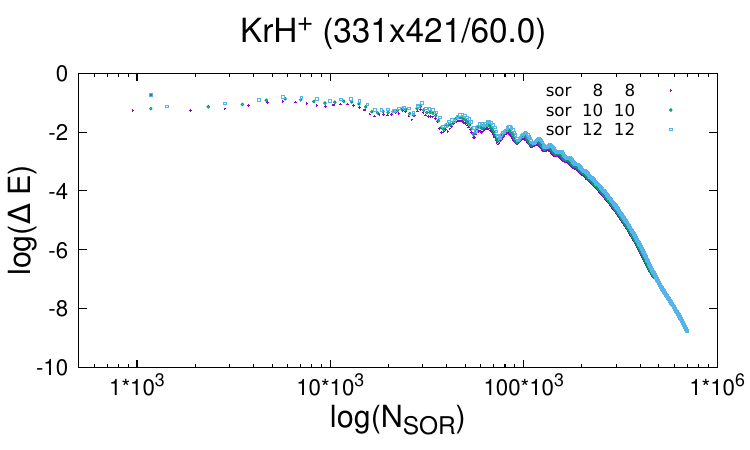}\\
\includegraphics[width=\scalea\textwidth]{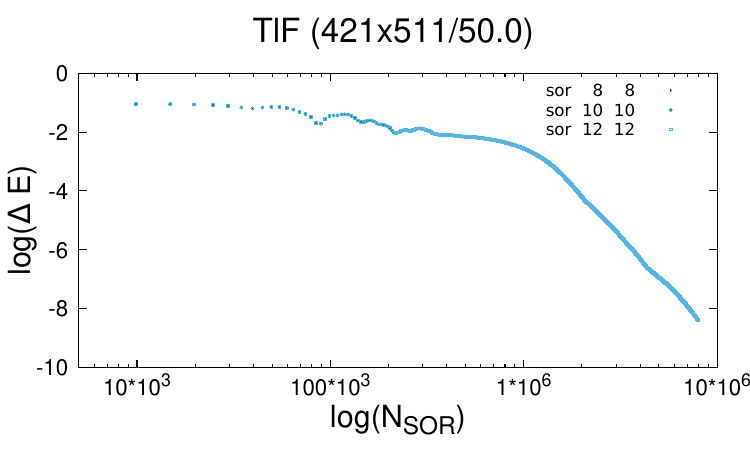}
\end{tabular}
\end{center}
}
\end{figure}

\subsubsection{Default number of (MC)SOR iterations \label{sec:sor-defaults}}

The default number of (MC)SOR microiterations for orbitals and potentials
during a single SCF iteration, $N_{\rm mSOR} = 10$, should be satisfactory
for ordinary cases, as the convergence of the FD HF solution usually
depends on the overall number of (MC)SOR microiterations needed to reach a
given orbital energy threshold, rather than on a judiciously chosen value
for this parameter.  This is illustrated in \cref{fig:sorDeltae}, which
shows the average change in the orbital energy between consecutive SCF
iterations as a function of the total number of SOR microiterations.
Various choices for $N_{\rm mSOR}$ lead to the same behaviour, confirming
that only the total number of SOR steps taken in the calculation is
relevant.

To add on the discussion in \cref{fig:sorDeltae}, we note that if
$N_{\rm mSOR}$ is too small (2--6), more time than necessary will be
spent on the calculation.  If $N_{\rm mSOR}$ is too large
(e.g. 14--18), instead, the overall convergence can be somewhat
faster, but the solution obtained may be distorted if the boundary
conditions get ``out of sync'' with the orbital energies
(see \cref{sec:scf-procedure} for discussion on the SCF algorithm).

The calculations in \cref{fig:sorDeltae} were carried out for
molecules with small, medium, and large number of orbitals and
potentials: FH, \ce{KrH+}, and TlF. Our analysis in
\cref{fig:sorDeltae} rests on the average change in the orbital energy
defined as
\begin{equation}
  \Delta E = \frac {1} {N_{\rm orb}} \sum_{i=1}^{N_{\rm orb}} |\Delta \varepsilon_{i}|
  \label{eq:ave-eorb-change}
\end{equation}
where $\Delta \varepsilon_{i}$ represents the relative change in the
orbital energy for a given orbital between two consecutive SCF iterations
and $N_{\rm orb}$ is the number of occupied orbitals.  It is worth noting
that according to \cref{fig:sorDeltae}, the orbitals and potentials are
improved at a virtually constant rate of $\Delta E\approx 10^{-2}$ per SCF
iteration during the first 15--20\% of SOR microiterations. The remaining
80--85\% of microiterations are needed to improve the accuracy of orbitals
and potentials, as measured by the $\Delta E$, by six orders of magnitude.

\subsubsection{Choice of overrelaxation parameter \label{sec:relaxation}}

The convergence of the SOR method depends critically on the value of the
overrelaxation parameter $\omega$. To investigate the performance of
various choices for the orbital overrelaxation parameter
$\omega_{\rm orb}$, we studied the \ce{Ne^{9+}} system, for which we
relaxed the six lowest orbitals of the $\sigma$, $\pi$, $\delta$, and
$\phi$ symmetries by the SOR and MCSOR methods on a $241\times391/60.0 a_0$
grid until the orbital energy threshold $10^{-10}$ was reached.  10 SOR
microiterations were carried out in each iteration.

Since \ce{Ne^{9+}} is a one-electron system, the sought-for orbitals have
well-known exact analytical expressions and it is not necessary to compute
Coulomb or exchange potentials; the SCF/SOR procedure boils down to
relaxing the orbitals, orthonormalizing them, and recomputing orbital
energies. As we still do want to study the efficiency of the numerical
algorithm, the orbitals were initialized to those of \ce{Na^{10+}}, which
is also a one-electron system.  The resulting guess is sufficiently close
to the pursued solution, and the SOR calculations converge without
difficulties for a wide range of overrelaxation parameters.

\def\scaleb{0.47}
\begin{center}
\begin{figure}
\begin{tabular}{cc}
\includegraphics[width=\scaleb\textwidth]{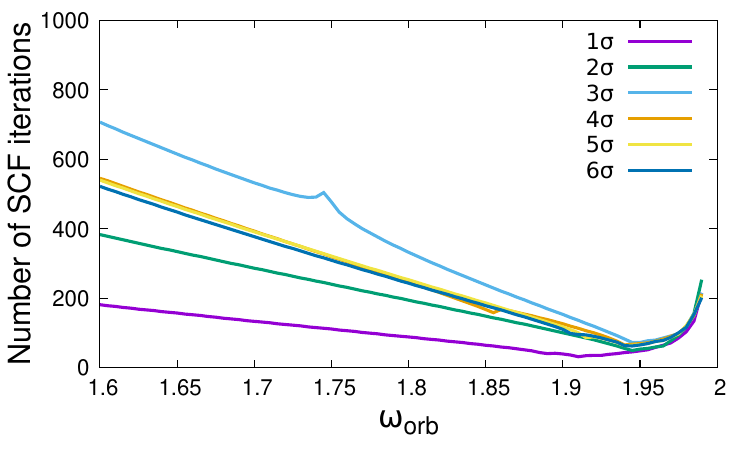}
&
\includegraphics[width=\scaleb\textwidth]{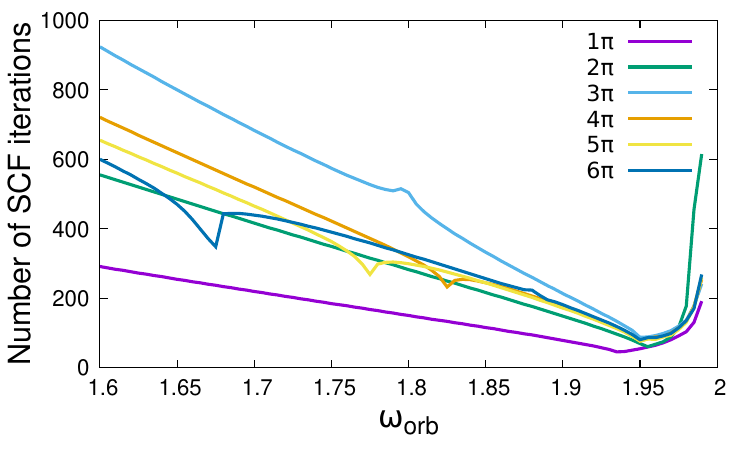}\\

\includegraphics[width=\scaleb\textwidth]{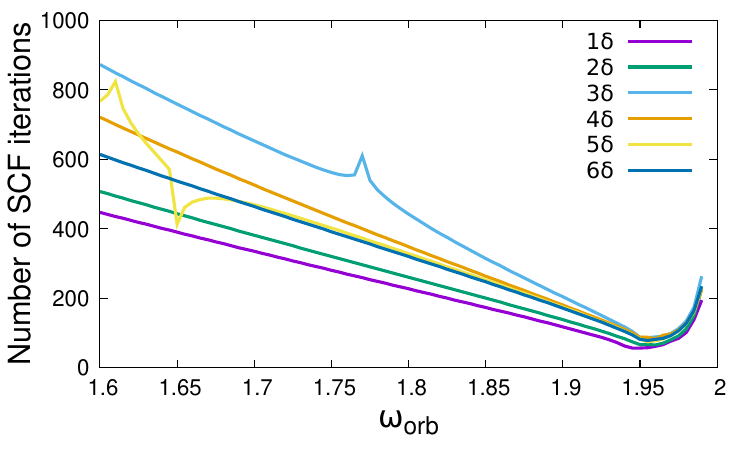}
&
\includegraphics[width=\scaleb\textwidth]{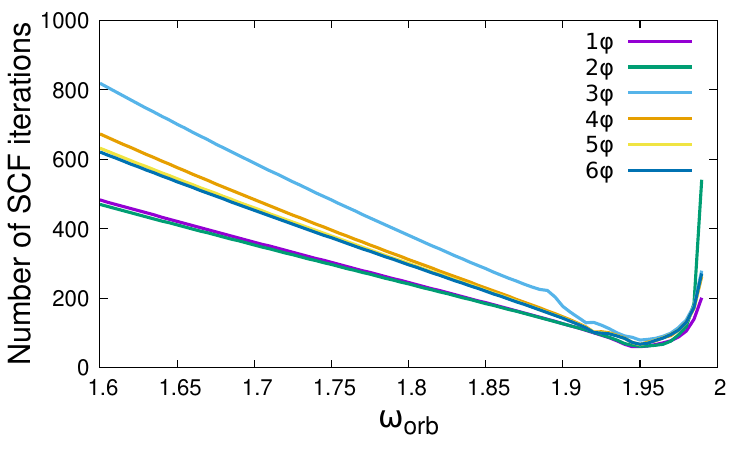}\\
\end{tabular}
\caption{The number of SCF/MCSOR iterations needed to converge the
  \ce{Ne^{9+}} orbitals as a~function of $\omega_{\rm orb}$.}
\label{fig:ne9psor}
\end{figure}
\end{center}
\begin{center}
\begin{figure}
\begin{tabular}{cc}
\includegraphics[width=\scaleb\textwidth]{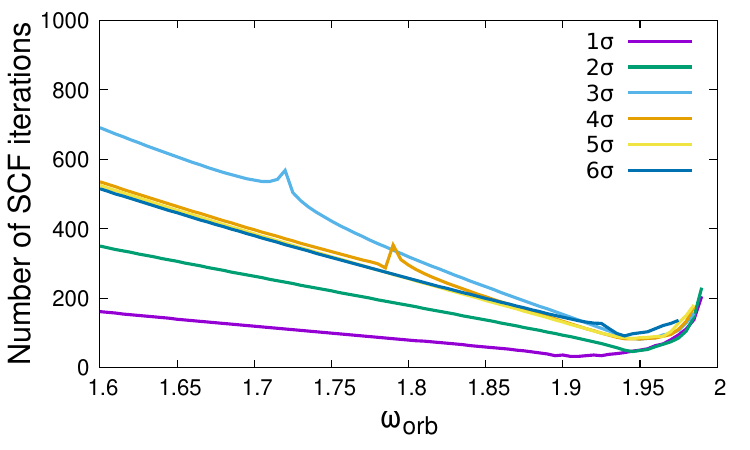}
&
\includegraphics[width=\scaleb\textwidth]{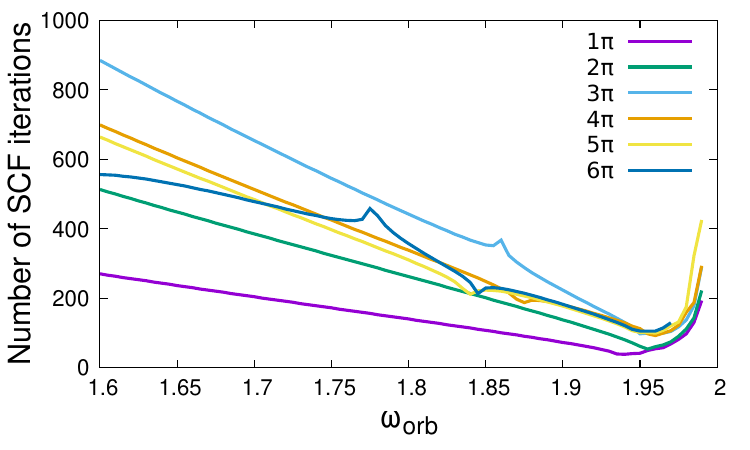}\\

\includegraphics[width=\scaleb\textwidth]{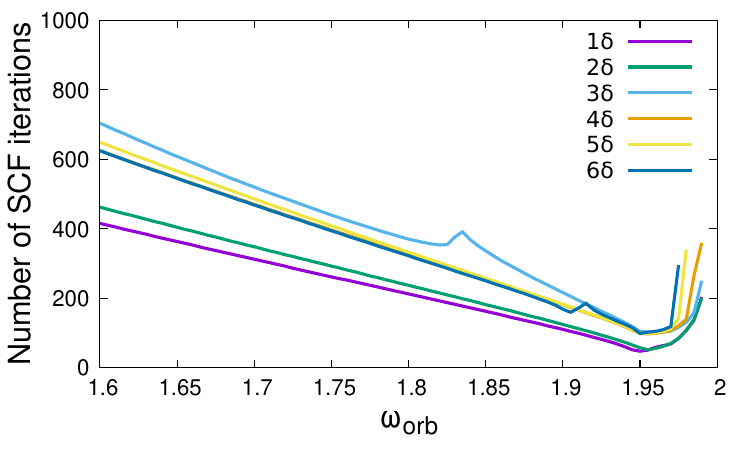}
&
\includegraphics[width=\scaleb\textwidth]{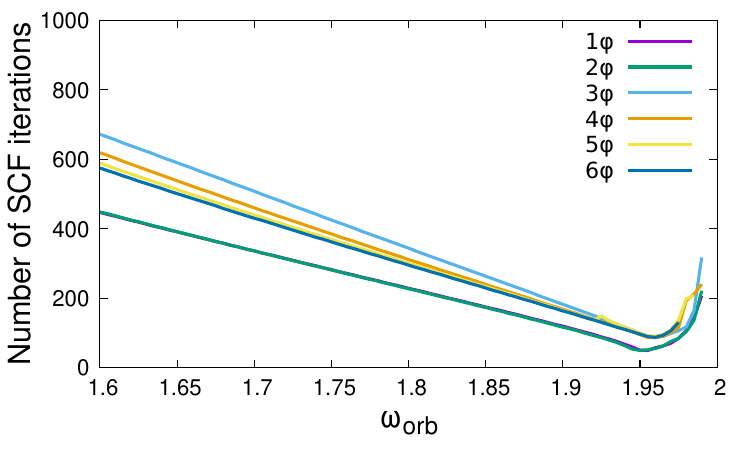}\\
\end{tabular}
\caption{The number of SCF/MCSOR iterations needed to converge the \ce{Ne^{9+}} orbitals as a function of $\omega_{\rm orb}$.}
\label{fig:ne9pmcsor}
\end{figure}
\end{center}

\Cref{fig:ne9psor,fig:ne9pmcsor} show the number of iterations needed
to converge the orbitals as a function of the employed orbital relaxation
parameter $\omega_{\rm orb}$ for the SOR and MCSOR methods, respectively.
We do not observe qualitative differences in the convergence behaviour
between orbitals of different symmetries, or the relaxation algorithms
used. In all cases, the optimal values of the orbital relaxation parameter
$\omega_{\rm orb}$ are virtually the same---around
$\omega_{\rm orb} \approx 1.95$---and the corresponding necessary number of
SCF iterations also turn out to be similar.

A FD HF calculation also requires a relaxation parameter for the
potential, $\omega_{\rm pot}$.  The same qualitative behaviour as
above is observed also when solving the HF equations of any
many-electron system with respect to both relaxation parameters,
$\omega_{\rm orb}$ and $\omega_{\rm pot}$: the required number of SCF
iterations decreases when the parameters approach their optimal values
from the left, and rapidly increases when the values become too large,
especially when they start to approach $\omega_{\rm orb} = 2$ or
$\omega_{\rm pot} = 2$. This behaviour is demonstrated in the
two-dimensional plots in \cref{fig:fh-omegas,fig:krh-omegas}, which
show the required number of SCF iterations for the FH and \ce{KrH+}
molecules, respectively, as a color bar on in the $(\omega_{\rm orb},
\omega_{\rm pot})$ plane.

\def\scaleb{0.65}
\begin{center}
\begin{figure}
\begin{tabular}{c}
\hspace*{2.5cm}\includegraphics[width=\scaleb\textwidth]{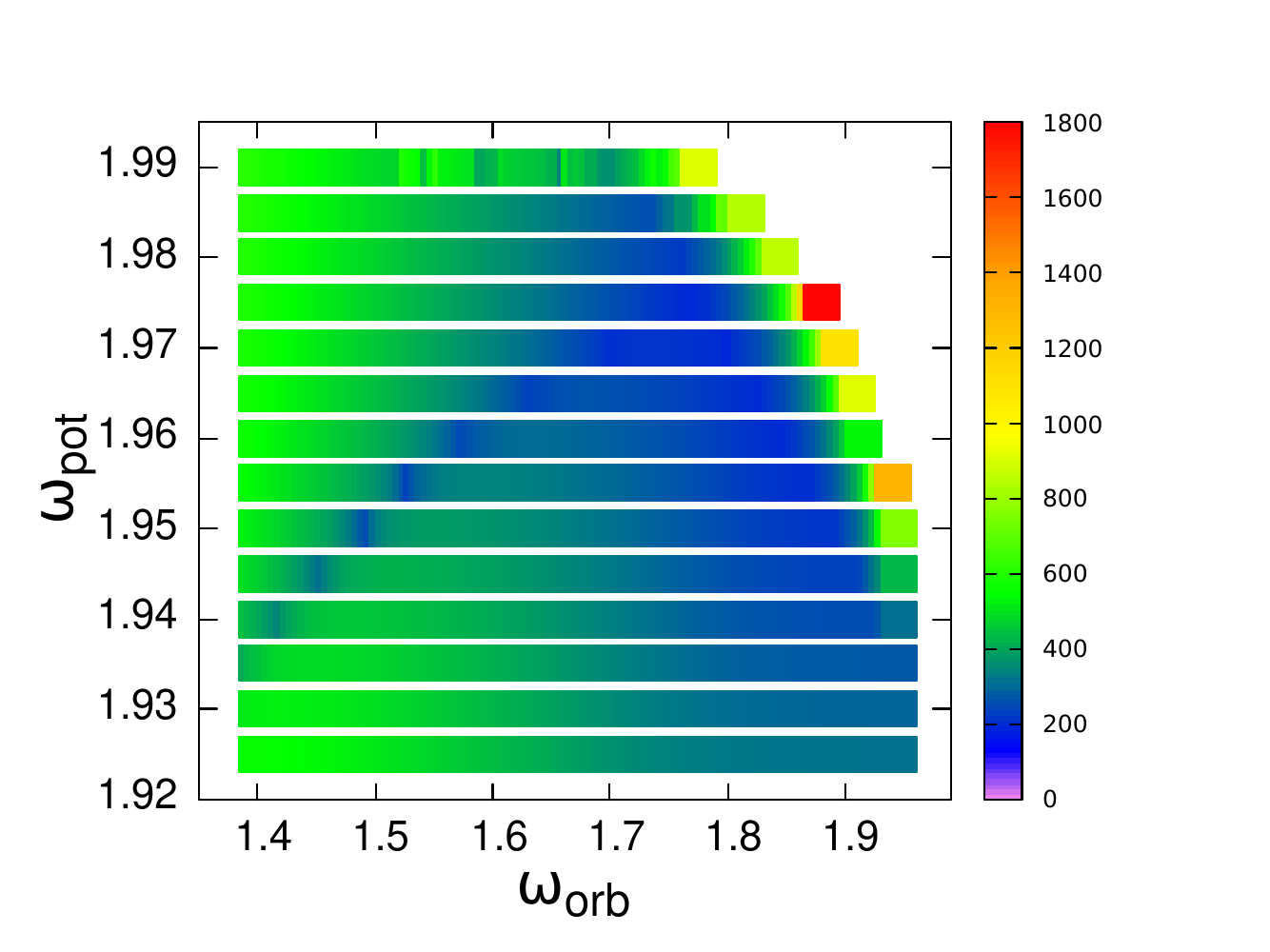}
\end{tabular}
\caption{The number of SCF iterations required to converge FH on a
  $[91\times181/65.0a_0]$ grid as a function of $\omega_{\rm orb}$ and
  $\omega_{\rm pot}$.  The default value $N_{\rm mSOR}=10$ was employed.}
\label{fig:fh-omegas}
\end{figure}
\end{center}
\begin{center}
\begin{figure}
\begin{tabular}{c}
\hspace*{2.5cm}\includegraphics[width=\scaleb\textwidth]{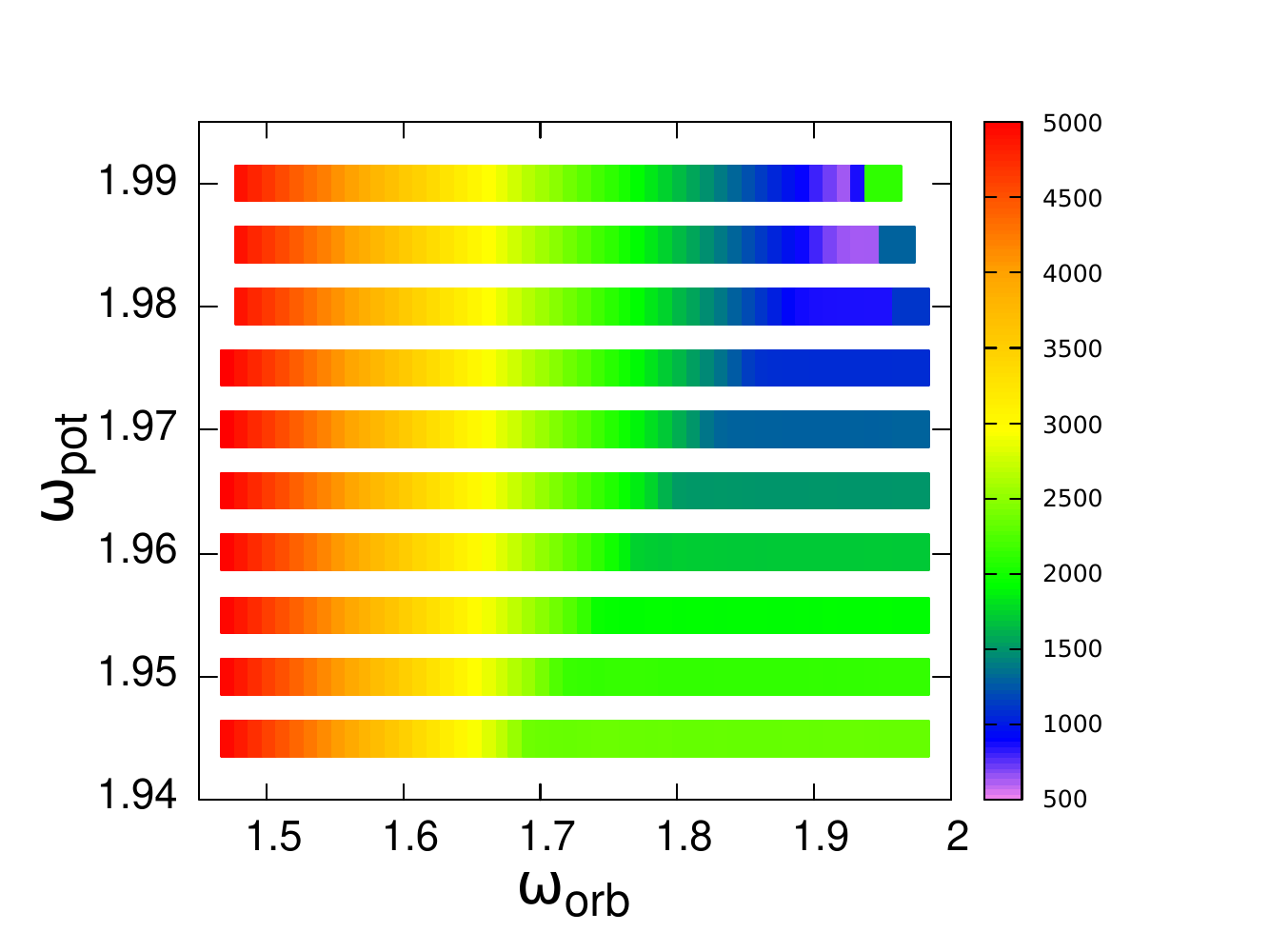}
\end{tabular}
\caption{The number of SCF iterations required to converge \ce{KrH+} on a $[349\times421/60.0a_0]$ grid
as a function of $\omega_{\rm orb}$ and $\omega_{\rm pot}$.
The default value $N_{\rm mSOR}=10$ was employed.}
\label{fig:krh-omegas}
\end{figure}
\end{center}

\subsubsection{Nearly optimal default values \label{sec:default-values}}

One can show that the optimal value $\omega^{\rm opt}_{\rm pot}$ of
the potential relaxation parameter for a model Poisson equation
discretized by the second-order stencil is given by the formula
\cite{StoerB:1980}
\begin{equation} \label{eq:omega-opt}
  \omega^{\rm opt}_{\rm pot}(N_\nu,N_\mu)=\frac{2}{1+\sqrt{1-\rho(N_\nu,N_\mu)^2}}
\end{equation}
where
\begin{equation} \label{eq:omega-opt-rho}
  \rho(N_\nu,N_\mu) = \frac 1 2 \left[\cos \frac {\pi} {N_\nu} + \cos \frac {\pi} {N_\mu} \right]
\end{equation}
is the spectral radius of the corresponding matrix. We are not aware
of analogous results for higher-order discretizations, or for the
general elliptical second-order PDEs like \cref{eq:gelliptic} that
arise in the FD HF method.  However, it has been shown that the
(nearly) optimal value of the overrelaxation parameter for the
potentials in this case can be approximated by the formula
\begin{equation}
\label{eq:omega-semiopt}
\omega^{\rm pot}_{\rm opt}(N_{\nu},N_{\mu})=\frac{2P_1}{1+\sqrt{1-\rho(N_{\nu},N_{\mu})^2}}+ P_2
\end{equation}
where $P_1\approx 0.603$ and $P_2\approx 0.79$ are constants
determined from numerical experiments; the current default values of
\xtwodhf{} were determined in \cite{Sobczak:2002en}.

We note that the (nearly) optimal values of $\omega_{\rm opt}^{\rm orb}$ are
always smaller than the corresponding $\omega_{\rm opt}^{\rm pot}$ values for a
given grid size \cite{Kobus:2013}.  This observation initially led us
to the automatic estimate
\begin{equation}
  \omega_{\rm opt}^{\rm orb}=\omega_{\rm opt}^{\rm pot}(\omega_{\rm opt}^{\rm pot}-1), \label{eq:omega-orbopt}
\end{equation}
where it is assumed that $1 < \omega_{\rm opt}^{\rm pot} < 2$.
However, this estimate turns out to be a bit too large for heavier
species like Kr or Rn, as the convergence of the resulting SCF process
is observed to be non-monotonic. As a remedy, we ended up further
decreasing $\omega_{\rm opt}^{\rm orb}$ from \cref{eq:omega-orbopt} to
\begin{equation}
 \omega_{\rm opt}^{\rm orb}=\omega_{\rm opt}^{\rm pot}(\omega_{\rm opt}^{\rm pot}-1) - 10^{-3} \max(Z_A,Z_B), \label{eq:omega-orb-revised}
\end{equation}
which is how the optimal value for the orbital relaxation parameter is
estimated in the current version of \xtwodhf{}.

\subsubsection{Potential for future improvements}
We note that there may be room to improve the convergence of the SOR
method. Bai and Chi \cite{Bai:2003} proposed a class of asymptotically
optimal SOR methods. Numerical tests with the five-point finite difference
discretization show that the new methods are more efficient and robust
than the classical SOR method. Li and Evans \cite{Li:1998} claim that
the convergence of the SOR method can be improved if the linear system
is preconditioned by Gauss--Jordan elimination. Such methods could be
investigated in future versions of the \xtwodhf{} program.

\section{Description of the program \label{sec:program-description}}

\subsection{Repository structure \label{sec:repository-structure}}

\xtwodhf{} is freely and openly available as a GitHub repository
\cite{x2dhf} under the GNU General Public License v2.0, or later
(GPL-2.0-or-later).  The repository contains README.md and INSTALL
files that describe how to build and run the program.  An in-depth
users' guide, which describes the program's input data can be found in
\texttt{docs/users-guide.pdf}. The \texttt{test-sets} subdirectory
contains over two hundred of examples of input data, \texttt{input$*$.data}, as
well as the corresponding reference output of the program in the
\texttt{re\-fe\-rence.lst}. The \texttt{bin/testctl} script should be
used to list and run the tests.

The \texttt{lda\_orbitals} and \texttt{hf\_orbitals} directories
contain tabulations of LDA and HF atomic radial orbitals which can be
used to initialize calculations (see \cref{sec:init-guess}).

The program is built with \texttt{CMake}.  The \texttt{bin} directory
contains the \xtwodhf{} executable(s) and several Bash and Perl
scripts to facilitate the usage of the program. The most important of
these is the \texttt{xhf} Bash script, which should be used to run the
program for a given set of input data (see \texttt{xhf -h}). The
directory also contains the \texttt{pecctl} script, which facilitates
calculations of potential energy curves, and the \texttt{elpropctl}
script that computes electrical properties out of total energies and
multipole moments evaluated at several finite field strengths.

\subsection{Flow of control via pseudocodes \label{sec:flow-control}}

\begin{singlespace}
\begin{algorithm}
  \caption{{\sc x2dhf} (main routine)}
  \label{alg:x2dhf}

  \DontPrintSemicolon % Some LaTeX compilers require you to use \dontprintsemicolon instead
  \KwIn{Input data defining the atomic/diatomic system to be solved by
    means of HF, DFT, or an independent particle model}

    \KwOut{At convergence, orbitals and corresponding orbital
      energies, Coulomb/exchange potentials, multipole moments, and
      CPU usage statistics}

  call setPrecision
  \tcp*[f]{calculate floating-point precision}\\
  \tcp*[f]{and lengths of integer/real variables}\\[\extraskip]
  
  call setDefaults
  \tcp*[f]{set default values of constants/variables}\\[\extraskip]
  
  call inputData
  \tcp*[f]{set up calculation according to input} \\[\extraskip]
  
  allocate
  \tcp*[f]{allocate memory for orbitals, potentials, etc}\\
  \tcp*[f]{whose sizes are defined by input data}\\[\extraskip]

  call zeroArray8
  \tcp*[f]{zero out arrays with integer*8 indices}\\[\extraskip]

  call initArrays
  \tcp*[f]{initialise common block arrays,}\\
  \tcp*[f]{one-electron potentials,}\\
  \tcp*[f]{integration and differentiation weights,}\\
  \tcp*[f]{Jacobians, address arrays, mesh arrays, etc}\\[\extraskip]

  call printCase
  \tcp*[f]{print out details about the calculation}\\[\extraskip]
  
  call initOrbPot
  \tcp*[f]{initialise orbitals and potentials}\\[\extraskip]

  call prepSCF
  \tcp*[f]{normalise and orthogonalise orbitals,}\\
  \tcp*[f]{calculate orbital/total energy}\\[\extraskip]

  call SCF \tcp*[f]{perform SCF/(MC)SOR iterations}\\

  call printResults
  \tcp*[f]{output total energy, orbital energies,}\\
  \tcp*[f]{norms, CPU usage statistics}\\[\extraskip]
  % \tcp*[f]{multipole moments and some CPU usage statistics}\\[\extraskip]

  stop 
\end{algorithm}

\end{singlespace}

The large-scale structure of the program is best seen by examining the
pseudocode of its main routine (see \cref{alg:x2dhf}).  The program
uses several parameters that are hard-coded to values defined in the
\texttt{params} module: for example, the maximum number of grid points
in the $\nu$ and $\mu$ variables and the maximum number of orbitals
are fixed at compile time to values given in \texttt{params.f90},
since some arrays are statically allocated for simplicity and
efficacy.

The program's execution begins by determining the employed precision
of the integer and floating-point variable data types (step 1 in \cref{alg:x2dhf}).  Next,
the default values of a host of variables used to control the
behaviour of the program are set (step 2 in \cref{alg:x2dhf}).  Some of these values can,
however, be modified by the input data read next from the
\texttt{input.data} file by the \texttt{inputData} routine and its
many subroutines (step 3 in \cref{alg:x2dhf}).

When the input data has been parsed, the memory requirements for the
various necessary data structures\footnote{Such as the orbitals and
  potentials, the differentiation and integration coefficients, the A, B,
  etc arrays of the Fock and Poisson equations, and the scratch memory. In
  addition, the \texttt{(mc)sor} routine employs two integer arrays to pick
  up subsequent values for relaxation from the primary and extended arrays
  of orbitals/potentials, for example.}  (\cref{sec:appendix-initArrays}
lists some of them) for the calculation can be determined, and the
necessary memory is allocated (step 4 in \cref{alg:x2dhf}). The fixed
length arrays are stored in common blocks.  Necessary arrays that have the
same content for any job type are then either initialized to zero (step 5
in \cref{alg:x2dhf}), or filled with data (step 6 in \cref{alg:x2dhf}).

\subsubsection{Initial guess \label{sec:init-guess}}
At this stage (step 7 in \cref{alg:x2dhf}), the program prints out
information on the type of calculation to run, and proceeds to the next
stage, which is to initialise the orbitals and potentials (step 8 in
\cref{alg:x2dhf}).

There are several options for starting the calculations from a linear
combination of atomic orbitals (LCAOs), which are an attractive form
for describing the molecular orbitals for a given system in compact
form.

In simple cases, the most straightforward way is to define the
molecular orbitals as a linear combination of hydrogenic functions
located on the centres $A$ and $B$.  The traditional setup of
\xtwodhf{} thus requires inputting the orbitals as linear combinations
of Slater-type orbitals on the two nuclei.

However, this approach becomes tedious for systems with many
orbitals. As an alternative, results from the \gaussian{} program can
also be used to initialize molecular orbitals: the basis set used and
the molecular orbital expansion coefficients are extracted from the
program's output, and values of the molecular orbitals at the grid
points are computed from these data.

The procedure to set up calculations has been further improved
and simplified in this version of \xtwodhf{} by providing ready-to-use LDA
(available for H--Og) and HF atomic orbitals (available for H--Ca, Ga,
Br--Sr, I--Ba, Au--Pb, At--Ra) as part of the \xtwodhf{} distribution.
Such good atomic basis functions allow for a compact representation of
the guess orbitals. If expansion coefficients have not been explicitly set
in the input file, \xtwodhf{} initializes the molecular orbitals with
the HF or LDA atomic orbitals ordered in energy; if the molecule is
homoatomic, the gerade and ungerade molecular orbitals are formed by
equal weighting of the atomic orbitals on the two centers.

The LDA orbitals have been determined in
exchange-only LDA calculations with the \helfem{} program
\cite{Lehtola2019_IJQC_25945, Lehtola2020_PRA_12516,
  Lehtola2023_JCTC_2502}, while the HF atomic orbitals were obtained from a
modified version of Froese-Fischer's program \cite{Fischer:1977}.

The special aspect of the FD HF approach is that in addition to the
orbitals, an initial guess must be provided for the potentials, as well; a
further complication here is that a guess for the total Coulomb and
exchange potential is not enough, but separate guesses have to be
formed for all the Coulomb potentials arising from single orbital
densities, as well as all the exchange potentials arising from the
various orbital products.

Coulomb potentials are initialized from a linear combination of
Thomas--Fermi potentials at the two centres; our implementation is based on
a routine from the atomic program of Desclaux
\cite{Desclaux:1975}. However, if LDA orbitals were used to
  initialize the calculation, the Coulomb potentials can also be
  initialised with the corresponding superposition of atomic Coulomb
  potentials. All exchange potentials are initialised as
$c_A/r_A+c_B/r_B$, where $c_A$ and $c_B$ are normalised LCAO coefficients.

\subsubsection{SCF procedure \label{sec:scf-procedure}}
Having completed the guess stage, the guess orbitals are
orthonormalised, and the orbital energies and the total energy are
calculated (step 9 in \cref{alg:x2dhf}). In the next stage the SCF
process is carried out (step 10 in \cref{alg:x2dhf}) and finally, its
results are printed out (step 11) before the execution of the program
ends (step 12 in \cref{alg:x2dhf}).

\begin{singlespace}
\newcommand{\forconda}{iscf=1 \KwTo maxscf}
\newcommand{\forcondb}{iorb=norb \KwTo 1}

\begin{algorithm}
  \caption{{\sc SCF} performs SCF/(MC)SOR iterations}
  \label{alg:scf}
  \DontPrintSemicolon % Some LaTeX compilers require you to use \dontprintsemicolon instead
  \KwIn{An initial set of ({norb}) orbitals and Coulomb/exchange potentials
    and corresponding orbital energies. Every {saveSCFdata} iterations
    total energy is recalculated and restart data are saved
    to disk.}

  \KwOut{On convergence (or otherwise) orbitals and Coulomb/exchange potentials and
    orbital energies.}

  \For{\forconda}{ \tcp*[f]{SCF loop} \\
  \For{\forcondb}{  \tcp*[f]{orbitals' loop; orbitals stored in reverse order} \\[\extraskip]    

    call potAsympt
    \tcp*[f]{set potentials at $R_{\infty}$} \\[\extraskip]    
    call relaxDriver
    \tcp*[f]{perform (MC)SOR iterations} \\
    \tcp*[f]{for potentials and orbitals}\\[\extraskip]
    
    call norm   \tcp*[f]{normalise orbitals} \\[\extraskip]    

    call ortho  \tcp*[f]{orthogonalise orbitals}    \\[\extraskip]
    
    call EaHF $||$ EaLXC $||$ EaDFT
    \tcp*[f]{calc orbital energy} \\[\extraskip]

    call EabHF $||$ EabLXC $||$ EabDFT
    \tcp*[f]{calc off-diagonal}\\
    \tcp*[f]{Lagrange multipliers}    \\[\extraskip]

    $\Delta$Ea(iorb)=Ea(iscf)-Ea(iscf-1)
    \tcp*[f]{monitor changes}\\
    \tcp*[f]{in orbital energies}\\[\extraskip]    
    
    $\Delta$Na(iorb)=Na(iscf)-Na(iscf-1)
    \tcp*[f]{monitor changes}\\
    \tcp*[f]{in orbital norms}\\[\extraskip]    
  }
  
  $\Delta${E}=max( $|\Delta$Ea(1)$|$,...,$|\Delta$Ea(norb)$|$ )\\
  $\Delta${N}=max( $|\Delta$Na(1)$|$,...,$|\Delta$Na(norb)$|$ )\\[\extraskip]
    
    \If{ iscf>Nscf2skip} {
      \tcp*[f]{at least Nscf2skip SCF iterations are performed}\\[\extraskip]
    
      \If{ $\Delta${E} $<$ E$_{\rm threshold}$ \KwSty{or} $\Delta${N} $<$
        N$_{\rm threshold}$} {  
        \Return{}
        \tcp*[f]{orbital energy/norm threshold reached}\\
      }
    }
  }
\end{algorithm}

\end{singlespace}

The SCF process is controlled by the SCF routine (see
\cref{alg:scf}). In each SCF iteration, the orbitals are processed in
an opposite order to the one used to define the electronic
configuration in \texttt{input.data}. For example, the input data for
the FH molecule at the experimental equilibrium bond length
$R=1.7328a_0$ is shown in \cref{fig:fh-input}.
\begin{figure}
\begin{verbatim}
title FH
method hf
nuclei 9.0 1.0 1.7328
config 0
1 pi
3 sigma end
grid 200 40.0
orbpot hf
scf 2000 10 10 14 3
stop
\end{verbatim}
  \caption{Input data for a HF calculation on the FH molecule
    ($R=1.7328a_0$) with a~$200 \times 200/40a_0$ grid.}
\label{fig:fh-input}
\end{figure}

With such a definition for the electronic configuration, the Fock equation
for the $1\sigma$ orbital, which is the orbital with the lowest energy, is
solved first.%
\footnote{This also requires calculating the orbital energy and the
off-diagonal Lagrange multipliers for the employed model. In DFT
calculations, the functionals can be provided either by the Libxc
library (\texttt{EaLXC} and \texttt{EabLXC}), or by the program itself
(\texttt{EaDFT} and \texttt{EabDFT}).}  The relaxation of the
$1\sigma$ orbital is preceded by the relaxation of the $V_C^{1\sigma}$
potential. One also needs the exchange potential $V_x^{1\sigma
  1\sigma}$; however, it is easy to see from
\cref{eq:coulomb,eq:exchange} that diagonal exchange potentials for
$\sigma$ orbitals coincide with their Coulomb potentials:
$V_x^{1\sigma 1\sigma} = V_C^{1\sigma}$.  Naturally, the $1\sigma$
orbital update also depends on the Coulomb and exchange potentials of
the other orbitals. However, at this stage they are treated as given.

Next, the Fock equation has to be solved for the $2\sigma$
orbital. However, the $V_x^{1\sigma2\sigma}$ and $V_C^{2\sigma} =
V_x^{2\sigma 2\sigma}$ potentials have to be determined first.  Again,
the potentials of the higher orbitals are used as-is.

In the same spirit, the relaxation of the $3\sigma$ orbital again
begins with the relaxation of the $V_x^{1\sigma3\sigma}$,
$V_x^{2\sigma3\sigma}$ and $V_C^{3\sigma} = V_x^{3\sigma 3\sigma}$
potentials, while fixed values are used for the $V_x^{3\sigma1\pi}$
and $V_C^{1\pi}$ potentials.

Finally, the relaxation of the $1\pi$ orbital is preceded by the relaxation
of the $V_C^{1\pi}$, $V_x^{1\pi1\pi}$, $V_x^{3\sigma1\pi}$,
$V_x^{2\sigma1\pi}$ and $V_x^{1\sigma1\pi}$ potentials. Note that the
Coulomb and exchange potentials do not coincide for non-$\sigma$ orbitals,
and therefore have to be computed separately. For example, the Coulomb
potential $V_C^{1\pi}$ arises from a cylindrically symmetric density, while
the exchange potential $V_x^{1\pi1\pi}$ contains both same-orbital
($1 \pi_{\pm1}$--$1\pi_{\pm1}$; $|\Delta m| = 0$) and other-orbital
($1 \pi_{\pm1}$--$1\pi_{\mp1}$; $|\Delta m| = 2$) contributions.%
\footnote{It is important to note that, in principle, the Coulomb and
exchange potentials could be evaluated directly. Nevertheless, the
present approach is preferred due to the expense associated with
calculating potentials via a series of integrations in two variables.}

\begin{singlespace}
\newcommand{\hs}{\hspace*{2em}}
\begin{algorithm}
  \caption{{\sc relaxDriver} controls SOR/MCSOR relaxation of orbitals and potentials}
  \label{alg:relaxDriver1}
  \DontPrintSemicolon % Some LaTeX compilers require you to use \dontprintsemicolon instead
  % \KwIn{}
  \KwIn{Orbital to relax (by default lpotmcsor=.FALSE. and lorbmcsor=.FALSE.).}

  \KwOut{Orbital and corresponding Coulomb and exchange potentials
    updated by maxsor1*maxsor2 (MC)SOR iterations.}

  \If {lpotmcsor }  {
    \KwSty{\#ifdef} OPENMP\\
    \hs call coulExchMCSOR
    \tcp*[f]{Coulomb/exchange potenitals}\\
    \tcp*[f]{relaxed in parallel in separate}\\    
    \tcp*[f]{OpenMP threads via MCSOR routine}\\
    \tcp*[f]{parallelised by another group of}\\
    \tcp*[f]{OpenMP threads}\\[\extraskip]    

    \KwSty{\#elif} PTHREAD $||$ TPOOL\\
    \hs call coulExchSORPT
    \tcp*[f]{Coulomb/exchange potentials}\\
    \tcp*[f]{relaxed in parallel in separate}\\
    \tcp*[f]{p-threads by MCSORPT routine}\\[\extraskip]
    
    \KwSty{\#else}\\
    \hs call coulExchMCSOR 
    tcp*[f]{Coulomb/exchange potenitals}\\
    \tcp*[f]{relaxed one by one via single-threaded MCSOR}\\[\extraskip]
    \KwSty{\#endif}\\  
  }
  \Else {
    \KwSty{\#ifdef} OPENMP\\
    \hs call coulExchSOR
    \tcp*[f]{Coulomb/exchange potenitals for}\\
    \tcp*[f]{a given orbital relaxed in separate}\\
    \tcp*[f]{OpenMP threads using SOR}\\[\extraskip]    

    \KwSty{\#elif} PTHREAD $||$ TPOOL\\
    \hs call coulExchSORPT
    \tcp*[f]{Coulomb/exchange potenitals}\\
    \tcp*[f]{for a given orbital relaxed in}\\
    \tcp*[f]{separate p-threads using SORPT}\\[\extraskip]        
    \KwSty{\#else} \\
    \hs call coulExchSOR
    \tcp*[f]{Coulomb/exchange potenitals for}\\
    \tcp*[f]{a given orbital relaxed one by}\\    
    \tcp*[f]{one within main execution thread}\\[\extraskip]        
  }
 \tcp*[f]{continued in \textbf{Pseudocode 4}}\\  
\end{algorithm}

\begin{algorithm}
  \caption{({Pseudocode 3} continued) {\sc relaxDriver} controls SOR/MCSOR
    relaxation of orbitals and potentials}
  \label{alg:relaxDriver2}
  \DontPrintSemicolon 

  \If {lorbmcsor }
  {  
    \KwSty{\#if} PTHREAD $||$ TPOOL\\
    \hs call orbMCSORPT
    \tcp*[f]{relax orbital using}\\
    \tcp*[f]{mcsor\_pthread or mcsor\_tpool C routines}\\[\extraskip]    
    
    \KwSty{\#else}\\
    \hs call orbMCSOR
    \tcp*[f]{relax orbital using}\\
    \tcp*[f]{single-threaded MCSOR}\\[\extraskip]        
    \KwSty{\#endif}\\  
  }
  \Else {
    call orbSOR \tcp*[f]{relax orbital using SOR}    
    }
    
    \Return{}
\end{algorithm}

\end{singlespace}

The \texttt{SCF} routine calls \texttt{relaxDriver} to perform
relaxations of the corresponding potentials for each orbital in turn;
the potentials can be relaxed in parallel (see discussion in
\cref{sec:parallel}).  The relaxation of the orbital itself is carried
out subsequently (see \cref{alg:relaxDriver1,alg:relaxDriver2}).

\begin{singlespace}
\newcommand{\forconda}{{nexchpot=1} \KwTo {nexchpots(iorb)}}

\begin{algorithm}
  \caption{{\sc coulExchSOR} prepares data for the SOR relaxations and
    performs the macro and micro SOR relaxations of Coulomb and exchange
    potentials for a given orbital. If pragma OPENMP is used each
    potential is relaxed in a separate thread. Poisson equation is
    abbreviated as PE.}
  \label{alg:coulExchSOR}

  \DontPrintSemicolon % Some LaTeX compilers require you to use \dontprintsemicolon instead
  % \KwIn{}
  \KwIn{Orbital number {iorb}.}

  \KwOut{Coulomb and exchange potentials updated by {maxsor1} macro and
    {maxsor2} micro SOR iterations.}

  \For{\forconda}{
    \tcp*[f]{loop over Coulomb/exchange potentials for}\\
    \tcp*[f]{a given orbital {iorb}} \\[\extraskip]      

    {ib1=i1b(ins1(iorb,nexchpot))} \\    
    {ib2=i1b(ins2(iorb,nexchpot))}
    \tcp*[f]{get location of orbitals}\\
    \tcp*[f]{within psi array needed for {nexchpot}}\\
    \tcp*[f]{Coulomb/exchange potential}\\[\extraskip]
    
    {ibexp=ibexcp(iorb,nexchpot)}
    \tcp*[f]{get location of}\\
    \tcp*[f]{Coulomb/exchange potential within {excp} array} \\[\extraskip]         
    %\tcp*[f]{potential within excp array that contains all Coulomb/exchange potentials} \\         

    {deltam=idelta(iorb,nexchpot)}
    \tcp*[f]{get difference of m} \\         
    \tcp*[f]{quantum numbers for the product of orbitals}\\[\extraskip]                      

    {isym=isyms(iorb,nexchpot)}
    \tcp*[f]{get the symmetry of {nexchpot}} \\
    \tcp*[f]{potential} \\[\extraskip]             

    {rhs=G*psi(ib1:)*psi(ib2:)}
    \tcp*[f]{prepare right-hand side of PE}\\

    {lhs=F3+deltam*E}
    \tcp*[f]{prepare left-hand side of PE}\\
    \tcp*[f]{with diagonal part of the diff. operator incl.}\\[\extraskip]             
    
    \tcp*[h]{perform maxsor1*maxsor2 SOR iterations}
    
    \renewcommand{\forconda}{{i=1} \KwTo {maxsor1}}

    \For{\forconda}{   
      call putin (isym,excp(ibexp:),work)
      \tcp*[f]{immerse}\\  
      \tcp*[f]{excp(ibexp:) in work array}\\
      \tcp*[f]{and add extra boundary values} \\[\extraskip]        

      call sor (isym,work,LHS,RHS,BPOT,D,...)
      \tcp*[f]{perform}\\
      \tcp*[f]{maxsor2 micro SOR iterations}\\[\extraskip]                        

      call putout (excp(ibexp:),work)
      \tcp*[f]{extract updated}\\
      \tcp*[f]{excp(ibexp:) values out of {work} array}\\[\extraskip]                                  
    }
  }
    
  \Return{}
\end{algorithm}

\end{singlespace}

\Cref{alg:coulExchSOR} shows the details of \texttt{coulExchSOR}
routine. Several arrays are prepared during the initialization phase of the
program, so that a simple loop can be used to prepare the right- and
left-hand sides of the Poisson equation for a particular Coulomb or
exchange potential, and relax the equation by means of \texttt{maxsor1}
macroiterations. The SOR routine is called in each macroiteration,
performing the \texttt{maxsor2} micro iterations.

\begin{singlespace}
\newcommand{\forconda}{{nexchpot=1} \KwTo {nexchpots(iorb)}}

\begin{algorithm}
  \caption{{\sc orbSOR} prepares data for the SOR relaxations and
    performs the macro and micro SOR relaxations of a given
    orbital. The program can be used to solve the HF equations, the
    DFT equations, or the HFS equations that involve $X\alpha$
    exchange where the $\alpha$ parameter is calculated according to
    the self-consistent multiplicative constant (SCMC) method. It can
    also be used to solve one-electron diatomic problems (OED) or the
    harmonium problem (TED). }
    \label{alg:orbSOR}

    \DontPrintSemicolon % Some LaTeX compilers require you to use \dontprintsemicolon instead
  % \KwIn{}
  \KwIn{Orbital number iorb.}

  \KwOut{Orbital updated by \texttt{maxsor1} SOR macroiterations and \texttt{maxsor2} SOR microiterations.}

  \tcp*[h]{Prepare LHS and RHS of Fock eq. for a given method}

  if (HF.or.OED) call fockHF (iorb)\\[\extraskip] 
  if (TED) call fockTED (iorb)\\[\extraskip] 
  if (DFT.or.HFS.or.SCMC) call fockDFT(iorb)\\[\extraskip] 
%    {\#ifdef LIBXC} 
    if (LXC) then \tcp*[f]{use xc functionals from libxc}\\
    \quad if (lxcPolar) then\\
          \quad \quad call fockLXCpol(iorb)\\
       \quad else\\
          \quad \quad {call fockLXCunpol(iorb)\\
      \quad endif\\ 
      endif\\[\extraskip] 

      isym=isymOrb(iorb) \tcp*[f]{get symmetry of orbital}\\[\extraskip]       
       
      \renewcommand{\forconda}{{i=1} \KwTo {maxsor1}}

      \tcp*[h]{perform maxsor1 macro SOR iterations}
      
      \For{\forconda}{  
        lhs=fock1+diag
        \tcp*[f]{add diagonal part of}\\
        \tcp*[f]{2$^{\rm nd}$ derivatives}\\[\extraskip] 

        call orbAsymptGet
        \tcp*[f]{get data needed to update orbital}\\
        \tcp*[f]{in asymptotic region after relaxation}\\[\extraskip] 
       
        call putin (isym,psi(iborb:),work)
        \tcp*[f]{immerse psi(iborb:)} \\  
        \tcp*[f]{into work array and add extra boundary values} \\[\extraskip] 

        call sor (isym,work,LHS,RHS,B,D,...)
        \tcp*[f]{perform}\\
        \tcp*[f]{maxsor2 micro SOR iterations}\\[\extraskip]                  
        
        call putout (psi(iborb:),work)
        \tcp*[f]{extract updated}\\        
        \tcp*[f]{psi(iborb:) values out of work array}\\[\extraskip]                   
        
        call orbAsymptSet
        \tcp*[f]{update tail region of orbital}\\[\extraskip]
    }
  }
    \Return{}

\end{algorithm}

\end{singlespace}

Once the potentials have been updated, the \texttt{orbSOR} routine (or
its equivalents \texttt{orbSORPT} or \texttt{orbMCSOR}) is called, and
the values of the orbitals are relaxed in a similar fashion to the
\texttt{coulExchSOR} routine (see \cref{alg:orbSOR}).%
\footnote{Depending on the choice of the method (see User's Guide),
the program can be used to solve the HF equations, the DFT equations,
or the HFS equations $X\alpha$ exchange where the $\alpha$
parameter is calculated with the self-consistent
multiplicative constant (SCMC) method. The program can also be used to
solve one-electron diatomic problems (OED) and the harmonium
problem (TED).} The current values of the orbitals and the potentials
are used by the proper version of Fock routine to prepare the right-
and left-hand sides of the Poisson equation for the chosen method
(\texttt{fockHF}, \texttt{fockDFT}, \texttt{fockLXC}, etc).  The
\texttt{orbAsymptGet} routine is invoked in step 14 of
\cref{alg:orbSOR} to get data that will be used to update the orbital
values in the tail region upon completion of the relaxation (see
\cref{sec:orbital-boundary}). The relaxation itself is achieved by the
call to the \texttt{SOR} routine in step 16 of \cref{alg:orbSOR}.

\subsection{Array storage}

As a major new feature in version 3.0 of \xtwodhf{}, the whole program
has been internally restructured in terms of Fortran modules, allowing
the Fortran compiler to check that function arguments are correctly
passed between functions in different compilation units. As a result
of taking advantage of the features of the Fortran 95 standard, the
code has also been simplified and streamlined considerably, making it much
easier to modify and to extend to new functionalities.

The refactoring has especially resulted in significant changes to the way
memory is passed around in the code, which has enabled eliminating many
potential bugs. The use of assumed-size arrays (\texttt{dimension(*)}
in Fortran) has been replaced wherever possible in favor of
assumed-shape arrays \linebreak[4] (\texttt{dimension(:)}), which also pass the
array size within function calls, thus eliminating possible
out-of-bounds errors.

While older versions of the program \cite{Kobus:2013, KobusLS:1996}
passed the arrays to other routines either in their entirety, or as
sub-arrays with particular data extracted, the current version of the
program employs Fortran \texttt{pointer}s to access the data
structures. As is usual with scientific code development
\cite{Lehtola2023_JCP_180901}, refactorings are usually never
complete, and some parts of the program still need further
cleanup and streamlining.

The function $f(\nu,\mu)$ is represented by a 2-dimensional array
$\gv{F}$ in \xtwodhf{}.  The grid points are readily mapped into the
corresponding elements of the two-dimensional array $\gv{F}$ used to
store the $f(\nu_i,\mu_j)$ values in the program; see also
\cref{fig:17ps-sor}.

To perform SOR iterations, the $\gv{F}$ array is immersed into a
larger $(N_{\nu}+8)\times(N_{\mu}+4)$ matrix, which has additional
values along the boundaries that have been computed with knowledge of
the symmetry of the function. The immersion and extraction is handled
by the \texttt{putin} and \texttt{putout} routines, respectively. As
discussed above, padding the solution onto the extended grid allows
for an easier application of the relaxation procedure for the interior
grid points.

The evaluation of the right-hand side of \cref{eq:discret3} for every
grid point $i, j$ can be carried out as a series of matrix-vector
multiplications of the submatrices $\gv{F}(1,j)$ and
$\gv{D}^{\mu}(1,j)$ for every $j$. The evaluation of the analogous
$\nu$ derivatives for a selected $\nu_i$ value and all the $\mu_j$
values can be done analogously by multiplying the corresponding submatrices
$\gv{F}^{\rm T}(1,i)$ and $\gv{D}^{\nu}(1,i)$. All differentiations of the
orbitals and the potentials can thus be carried out efficiently on modern
hardware with a call to the \texttt{dgemv} routine included in the
basic linear algebra subprograms (BLAS) library \cite{blas:2002}.

\subsection{Language, unusual features and limitations \label{sec:language}}

The program has been (re)written in Fortran 95. The program can be
compiled as stand-alone, as it contains simplified replacements of the
employed BLAS routines. Compilation is carried out with
CMake. Optimized BLAS libraries should be used whenever possible (see
\texttt{x2dhfctl -B}).

The command and data file structure is described in a separate document
(see the User's Guide in the repository \cite{x2dhf}). Over two
hundred examples with corresponding inputs and outputs are provided by the
various test sets (see \texttt{testctl -h}).

Several C routines have been added in version 3.0 to facilitate
parallelisation of the SCF process and MCSOR routine via Portable
Operating System Interface threads (pthreads; see
\cref{sec:parallel}). If the multi-threaded version employing
pthreads is requested, a C compiler is therefore also necessary to
build the necessary extensions. OpenMP parallellism is handled with
the standard Fortran extension.

The program can be compiled in quadruple precision with e.g. the
\texttt{-freal\--8-real-16} option of the gfortran compiler (see
\texttt{x2dhfctl -r 16}). As quadruple precision is not supported by
commonly available implementations of the BLAS standard at the moment,
the bundled implementations have to be used for calculations in
quadruple precision, instead.

\subsection{Parallellization speedups \label{sec:parallel}}

As was already discussed above in \cref{sec:scf-procedure}, an SCF
iteration begins by the relaxation of the Coulomb potential of the first
orbital---which should be the one lowest in energy---and the orbital
itself. The relaxation procedure then continues with higher lying
orbitals. When such an orbital is to be relaxed, all the Coulomb and
exchange potentials that involve the lower, already relaxed orbitals, must
be updated. The higher the orbital is, the more exchange potentials must be
relaxed. Therefore, if this work can be done in parallel with the
relaxation of the Coulomb potential for the given orbital, a reasonable
(but system dependent) speedup can be expected.

The \texttt{x2dhfctl} script that is used to build the executable binary
offers three options to handle parallellization of the potential
relaxation. The options \texttt{-o}, \texttt{-p} and \texttt{-t} switch on
OPENMP, PTHREAD, and the TPOOL directive, respectively, which select
separate routines to relax potentials with OpenMP or pthreads.

The Fortran version of the (MC)SOR routine is used when parallellism
is not employed, or when the program is built with OpenMP support.
Parallel relaxation of the potentials with pthreads is carried out by
a C version of the (MC)SOR routine.  If the program has been compiled
with the pragma PTHREAD, pthreads are created whenever the routine
\texttt{coulExch\_pthread} is called and destroyed when the routine
finishes. In contrast, if the pragma TPOOL is used, several pthreads
are created when the program is started, they are used by any call to
the \texttt{coulExch\_tpool} routine, and only destroyed when the
program ends.

The time needed to calculate the multipole moments and to relax the
orbitals and potentials for the same FH, \ce{KrH+} and TlF systems
discussed above in \cref{fig:sorDeltae} is given in \cref{table:sopt}
for the various parallellisation options. These data show that the
relaxation of the potentials can be sped up by a factor of 3 for the
smallest system, and by factor of 15-20 for the largest system
when sufficiently many parallel cores are available.

Since the relaxation of a particular orbital is a single-threaded process,
the speed-up factors for the relaxation of orbitals and potentials together
is limited by Amdahl's law to between 2 and 8. Since the relaxation of the
orbitals and potentials is the dominant step in a FD HF calculation, these
ratios are also a fair estimate of the resulting wall-time speed-ups of the
whole calculation.

\begin{center}
  \begin{table}
   \captionsetup{width=.90\textwidth}
  \caption{\label{table:sopt}%
    Effects of code parallelisation for the FH, \ce{KrH+} and TlF systems.
    The system clock timings (in seconds) are given for the evaluations of
    the multipole moments (\texttt{mm}), relaxation of orbitals
    (\texttt{orbs}), relaxations of the Coulomb and exchange potentials
    (\texttt{pots}) and their sum (\texttt{orb+pots}). In the column
    labelled \texttt{-s} the timings for the single-threaded version of the
    code were given. The columns labelled \texttt{-o}, \texttt{-p} and
    \texttt{-t} give the timings obtained when the pragmas OPENMP, PTHREAD,
    and TPOOL were used during the compilation, respectively. The values in
    parentheses give the speedup.}

  \vspace{10pt}
\begin{center}
  \begin{tabular}{crrrr}
\hline
\toprule
& \multicolumn{1}{c}{\texttt{-s}}
& \multicolumn{1}{c}{\texttt{-o}\ph{aa}}
& \multicolumn{1}{c}{\texttt{-p}\ph{aa}}
& \multicolumn{1}{c}{\texttt{-t}}\ph{aa}\\
& \multicolumn{1}{c}{\texttt{ST}}
& \multicolumn{1}{c}{\texttt{OpenMP}\ph{aa}}
& \multicolumn{1}{c}{\texttt{PTHREAD}\ph{aa}}
& \multicolumn{1}{c}{\texttt{TPOOL}}\ph{aa}\\

\hline\\[-5pt]
& \multicolumn{4}{c}{FH} \\\cmidrule{2-5}
\text{mm} & 3.8 & 1.9 \ph{(3.1)} & 3.8 \ph{(3.1)} & 3.8 \ph{(3.7)}\\
orbs & 138.3 & 116.9 \ph{(3.1)} & 139.8 \ph{(3.1)} & 140.1 \ph{(3.7)} \\
pots & 351.3 & 112.3 (3.1) & 160.0 (2.2) & 128.6 (2.7) \\
orbs+pots & 489.7 & 229.2 (2.1) & 299.8 (1.6) & 268.7 (1.8) \\
\midrule\\[-5pt]
& \multicolumn{4}{c}{KrH$^+$} \\\cmidrule{2-5}
\text{mm} & 9.2 & 1.9 \ph{(3.1)} & 9.3 \ph{(3.1)} & 9.1 \ph{(3.1)} \\
orbs & 81.8 & 83.5 \ph{(3.1)} & 84.1 \ph{(3.1)} & 83.6 \ph{(3.1)} \\
pots & 517.9 & 69.5 (7.5) & 81.5 (6.4) & 65.3 (7.9) \\
orbs+pots & 599.7 & 153.0 (3.9) & 165.6 (3.6) & 148.9 (4.0) \\
\midrule\\[-5pt]

& \multicolumn{4}{c}{TlF} \\\cmidrule{2-5}
\text{mm} & 39.9 & 3.9 \ph{(15.3)} & 39.96 \ph{(15.3)} & 32.88 \ph{(15.3)} \\
orbs & 80.0 & 87.2 \ph{(15.3)} & 82.09 \ph{(15.3)} & 72.22 \ph{(15.3)} \\
pots & 922.6 & 60.5 (15.3) & 63.79 (14.5) & 47.29 (19.5) \\
orbs+pots & 1002.6 & 147.7 \ph{1}(6.8) & 145.89 \ph{1}(6.9) & 119.52 \ph{1}(8.4) \\
\bottomrule
\end{tabular}
\end{center}
\end{table}
\end{center}

\section{A complementary review of related literature \label{sec:review}}

Having discussed the theory behind \xtwodhf{} in \cref{sec:theory}, the
numerical discretization and relaxation procedure in
\cref{sec:solvingPDEs}, and the internal layout of the program in
\cref{sec:program-description}, we proceed with a brief review of related
literature. The main purpose of this section is again to augment the recent
extensive review \cite{Lehtola2019_IJQC_25968} by rediscussing pertinent
work carried out with the \xtwodhf{} program, as well as to discuss the
most recent advances in the field of fully numerical calculations on
diatomic molecules.

\subsection{Early years}

For many years, the \xtwodhf{} program was mainly used to assist the
development and calibration of sequences of universal even-tempered basis
sets.  This effort was initiated by Moncrieff and Wilson in the early 1990s
\cite{MoncrieffW:1993a, MoncrieffW:1993b, KobusMW:1994a, KobusMW:1994b,
  KobusMW:1995, MoncrieffKW:1995, MoncrieffW:1996, MoncrieffKW:1998,
  KobusMW:1999a, KobusMW:2000, KobusMW:2000a, KobusMW:2002}. In the course
of that work, the FD HF method proved to be a reliable source of reference
values of total energies, multipole moments, static polarizabilities and
hyper\-polarizabilities ($\alpha_{zz}$, $\beta_{zzz}$, $\gamma_{zzzz}$,
${A}_{z,zz}$ and ${B}_{zz,zz}$) for atoms, diatomic molecules, and their
ions \cite{KobusMW:2001b, KobusMW:2004, KobusMW:2007, Kobus:2007,
  Kobus:2009, Kobus:2015}.

\subsection{Basis set convergence of total energies}

\xtwodhf{} was also used in the literature to provide HF-limit values for
examining the convergence patterns of properties calculated using the
cc-pVXZ correlation-consistent basis sets of Dunning and coworkers
\cite{Dunning1989_JCP_1007, Woon1993_JCP_1358, Woon1994_JCP_2975,
  Woon1995_JCP_4572, Wilson1996_JMST_339} within the context of CBS models
\cite{Halkier:1999, Halkier:1999a, Jensen:1999a, ChristensenJ:2000,
  Jensen:2000}.

Important studies by Helgaker \etal{} \cite{Helgaker:1997} and Halkier
\etal{} \cite{Halkier:1999a} demonstrated that HF energies exhibit
exponential convergence both with respect to the total number of the
basis functions of a given type, as well as with respect to the
maximum angular momentum of the basis set, while correlation energies
only converge according to an inverse power law.

The differing convergence patters of HF and post-HF calculations have
ramifications for basis set design: the cc-pVXZ basis sets were constructed
to allow the extrapolation of correlation energies to the CBS limit, and
are thereby not optimal for CBS limit extrapolation of HF or DFT total
energies. This was the rationale behind Jensen's project to build
hierarchies of polarization-consistent basis sets specifically tailored to
facilitate CBS limit extrapolation of HF and DFT energies, dipole moments,
and equilibrium distances \cite{Jensen2001_JCP_9113, Jensen2002_JCP_3502,
  Jensen2002_JCP_7372, Jensen2002_JCP_9234, Jensen2003_JCP_2459,
  Jensen2004_JCP_3463, Jensen2007_JPCA_11198, Jensen2012_JCP_114107,
  Jensen2013_JCP_14107}.  Related to this effort, Jensen described
obtaining significantly different or even lower HF limit energies with his
new Gaussian basis sets than the FD HF values reported in the literature
\cite{Jensen2005_TCA_187}.

In an important contribution, Jensen examined the dependence of FD HF total
energies on the grid size for 42 diatomic species composed of first and
second-row elements and reported their energies to better than $\upmu \Eh$
accuracy \cite{Jensen2005_TCA_187}.  Jensen pointed out that large values
for the practical infinity $r_\infty$ (up to $r_\infty=400a_0$ in Jensen's
study) are sometimes necessary to obtain the HF limit in FD HF
calculations, as even though the orbitals have finite range, the potentials
may approach their asymptotic limits only slowly.

The FD HF method was also used by Weigend \etal{} in the construction
of Gaussian basis sets of quadruple zeta valence quality with a
segmented contraction scheme for the H--Kr atoms \cite{Weigend:2003},
and by Petersson and coworkers in the development of the
$n$-tuple-$\zeta$ augmented polarized family of basis sets ($n$ZaP,
$n=1$--6) designed to allow extrapolating both the SCF energy and the
correlation energy to their corresponding CBS limits \cite{ZhongBP:2008}.

The FD HF energies accurate to at least 1$\upmu \Eh$ were calculated
with \xtwodhf{} for 27 diatomic transition-metal-containing species to
investigate the convergence of the HF energies upon increasing the
sizes of correlation-consistent basis sets and augmented basis sets
developed for the transition atoms by Balabanov and Peterson
\cite{WilliamsYW:2008, BalabanovP:2005}. Some of these values were
later revised by Lehtola \cite{Lehtola2019_IJQC_25944} with
calculations with the \helfem{} program.

Sheng \etal{} determined FD HF values for \ce{He2} to study CBS limit
extrapolations of the correlation energy with the aug-cc-pVXZ orbital
basis sets \cite{Sheng:2018}.

\subsection{Basis set convergence of molecular properties}

Halkier and Coriani studied the electric quadrupole moment of the FH
molecule \cite{HalkierC:2001} by estimating the CBS limit of full
configuration interaction (FCI) calculations in Gaussian basis sets.
FD HF calculations were used to check that the largest employed
Gaussian basis sets reproduced the HF quadrupole moment accurately.

Likewise, Pawlowski \etal{} \cite{PawlowskiJH:2004} used the FD HF
method to compute the HF limit dipole polarizability and second
hyperpolarizability of the Ne atom to determine basis set truncation
errors in Dunning's cc-pVXZ basis sets augmented with diffuse
functions (aug-cc-pVXZ). Specifically, the study considered double
(d-aug), triple (t-aug), and quadruple augmentation (q-aug), finding
that at least triple augmentation is required to converge the
hyperpolarizability to 0.05 a.u. from the FD HF value with a 6-$\zeta$
level basis set (t-aug-cc-pV6Z). This knowledge was then used to build
a Gaussian basis with an even closer agreement to the FD HF value.

Roy and Thakkar \cite{RoyT:2002} computed the leading coefficients of the
MacLaurin expansion of the electron momentum density from wave functions
computed with the \xtwodhf{} program, finding large differences from
literature values computed earlier with Slater-type orbital basis sets.

The \xtwodhf{} program was also employed by Shahbazian and Zahedi
\cite{ShahbazianZ:2005} to compare the convergence patterns of total
energies and spectroscopic parameters in the correlation-consistent
and polarization-consistent basis sets in HF calculations on a set of
first-row diatomic molecules, also considering extrapolations to the
CBS limit.

\subsection{Miscellaneous}

Accurate HF values for a set of diatomic molecules were used to proposed
quality measures for Gaussian basis sets \cite{Rappoport:2011} and to test
whether the counterpoise method can be used to correct basis set
superposition effects \cite{Mentel:2013}.

Codes to evaluate the prolate spheroidal harmonics has been reported by Gil
and Segura \cite{Gil1998_CPC_267} and by Schneider and coworkers
\cite{Schneider2010_CPC_2091, Schneider2018_CPC_192}; the latter is used in
\helfem{} for the analytical evaluation of potentials
\cite{Lehtola2019_IJQC_25944}.  Mendl proposed an efficient algorithm for
two-center Coulomb and exchange integrals of prolate spheroidal orbitals
\cite{Mendl2012_JCP_5157}.

Bodoor \etal{} investigated the numerical solution of two-electron pair
equations of diatomic molecules, which turn out to be second order partial
differential equations of 5 variables \cite{BodoorKM:2015}.

\subsection{Steps towards relativistic calculations}

A separate problem is the assessment of relativistic basis functions
used for solving the Dirac--Fock equations. Despite some early
attempts on small systems like \ce{H2+} \cite{Laaksonen1984_CPL_157,
  Laaksonen1984_CPL_485, Sundholm1987_PS_400}; \ce{HeH^{2+}}
\cite{Laaksonen1984_CPL_157}; \ce{HeH+} \cite{Laaksonen1984_CPL_485};
and LiH, \ce{Li2}, BH, and \ce{CH+} \cite{Sundholm1988_CPL_251}; no
general-use relativistic version of the FD HF method has yet been
reported to the best of our knowledge. However, analogous relativistic
finite element variants have been described in the literature
\cite{Lehtola2019_IJQC_25968}.

Still, the two-dimensional, fully numerical finite-difference approach
to the second-order Dirac equation for one-electron diatomics might be
seen as the first step towards the Dirac--Fock method
\cite{KobusK:2022}, and FD HF can be used to indirectly assess the
quality of the basis sets used for relativistic calculations.

Styszy\'nski studied the influence of the relativistic core-valence
correlation effects on total energies, bond lengths and fundamental
frequencies for a series of hydrogen halide molecules (HF, HCl, HBr,
HI and HAt) \cite{Styszynski:2000}. The results of non-relativistic HF
calculations in Gaussian basis sets were compared by Styszy\'nski and
collaborators with those of FD HF calculations to assess the quality
of basis sets and the dependence of spectroscopic constants on the
basis set truncation errors \cite{Styszynski:2000, StyszynskiK:2003,
  MatitoSK:2005}.

\subsection{Modeling irradiation processes}

Calculations of the stopping powers and ranges of energetic ions in
matter within the semiempirical approach developed by Ziegler,
Biersack, and Littmark require the knowledge of interatomic potentials
\cite{ZieglerBL:85}. Nordlund \etal{} \cite{NordlundRS:97} studied
repulsive potentials for C-C, Si-Si, N-Si, and H-Si systems,
pioneering fully numerical HF and LDA calculations as well as
calculations with numerical atomic orbital (NAO) basis sets for these
systems.

Pruneda and Artacho \cite{PrunedaA:2004} extended the analysis to C-C,
O-O, Si-Si, Ca-O, and Ca-Ca systems with fully numerical HF
calculations, which were supplemented with DFT calculations carried
out in a NAO basis set with pseudopotentials.

Kuzmin examined relativistic effects in the potentials of the Kr-C,
Xe-C, Au-C, and Pb-C diatomics with Gaussian basis sets, employing the
FD HF method to validate the accuracy of the used Gaussian basis set
in non-relativistic calculations \cite{Kuzmin:2006,
  Kuzmin:2007}.

More recently, Lehtola \cite{Lehtola2020_PRA_32504} studied
all-electron HF calculations of repulsive potentials of the He-He,
He-Ne, Ne-Ne, He-Ar, Mg-Ar, Ar-Ar, and Ne-Ca systems with the
\helfem{} program, and investigated the accuracy of Gaussian basis set
approaches.

Numerical orbitals obtained from the FD HF method proved to be useful
in the development of a model of high-harmonic generation in diatomic
molecules \cite{MadsenM:2006}.

It turns out that the FD HF method (within the local exchange
approximation) can be of some help when the multiple interatomic
coulombic decay model is employed to analyse the process of
neutralization and deexcitation of highly-charged ions being splashed
from graphene \cite{Wilhelm:2017}. The method was used to calculate
the orbital energies for C-ion separations down to a thousandth of the
atomic unit. Solutions of the HF equations can also be used to
describe the tunnelling ionisation of molecules within a model which
relies on precise values of the asymptotic form of the valence
(tunnelling) orbital \cite{Tong:2002, Madsen:2012, Madsen:2013,
  Madsen:2014, Kornev:2014, Wang:2014, Kornev:2015, Kornev:2016,
  Endo:2016, Kopytin:2019, Romashenko:2020}.

\subsection{Warning about confinement}

Diatomic molecules can be studied in elliptical confinement that
simulates a high pressure environment by enforcing the wave function
to vanish at $r_\infty$ \cite{LeyKoo1981_JCP_4603,
  OlivaresPilon2017_IJQC_25399, OliveiraBatael2020_TCA_129}.  The FD
HF method was recently employed to study the \ce{H2+} and \ce{H2}
systems in their ground states in hard-wall confinement
\cite{Mukherjee:2023}. However, we warn that since \xtwodhf{} relies
on solving the Poisson equation for the orbitals and potentials by
imposing the boundary conditions by asymptotic expansions for the
united atom in gas phase close to $r_\infty$, the use of smaller
values for $r_\infty$ does not yield the correct confined solution, as
incorrect boundary values are being imposed. Proper calculations of
atoms and diatomic molecules in confinement are possible with
\helfem{}, however, as that program uses Green's functions for an
exact solution to the Poisson equation for the potentials, and solves
orbitals with the posed boundary conditions without imposing any
assumed asymptotic behaviour. As a result, \helfem{} yields the correct
result even when hard-wall or soft confinement potentials are
employed.

\subsection{Density functional calculations}

The FD HF approach proved useful in the DFT
context to construct and test various functionals \cite{Liu:1998,
  Ludena:1999, Karasiev:1999, KarasievLA:2000, KarasievL:2002a,
  KarasievL:2002b, Karasiev:2003, KarasievTH:2006} including exchange
energy functionals for excited-states \cite{Hemanadhan:2014}. Recently
the FD HF results together with the machine-learning techniques have
been used to test semi-local kinetic energy functionals on atoms and
diatomics \cite{Wang:2024}.

Makmal \etal{} described the fully numerical all-electron solution of
the optimized effective potential equation for diatomic molecules in
the \texttt{DARSEC} program in 2009 \cite{Makmal2009_JCTC_1731,
  Makmal2009_JCTC_1731}, following a similar finite differences
approach to \xtwodhf{}. The DARSEC implementation has since been
employed in a number of studies on fundamental DFT
\cite{Makmal2009_PRB_161204, Makmal2011_PRA_62512,
  Kraisler2013_PRL_126403, Schmidt2014_PCCP_14357, Schmidt2014_JCP_18,
  Kraisler2015_JCP_104105, Kraisler2015_PRA_32504,
  Schmidt2016_PRB_165120, Schmidt2016_C_33, Banafsheh2017_IJQC_25410,
  Aschebrock2017_PRB_75140, Aschebrock2017_PRB_245118,
  Aschebrock2019_PRR_33082, Aschebrock2019_JCP_154108,
  Garrick2020_PRX_21040, Banafsheh2022_PRA_42812,
  Lebeda2022_PRR_23061, Richter2023_JCP_124117}.

Ch\'{a}vez \etal{} \cite{Chavez2022_JoOSS_4459} described a Python
module for embedding calculations in the prolate spheroidal coordinate
system, which also employs a discretization similar to that in
\xtwodhf{}.

\subsection{Recent work with \helfem{}}

Kraus examined basis set extrapolations in DFT \cite{Kraus2020_JCTC_5712,
  Kraus2021_JCTC_5651} with fully numerical calculations with the \helfem{}
program.

Lehtola studied the numerical behaviour of recent density functionals
in non-self-consistent \cite{Lehtola2022_JCP_174114} as well as
self-consistent atomic calculations \cite{Lehtola2023_JCTC_2502,
  Lehtola2023_JPCA_4180}. Many functionals were found to be
numerically ill-behaved already at fixed electron density
\cite{Lehtola2022_JCP_174114}, while some functionals---such as most
members of the Minnesota family---only exhibit pathological behaviour
in self-consistent calculations, with the issues sometimes arising
already for hydrogen \cite{Lehtola2022_JCP_174114}, or on heavier
atoms, with the Li and Na atoms causing the most problems
\cite{Lehtola2023_JCTC_2502}.

Lehtola \etal{} studied diatomic molecules in strong magnetic fields
\cite{Lehtola2020_MP_1597989} with \helfem{}, finding large basis set
truncation errors in the standard Dunning basis set series in extreme
conditions. In a~follow-up study, Lehtola and \AA{}str\"om
\cite{Aastroem2023_JPCA_10872} identified large errors for also atoms
in strong magnetic fields. Lehtola has also recently discussed the use of
fully numerical wave functions for diatomics to fit atomic orbital
basis sets by the maximal overlap method \cite{Lehtola2024_ES_15015}.

\section{Example results \label{sec:example-results}}

As the review in \cref{sec:review} illustrates, it has been demonstrated many
times over the years that the FD HF method can produce HF-limit
values of total energies, orbital energies, multipole moments, as well as
(hyper)\-po\-lar\-is\-abil\-ities for various atomic and
diatomic systems. As was also discussed in \cref{sec:review}, such results have
been often used to assess the quality of basis sets and to test various schemes
to improve them.

\subsection{Basis set truncation errors in \ce{He2} and other diatomics \label{sec:bastrunc}}

Although Gaussian basis set (GBS) calculations can reach high accuracy
for small molecules \cite{Lehtola2020_JCP_134108}, typically employed
basis sets suffer from truncation errors and basis set superposition
errors, which are highly dependent on the property that is being
studied. As discussed in \cref{sec:review}, various schemes have been
suggested for reducing the error of the total energy, such as
counterpoise (CP) methods \cite{Boys1970_MP_553} to reduce the basis
set superposition error, as~well as various CBS limit extrapolation
techniques.

We exemplify these kinds of investigations in \cref{tab:he2} which
shows the HF interaction energy in \ce{He2} as a function of the
internuclear distance. Results are shown for fully numerical
calculations with \xtwodhf{}, as well as two sets of GBS calculations:
an old calculation of Gutowski \etal{} \cite{GutowskiDCP:1987}, and
a newer one of Varandas \cite{Varandas:2008} that includes CBS and CP corrections.
The data in \cref{tab:he2}
demonstrates the usefulness of having access to fully numerical data:
while the raw, CP corrected, or CBS extrapolated data are sometimes in
excellent agreement with results of fully numerical calculations, the basis set
truncation error varies significantly along the studied geometries.

\Cref{fig:te-err,fig:mm-relerr} demonstrate the accuracy of
distributed, universal, even-tempered basis sets (DUET) in reproducing
total energy and multipole moments for a group of small, medium and
large diatomic molecules \cite{Kobus:2000}.  Again, the numerical
solution offers a way to assess the accuracy of the employed DUET
basis sets, which show considerable differences across molecules.

\begin{table}
  \caption{HF interaction energies of the helium dimer in
    $\upmu \Eh$ from FD calculations \cite{Kobus:1993}. For comparison,
    results of two GBS calculations are also shown: the raw
    $10s4p3d1f$ GBS values of Gutowski \etal{} \cite{GutowskiDCP:1987}
    and GBS values from the doubly augmented cc-pVXZ basis set series
    (d-aug-cc-pVXZ) that have either also been extrapolated to the CBS
    limit with (5$\zeta$,6$\zeta$) extrapolation \cite{Varandas:2008},
    or extrapolated to the CBS limit while also including counterpoise
    (CP) corrections \cite{Boys1970_MP_553}. The last three columns
    show the differences between the values in columns 3, 4, and 5 and
    the FD values in column 2, respectively.
  }
  \label{tab:he2}
%  {\footnotesize
\begin{tabular}{crccccccc}
\toprule\\[-10pt]
  & \multicolumn{4}{c}{interaction energy}
  &
  & \multicolumn{3}{c}{basis set truncation error} \\
   \cmidrule{2-5}    \cmidrule{7-9}    \\[-10pt]
   \multicolumn{1}{c}{R(au)}
  & \multicolumn{1}{c}{FD$^{\rm a}$}
  & \multicolumn{1}{c}{GBS1$^{\rm b}$} %\cite{gut87}}
  & \multicolumn{1}{c}{GBS2$_{\rm CBS}^{\rm c}$} %\cite{varadas2008}}
  & \multicolumn{1}{c}{GBS2$_{\rm CBS+CP}^{\rm c}$} %\cite{varadas2008}}
  &
  & \multicolumn{1}{c}{$\Delta_{}$}
  & \multicolumn{1}{c}{$\Delta_{\rm CBS}$}
  & \multicolumn{1}{c}{$\Delta_{\rm CBS+CP}$}  \\
\hline\\[-7pt]
  3.0  & 13517.08  &          13518.21  &          13517.50   &        13517.02
                  &&  \phantom{-}1.13  &  \phantom{-1}0.42     &      {-}0.06  \\
  3.5  &  4335.76  & \phantom{1}4336.16 & \phantom{1}4336.27  & \phantom{1}4335.79
                  &&    \phantom{-}0.40 & \phantom{-1}0.51     & \phantom{-}0.03  \\
  4.0  &  1357.88  & \phantom{1}1358.03 & \phantom{1}1358.24  & \phantom{1}1357.91
                  && \phantom{-}0.15    & \phantom{-1}0.36     & {-}0.14\\
  4.5  &   416.54  & \phantom{11}416.60 & \phantom{11}416.75  & \phantom{11}416.55
                  && \phantom{-}0.06    & \phantom{-1}0.21     & {-}0.05\\
  5.0  &   125.55  & \phantom{11}125.57 & \phantom{11}125.72  & \phantom{11}125.53
                  && \phantom{-}0.02    & \phantom{-1}0.17     &  {-}0.04   \\
  5.4  &    47.58  & \phantom{115}47.59 & \phantom{115}47.71  & \phantom{115}47.57
                  && \phantom{-}0.01    & \phantom{-1}0.13     & {-}0.02  \\
  5.6  &    29.20  & \phantom{135}29.21 & \phantom{135}29.31  & \phantom{135}29.19
                  && \phantom{-}0.01    & \phantom{-1}0.11     &  {-}0.02  \\
  5.8  &    17.88  & \phantom{135}17.89 & \phantom{135}17.97  & \phantom{135}17.88
                  && \phantom{-}0.01    & \phantom{-1}0.08     &  {-}0.01  \\
  6.0  &    10.94  & \phantom{135}10.94 & \phantom{1135}11.0  & \phantom{135}10.93
                  && \phantom{-}0.00    & \phantom{-1}0.06     & {-}0.01   \\
  6.5  &     3.17  & \phantom{1351}3.18 & \phantom{1135}3.23  & \phantom{1351}3.17
                  && \phantom{-}0.01    & \phantom{-1}0.06     & {-}0.01  \\
  7.0  &     0.92  & \phantom{1151}0.92 & \phantom{1135}0.98  & \phantom{1151}0.91
                  && \phantom{-}0.00    & \phantom{-1}0.06     &         {-}0.01  \\
  7.5  &     0.26  & \phantom{1151}0.26 & \phantom{1135}0.30  & \phantom{1151}0.26
                  && \phantom{-}0.00    & \phantom{-1}0.04     & \phantom{-}0.00  \\
  8.0  &     0.08  & \phantom{1151}0.06 & \phantom{1135}0.07  & \phantom{1151}0.08
                  &&           -0.02    & \phantom{1}-0.01    & \phantom{-}0.00  \\
  \hline
  \multicolumn{9}{l}{$^{\rm a}$\citeref{Kobus:1993},
                     $^{\rm b}$\citeref{GutowskiDCP:1987},
                     $^{\rm c}$\citeref{Varandas:2008}, the (5,6) extrapolation values}\\
  \bottomrule
\end{tabular}
\end{table}

\def\scalea{0.52}
\begin{center}
\begin{figure}
\begin{tabular}{c}
  \includegraphics[width=\scalea\textwidth]{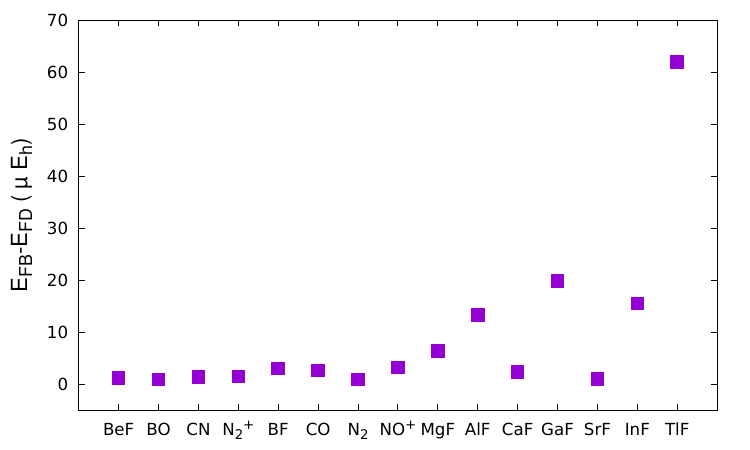}\hspace{0.1cm}
  \includegraphics[width=\scalea\textwidth]{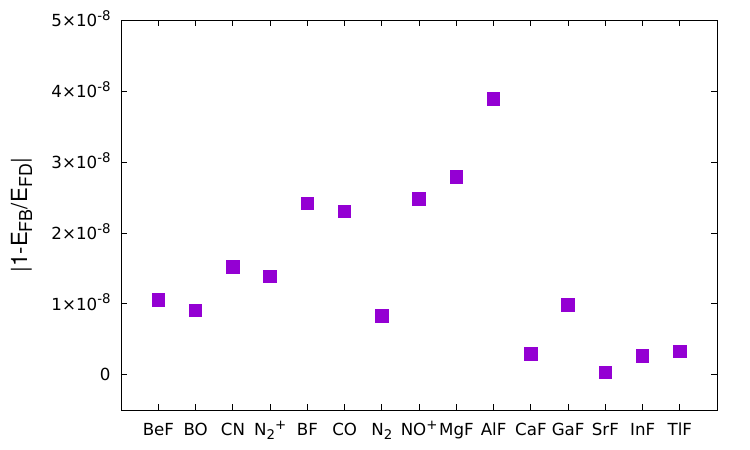}\\
\end{tabular}
\caption{Absolute ($\upmu \Eh{}$, left plot) and relative errors of the
  total energy calculated using DUET basis sets for a range of diatomic
  molecules \cite{Kobus:2000}.}
\label{fig:te-err}
\end{figure}
\end{center}

\def\scalea{0.92}
\begin{center}
\begin{figure}
\begin{tabular}{c}
  \includegraphics[width=\scalea\textwidth]{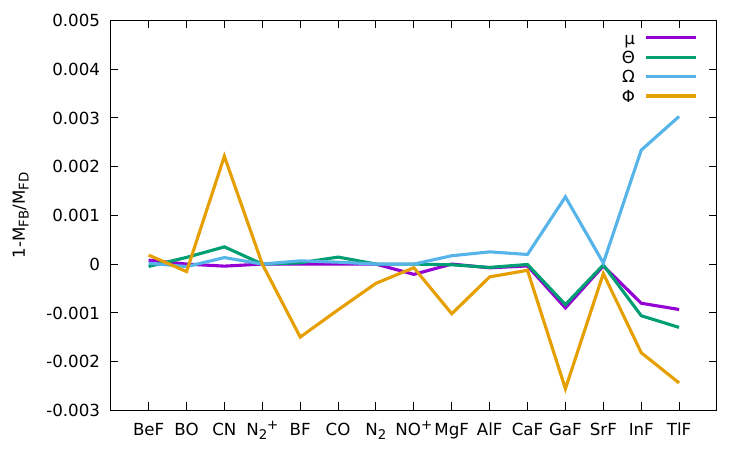}
\end{tabular}
  \caption{Relative errors of $\mu$, $\Theta$, $\Omega$ and $\Phi$ multipole
    moments calculated using DUET basis sets (${\rm M}_{\rm FB}$) and FD HF
    method (${\rm M}_{\rm FD}$)  \cite{Kobus:2000}.}
\label{fig:mm-relerr}
\end{figure}
\end{center}

\subsection{Harmonium atom \label{sec:harmonium-results}}

\begin{figure}
%\caption{The logarithm of the $\sigma$ ground-state energy $E$ of the intracule
%  as a function of the logarith of $\omega_r$ parameter.}
  \caption{The log-log plot of the $\sigma^2$ ground-state energy, $E$ ($\Eh{}$), of
    the intracule as a~function of the $\omega_r$ parameter.}

\label{fig:intracule}
\begin{center}
\begin{tabular}{c}
\includegraphics[width=\scalea\textwidth]{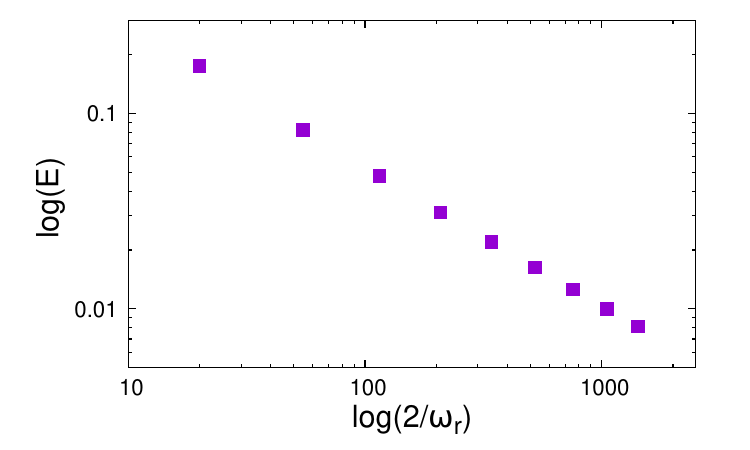}
\end{tabular}
\end{center}
\end{figure}

The intracule equation (\cref{eq:intracule}) can be solved by the
present version of the program.%
\footnote{The extracule equation (\cref{eq:extracule}) could also be
solved with the FD HF method, but recovering well-known eigenvalues is
of no special interest. However, the scheme could be easily used to
find solutions of some distorted harmonic potentials.}
To verify the implementation, we have solved the intracule equation
for various values of $\omega_r$ and collected the results in
\cref{tab:intracule}, where we have also included results of Taut
\cite{Taut:1993} for comparison. The dependence of the $\sigma$
ground-state energy on the $\omega_r$ parameter is also shown in
\Cref{fig:intracule}.

The Hamiltonian in \cref{eq:intracule} is a combination of a repulsive
Coulomb interaction with charge $1/2$ and a harmonic confinement
potential, whose strength is controlled by $\omega_r$. Unsurprisingly,
the smaller the $\omega_r$ parameter is, the more extended the
solution becomes, which is reflected in a poorer accuracy of the
numerical solution observed in \cref{tab:intracule}.

Supplementing the results in \cref{tab:intracule}, we note that the
program easily reproduces the correct solution for $\omega_r=1/2$
($2/\omega_r = 4$), with the energy equalling $5/8$. Higher states can
also be calculated with \xtwodhf{}, and the energies of the first two
excited $\sigma$ states are obtained as $0.804\,828\,530 \Eh$ and
$1.021\,806\,949 \Eh$.

\begin{table}
  \caption{The $\sigma^2$ ground-state energy, $E$, of the intracule
    (\cref{eq:intracule}) as a function of the $\omega_r$
    parameter. $|1-N|$ denotes the norm error. The results of Taut
    \cite{Taut:1993} are also given for reference.}
\label{tab:intracule}
\begin{center}
\begin{tabular}{ccccr}
\toprule
\multicolumn{1}{c}{$2/\omega_r$}
& \multicolumn{1}{c}{$E/\Eh$}
& \multicolumn{1}{c}{$|1-N|$}
& \multicolumn{1}{c}{$E[{\rm Taut}]/\Eh$}
& \multicolumn{1}{c}{Grid} \\
\hline
$\ph{12}20.000\,0$ & $0.175\,000\,000\,000$ & $8\times10^{-13}$ & $0.175\,0$ &
$151\times\ph{1}241/100.0a_0$\\

$\ph{12}54.738\,6$ & $0.082\,208\,885\,019$ & $4\times10^{-12}$ & $0.082\,2$ &
$151\times\ph{1}241/100.0a_0$\\

$\ph{1}115.229\ph{\,6}$ & $0.047\,723\,028\,034$ & $5\times10^{-11}$ & $0.047\,7$ &
$151\times\ph{1}271/200.0a_0$\\
$\ph{1}208.803\ph{\,6}$ & $0.031\,129\,840\,981$ & $3\times10^{-11}$ & $0.031\,1$ &
$151\times\ph{1}301/350.0a_0$\\
$\ph{1}342.366\ph{\,6}$ & $0.021\,906\,383\ph{\,444}$ & $8\times10^{-8\ph{1}}$ & $0.021\,9$ &
$241\times\ph{1}511/550.0a_0$\\

$\ph{1}523.102\ph{\,6}$ & $0.016\,249\,219\ph{\,951}$ & $6\times10^{-8\ph{1}}$ & $0.016\,2$ &
$631\times1441/750.0a_0$\\
$\ph{1}758.124\ph{\,6}$ & $0.012\,541\ph{\,701\,402}$ & $5\times10^{-6\ph{1}}$ & $0.012\,5$ &
$631\times1351/450.0a_0$\\

$1054.54\ph{0\,6}$ & $0.009\,956\,947\ph{\,919}$ & $6\times10^{-8\ph{3}}$ & $0.010\;0$ &
$1261\times2851/650.0a_0$\\

$1419.47\ph{0\,6}$ & $0.008\,106\ph{\,343\,906}$ & $3\times10^{-6\ph{1}}$ & $0.008\,1$ &
$1261\times2851/650.0a_0$\\
\bottomrule
% \multicolumn{4}{l}{$^{\rm h}${Grid: $871\times1711/260.0$}}\\
\end{tabular}
\end{center}
\end{table}

Having established that the program finds the correct ground state for
various values of $\omega_r$, we proceed by studying higher-lying states of
harmonium for $\omega=1/2$. The orbital and total energies of the 12
lowest states of the $\sigma^2$, $\pi^2$, $\delta^2$, or $\phi^2$
configurations at the HF level of theory are given in \cref{tab:hooke1}.

All but four of the orbitals are well converged. The outliers---the
$2\sigma_g$, $2\pi_u$, $2\delta_g$ and $2\phi_u$ orbitals---have gaps
of only some m$\Eh$ to their higher lying neighbours (the $3\sigma_g$,
$3\pi_u$, $3\delta_g$ and $3\phi_u$ orbitals).  These differences are
too small to be properly treated by the present implementation of the
SOR method, as performing the calculations on more refined grids or in
quadruple precision did not resolve the problem.

According to Kais \etal{} \cite{Kais:1993}, the orbital and total
energy for the lowest state are $\varepsilon_{\rm HF}=1.227\,13 \Eh$
a.u. and $E_{\rm HF}=2.039\,325 \Eh$. Kais \etal{} did not specify how
their calculations were carried out; only that the \texttt{CADPAC}
program [The Cambridge Analytic Derivatives Package] was used. As
\texttt{CADPAC} employs Gaussian basis sets \cite{Amos1989_CPR_147},
it is not surprising to find that our total and orbital energies
differ: our total energy is 0.89 m$\Eh$ lower than that of Kais
\etal{}, while our orbital energy (for which the variational principle
does not apply since the potential of the orbital is different) is
49.5 m$\Eh$ higher.

\begin{table}
\caption{Orbital $\epsilon_a$ and total energies $E$ of the lowest
  three singlet states of the $\sigma^2$, $\pi^2$, $\delta^2$ and $\phi^2$
  configurations of the harmonium atom ($\omega=1/2)$ on
  $241\times271/25.0a_0$ grid. $|1-N|$ denotes the error of the
  orbital norm.}
\label{tab:hooke1}
\begin{center}
\begin{tabular}{cccc}
\toprule
\multicolumn{1}{c}{Orbital}
& \multicolumn{1}{l}{$\varepsilon_a$}
& \multicolumn{1}{l}{$E/\Eh$}
& \multicolumn{1}{c}{$|1-N|$}\\
\hline\\[-5pt]

$2\phi_u$ & $3.513\ph{\,061\,635}$ & 6.763\ph{\,367\,189} & $1\times10^{-04}$ \\
$1\phi_g$ & $3.037\,850\,636$ & 5.788\,811\,079 & $4\times10^{-16}$ \\
$2\delta_g$ & $3.033\ph{\,367\,189}$ & 5.783\ph{\,367\,189} & $2\times10^{-05}$ \\[5pt]
$1\phi_u$ & $2.580\,510\,831$ & 4.832\,264\,681 & $8\times10^{-15}$ \\
$1\delta_u$ & $2.565\,227\,220$ & 4.816\,620\,939 & $9\times10^{-16}$ \\
$2\pi_u$ & $2.563\ph{\,344\,935}$ & 4.815\ph{\,367\,189} & $5\times10^{-05}$ \\[5pt]
$1\delta_g$ & $2.114\,837\,213$ & 3.867\,436\,248 & $3\times10^{-15}$ \\
$2\sigma_g$ & $2.109\ph{\,762\,229}$ & 3.863\,979\,716 & $5\times10^{-04}$ \\[5pt]
$1\pi_g$ & $2.107\,519\,031$ & 3.859\,874\,164 & $1\times10^{-15}$ \\
$1\sigma_u$ & $1.692\,631\,671$ & 2.948\,477\,552 & $1\times10^{-15}$ \\
$1\pi_u$ & $1.668\,530\,193$ & 2.923\,072\,506 & $2\times10^{-15}$ \\
$1\sigma_g$ & $1.276\,676\,881$ & 2.038\,438\,872 & $9\times10^{-16}$ \\[5pt]

\bottomrule
\end{tabular}
\end{center}
\end{table}

\subsection{Ar-C at small internuclear distances \label{sec:arc}}

The great benefit of a fully numerical approach is that it can also be
used to compute total and orbital energies for internuclear distances
relevant in atom-atom or atom-ion collisions, where the behaviour at
extremely small internuclear distances is of great importance.
Although some important advances have been made recently
\cite{Lehtola2020_PRA_32504}, the reliable modeling of such bizarre
systems with atomic basis set approaches remains extremely difficult.

First, the inner electronic orbitals of atoms can hybridize at small
internuclear distances, while standard electronic basis sets do not
contain functions to describe core orbital polarization. Moreover, the
character of the atomic orbitals themselves can change: for example,
at the limit $R\to 0$, the Ar-C system approaches the Cr atom, which
is well known to exhibit a ground state with an occupied $3d$ orbital
while Ar and C only feature occupied $s$ and $p$ orbitals.

Second, various electronic states are relevant when examining the
range of internuclear distances relevant in collisions. Since the
ground states of the Ar and C atoms are a singlet and a triplet,
respectively, while the Cr atom has a septet ground state, it is clear
that at least the triplet, quintet, and septet states need to be
included in the consideration when studying the Ar-C collision.

The third major issue is that the linear dependencies in the atomic
basis change considerably when the internuclear distance spans many
orders of magnitude, giving rise to unknown truncation errors in the
calculations.

In the following, we consider exchange-only LDA
\cite{Bloch1929_ZfuP_545, Dirac1930_MPCPS_376} calculations on a
single configuration of the triplet state of Ar-C, in which two $\pi$
orbitals are fully occupied, two electrons occupy a third $\pi$
orbital, and the rest of the electrons are placed on doubly occupied
$\sigma$ orbitals.

\def\scalea{0.88}
\begin{figure}
  \caption{DFT orbital energies of Ar-C as a function of the
    internuclear distance $R$ given in $a_0$. All orbitals are shown
    in the upper panel. Since the large changes of the $1\sigma$
    orbital dominate the plot, the zoomed in behaviour of the other
    orbitals is shown in the lower panel.}
  \label{fig:arc-1}
\begin{center}
\begin{tabular}{c}
\includegraphics[width=\scalea\textwidth]{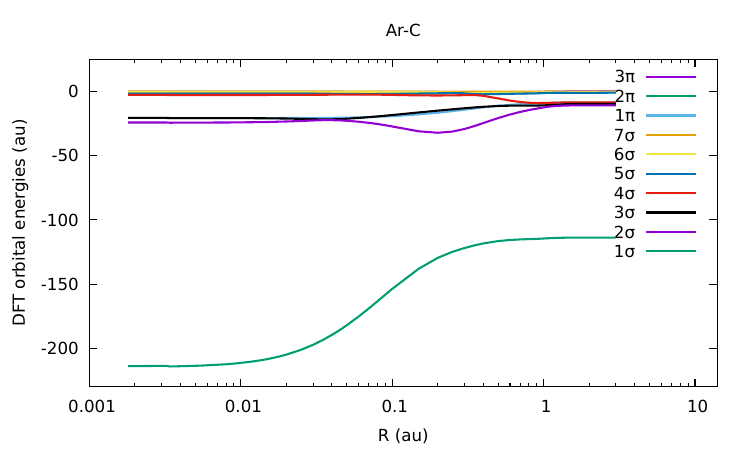}\\
\includegraphics[width=\scalea\textwidth]{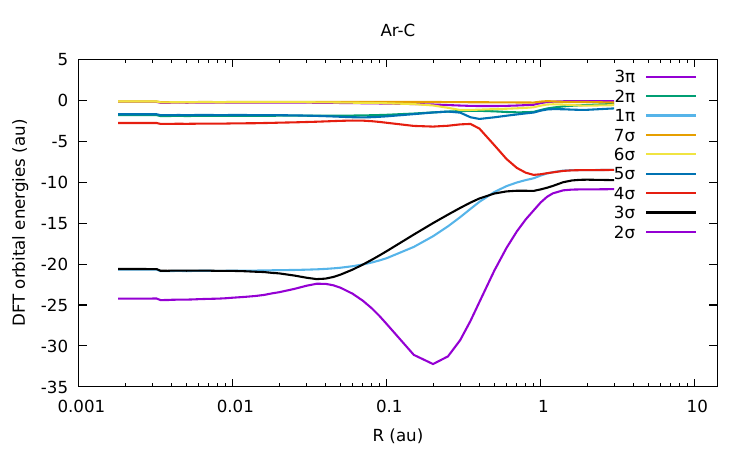}\\
\end{tabular}
\end{center}
\end{figure}

\Cref{fig:arc-1} shows how the orbital energies in the Ar-C system depend
on the internuclear distance $R \in [10^{-3},2]a_0$. The major feature to
observe in the upper panel of \cref{fig:arc-1} is how the energy of the
$1\sigma$ orbital undergoes a~major change. At large $R$, the $1\sigma$
orbital corresponds to the $1s$ orbital of the Ar atom
($\varepsilon_{1s}^{\rm Ar} \approx -113.7 \Eh$). Around $R\approx 0.3a_0$,
the orbital appears to start experiencing the attraction of the C nucleus,
as can be observed from the visibly non-zero slope of the orbital energy
curve. The orbital energy decreases sharply around $R\approx 0.1 a_0$,
which is around the size of the Ar $1s$ orbital. The united atom limit
($\varepsilon_{1s}^{\rm Cr} \approx -213.8 \Eh$) is finally reached for
$R < 0.01 a_0$, which is close to the size of the Cr $1s$ orbital.

Also the lower panel of \cref{fig:arc-1} contains interesting features:
significant changes are especially observed in the energies of the $2\sigma$,
$3\sigma$, $4\sigma$, and $1 \pi$ orbitals. The first three of these correspond
to the Ar $2s$, C $1s$, and the Ar $2p_0$ core orbitals at large $R$, while the
$1\pi$ orbital in Ar-C at large $R$ contains the Ar $2p_{\pm 1}$ core orbitals.
When the internuclear distance is decreased, the energies of these orbitals are
strongly affected, and it is clear that hybridization is again going on between the
deep-lying orbitals of Ar and C.

The highly non-monotonic behaviour of the $2\sigma$ energy in the region
$R \in [0.01a_0, 1a_0]$ is especially worthwhile to notice, and it is
likely an ``after-effect'' of the strong changes in the $1s$ orbital in
that region. The $3\sigma$ and $1\pi$ orbitals show more muted changes.  At
small $R$, the $1\sigma$ and $2\sigma$ orbitals become the Cr united atom
$1s$ and $2s$ orbitals, respectively, while the $3\sigma$ and $1\pi$
orbitals form the $2p$ orbital of the Cr united atom.

Many of the higher-lying orbital energies do not appear to change much
in the studied region, as makes perfect sense: the outermost electrons
are sufficiently far removed from the nuclei that they only experience
a screened interaction.

While this calculation is performed for a fixed electronic
configuration for a spin triplet, it is sufficient to highlight that
calculations can be performed with \xtwodhf{} at small internuclear
separations, although they can be troublesome to converge.
A more in-depth examination of the Ar-C system would include additional
spin states and configurations, such that the united Cr atom limit
that also features occupied $\delta$ orbitals would also
be~considered, for instance.

\subsection{Tests of Libxc functionality \label{sec:libxc}}

The present version of the program allows one to perform DFT
calculations with many functionals available through the Libxc library
\cite{Lehtola2018_S_1}. In the following, we compare total energies of
the He, Be, Ne, and Kr atoms obtained by the \xtwodhf{} program to
calculations performed in two ways with the \helfem{}
\cite{Lehtola2019_IJQC_25945, Lehtola2020_PRA_12516,
  Lehtola2023_JCTC_2502} program, while a third independent reference
is provided by Engel's atomic \opmks{} program \cite{EngelVosko:1993}.

Both \xtwodhf{} and \helfem{} make use of Libxc
\cite{Lehtola2018_S_1}, which is why in the following, we will
exclusively employ Libxc's functional identifiers to uniquely specify
the employed functionals. However, independent implementations of
density functionals are employed in Engel's \opmks{} program.

Comparisons for calculations that only include a correlation
functional are not carried out for \opmks{}, as the program appears to
set the Coulomb interaction between the electrons to zero when no
exchange potential has been specified. The calculations in \opmks{}
employed the parameters \texttt{IMAX=800}, \texttt{RWALL=100.0}, and
\texttt{TOLPOT=1.d-9}.

Atomic calculations were carried our with the \texttt{gensap} program
of \helfem{}, which is specialized for calculations with spherical
symmetric electron densities \cite{Lehtola2019_IJQC_25945,
  Lehtola2020_PRA_12516, Lehtola2023_JCTC_2502}. The calculations were
carried out with the default 15-node Lagrange interpolating polynomial
basis, employing 10 radial elements and the default value $r_\infty =
40a_0$. Exploratory calculations suggest that such a numerical basis
set affords total energies converged to n$\Eh$ for the studied He,
Be, Ne, and Kr atoms.

Atomic calculations can be carried out in the diatomic approaches
employed in \xtwodhf{} and the \texttt{diatomic} program of \helfem{}
by setting $Z_B =0$.  The internuclear distance then becomes an
arbitrary parameter; the calculations were run for $R = 1.7328 a_0
\approx 0.917$ \AA{}, which is the experimental equilibrium bond
length of the FH molecule that was also used above in
\cref{fig:fh-input}.

As we wanted to compare the accuracy of the two methods especially
devised for diatomic systems, another set of results were obtained
using the \texttt{diatomic} program of \helfem{}
\cite{Lehtola2019_IJQC_25944}. The numerical basis set for diatomic
molecules in \helfem{} is of the form of \cref{eq:pwscfexp} with an
$m$ dependent truncation parameter $l_{\rm max}$ for the partial wave
expansion, while $X^{ml}(\mu)$ was described by 15-node LIP elements
\cite{Lehtola2019_IJQC_25944}.  The \helfem{} default value of the
practical infinity of $r_\infty = 40 a_0$ was likewise employed for
these calculations.  The proxy method \cite{Lehtola2019_IJQC_25944}
with $\epsilon= 10^{-10}$ was used to generate a suitable numerical
basis for the diatomic calculations with \helfem{} using the
\texttt{diatomic\_cbasis} program.  This procedure led to the
following numerical basis sets: $N_{\rm elem}=3$ ($N_{\rm bf}^{\rm
  rad}=42$) and $l^\sigma_{\rm max}=8$ for He; $N_{\rm elem}=3$
($N_{\rm bf}^{\mu}=42$) and $l^\sigma_{\rm max}=12$ for Be; $N_{\rm
  elem}=3$ ($N_{\rm bf}^{\mu}=42$), $l^\sigma_{\rm max}=17$ and
$l^\pi_{\rm max}=13$ for Ne; and $N_{\rm elem}=5$ ($N_{\rm
  bf}^{\mu}=70$), $l^\sigma_{\rm max}=32$, $l^\pi_{\rm max}=24$, and
$l^\delta_{\rm max}=22$ for Kr; $N_{\rm elem}$ being the number of
elements in $\mu$ and $N_{\rm bf}^{\mu}$ the resulting number of
numerical basis functions in $\mu$.

There is no automatic or easy way to decide what size of a grid is
needed in \xtwodhf{} for a particular system. The required size of the
grid and the value of the practical infinity are dependent on the
target accuracy of the solution, the heaviness of the individual
nuclei, the overall charge state, and the examined electronic
configuration. The examples in the \texttt{test-sets} subdirectory
should be consulted to perform grid convergence studies. For these
example calculations, the $181 \times 271 / 65.0 a_0$ grid was used for
He, Be, and Ne, and the $331 \times 511 / 150.0 a_0$ grid was used for
Kr.

A comparison of the resulting data is shown in \cref{tab:lxc1}.  Since
both \xtwodhf{} and \helfem{} use the same implementation of density
functionals provided by Libxc, the results from the atomic program of
\helfem{} are expected to be the most reliable, as the one-center
expansion makes the most sense for these calculations, and the atomic
calculations were easy to converge to the CBS limit. The use of the
same density functional implementation is known to be extremely
important in studies performed at this level of accuracy, since small
changes in the numerical parameters of density functionals arising
from e.g. ambiguities in the original literature often result in total
energy differences in the $\upmu \Eh$ range
\cite{Lehtola2023_JCP_114116}.

The agreement between all four approaches is in general excellent, and even
for the krypton atom an agreement with 7-8 significant figures is reached.
However, since the total energy varies by four orders of magnitude between
He and Kr, sub-$\upmu \Eh$ precision has not been reached in all of the
diatomic calculations.  

The proxy approach used in the \texttt{diatomic} calculations in
\helfem{} was originally designed and tested for Hartree--Fock
calculations \cite{Lehtola2019_IJQC_25944}. We observe from the data
in \cref{tab:lxc1} that although the claimed sub-$\upmu \Eh$ precision
for the choice of the numerical basis is achieved by this method in
Hartree--Fock calculations, some density functional calculations with
the \texttt{diatomic} program show differences of the order of 1--3
$\upmu \Eh$ for the Ne and Kr atoms to the reference values obtained
with the \texttt{gensap} atomic program.

We attribute this difference to the non-linearity of density
functional approximations. However, we also note that the approach in
\helfem{} is still variational, as evinced by the \texttt{diatomic}
total energies always being above the CBS limit values produced with
the \texttt{gensap} program. Further calculations performed with
larger numerical basis sets\ie{} larger values of the $l_{\rm max}$
truncation parameters demonstrate that excellent agreement with
results from the \texttt{gensap} program can be obtained (not shown).

Because \xtwodhf{} solves the Poisson equation with SOR, it can
approach the CBS limit value from above or below. This behaviour is
also observed in the data in \cref{tab:lxc1}. For example, with the
\texttt{GGA\_X\_B88} functional, the total energy reproduced by
\xtwodhf{} is 26 n$\Eh$ above the CBS limit value from
\texttt{gensap} for Ne, while it is 21 $\upmu \Eh$ below the CBS limit
value for Kr. It is thus clear that a larger grid is needed to get rid
of the discrepancies observed for the krypton atom. Using a $631\times
991/150.0 a_0$ grid one gets the total energies $-2752.100\,620 \Eh$,
$-2748.627\,895 \Eh$ and $-2753.851\,525 \Eh$, for the
\texttt{GGA\_X\_B88}, \texttt{LDA\_X-GGA\_C\_PBE}, and
\texttt{HYB\_GGA\_XC\_B3LYP} functionals, respectively, which are in
$\upmu \Eh$ agreement with the reference values from the
\texttt{gensap} program.
  
The results of this subsection demonstrate that \xtwodhf{} can be used
to run DFT calculations with semi-local or global hybrid LDA and GGA
functionals. However, the results also exemplify the need to carefully
converge the calculations with respect to the grid size and the value
of the practical infinity $r_\infty$ to achieve the CBS limit in high
precision. We again remind here about the double role of the practical
infinity $r_\infty$ in \xtwodhf{} calculations, which was discussed
above in \cref{sec:grid-specification}.

\newcommand{\fss}{\fontsize{8}{8}\selectfont}
\begin{center} {\footnotesize \begin{longtable}{ccrrrr} \caption {Total SCF
        energies of the He, Be, Ne, and Kr atoms calculated using exchange
        and correlation functionals from Libxc \cite{Lehtola2018_S_1}. The
        results of the \xtwodhf{} (the first lines) are given together with
        the deviations, $E-E(\xtwodhf{})$, from the \texttt{diatomic}
        \cite{Lehtola2019_IJQC_25944} program of \helfem{}, fully numerical
        results from the atomic \texttt{gensap}
        \cite{Lehtola2019_IJQC_25945, Lehtola2020_PRA_12516,
          Lehtola2023_JCTC_2502} and from the \opmks{} program
        \cite{EngelVosko:1993}. The deviations are given in
        n$\Eh{}$ units for He, Be, Ne and
        $\upmu\Eh{}$ for Kr.}
  
  \label{tab:lxc1}\\
\hline \hline
\multicolumn{1}{c}{Functional(s)}
& \multicolumn{1}{c}{Approach}
& \multicolumn{1}{c}{He}
& \multicolumn{1}{c}{Be}
& \multicolumn{1}{c}{Ne}
& \multicolumn{1}{c}{Kr} \\
\hline
\endfirsthead

\multicolumn{6}{c}{{\tablename\ \thetable{} -- continued from previous page}} \\
\hline \hline
\multicolumn{2}{c}{Functional(s)}
& \multicolumn{1}{c}{He}
& \multicolumn{1}{c}{Be}
& \multicolumn{1}{c}{Ne}
& \multicolumn{1}{c}{Kr} \\
\hline
\endhead

\hline \multicolumn{6}{r}{{Continued on next page}} \\ \hline \hline

\endfoot

\hline \hline
\endlastfoot
    {\fss HF} & \xtwodhf{}
& -2.861\,679\,996
& -14.573\,023\,168
& -128.547\,098\,052
& -2\,752.054\,976 \\
& \texttt{diatomic}
&  2
&  2
&  -5
&  0\\
& \texttt{atomic}
&  0
&  0
& -57
& -1\\
\hline
{\fss LDA\_X} \cite{Bloch1929_ZfuP_545, Dirac1930_MPCPS_376} & \xtwodhf{}
& $-2.723\,639\,793$
& -14.223\,290\,827
& -127.490\,740\,825
& -2\,746.866\,100 \\
& \texttt{diatomic}
&  9
& 12
& 232
& 0\\
& \texttt{atomic}
&  0
&  0
&  -6
&  0\\
& \opmks{}
& 0
& 0
& -6
& -1\\
\hline
{\fss GGA\_X\_PBE} \cite{Perdew1996_PRL_3865, Perdew1997_PRL_1396}  & \xtwodhf{}
& -2.852\,037\,536
& -14.545\,039\,137
& -128.520\,129\,768
& -2\,751.653\,674 \\
& \texttt{diatomic}
& 11
& 561
& 996
& 4\\
& \texttt{atomic}
&  0
&  33
& -67
& $1$\\
& \opmks{}
&  0
& 33
& -67
& 1\\
\hline
{\fss GGA\_X\_B88} \cite{Becke1988_PRA_3098}  & \xtwodhf{}
& -2.863\,379\,361
& -14.566\,364\,985
& -128.590\,092\,756
& -2\,752.100\,640 \\
& \texttt{diatomic}
& 11
& 143
& 622
& 23\\
& \texttt{atomic}
& 0
& -5
& -26
& 21\\
& \opmks{}
& -79
& -198
& -644
& 17\\
\hline
{\fss LDA\_C\_VWN} \cite{Vosko1980_CJP_1200} & \xtwodhf{}
& -2.052\,843\,324
& -12.273\,216\,433
& -117.699\,882\,594
& -2\,663.000\,586\\
& \texttt{diatomic}
& 1
& 5
& 152
& 2\\
& \texttt{atomic}
& 0
& 0
& 0
& 0\\
\hline
{\fss LDA\_C\_PW} \cite{Perdew1992_PRB_13244} & \xtwodhf{}
& -2.052\,576\,253
& -12.272\,556\,485
& -117.696\,659\,062
& -2\,662.986\,395\\
& \texttt{diatomic}
& 2
& 4
& 152
& 30\\
& \texttt{atomic}
& 0
& 0
& 0
& 28\\
\hline
{\fss GGA\_C\_PBE} \cite{Perdew1996_PRL_3865, Perdew1997_PRL_1396} & \xtwodhf{}
& -1.988\,501\,616
& -12.140\,977\,369
& -117.314\,905\,038
& -2\,661.499\,759\\
& \texttt{diatomic}
&  1
&  472
&  302
&  2\\
& \texttt{atomic}
& 0
& -7
& 0
& 0\\
\hline
{\fss GGA\_C\_LYP} \cite{Lee1988_PRB_785, Miehlich1989_CPL_200} & \xtwodhf{}
& -1.992\,266\,386
& -12.150\,994\,762
& -117.357\,415\,352
& -2\,661.500\,308\\
& \texttt{diatomic}
& 2
& 8
& 150
& 2\\
& \texttt{atomic}
& 0
& 0
& 0
& 0\\
\hline
{\fss LDA\_X} \cite{Bloch1929_ZfuP_545, Dirac1930_MPCPS_376} & \xtwodhf{}
& -2.834\,835\,624
& -14.447\,209\,474
& -128.233\,481\,269
& -2\,750.147\,940\\
{\fss LDA\_C\_VWN} \cite{Vosko1980_CJP_1200} & \texttt{diatomic}
& 9
& 12
& 240
& 1\\
& \texttt{atomic}
& 0
& 0
& 0!!!
& 0\\
& \opmks{}
& 148
& 285
& 791
& 2\\
\hline
{\fss LDA\_X} \cite{Bloch1929_ZfuP_545, Dirac1930_MPCPS_376} & \xtwodhf{}
& -2.834\,455\,181
& -14.446\,473\,478
& -128.229\,917\,209
& -2\,750.133\,306\\
  {\fss LDA\_C\_PW} \cite{Perdew1992_PRB_13244}
& \texttt{diatomic}
& 10
& 13
& 235
& 1\\
& \texttt{atomic}
& 0
& 1
& -6
& 0\\
& \opmks{}  
& 320!!!
& 701!!
& 2833!!!
& 15\\
\hline
{\fss LDA\_X} \cite{Bloch1929_ZfuP_545, Dirac1930_MPCPS_376} & \xtwodhf{}
& -2.764\,587\,670
& -14.308\,185\,377
& -127.836\,853\,037
& -2\,748.627\,924\\
{\fss GGA\_C\_PBE} \cite{Perdew1996_PRL_3865, Perdew1997_PRL_1396} & \texttt{diatomic}
& 8
& 683
& 387
& 30\\
& \texttt{atomic}
& 0
& 75
& -81
& 29\\
& \opmks{}
& 0
& 75
& -81
& 28\\
\hline
{\fss LDA\_X} \cite{Bloch1929_ZfuP_545, Dirac1930_MPCPS_376} & \xtwodhf{}
& -2.767\,057\,515
& -14.318\,062\,460
& -127.873\,287\,501
& -2\,748.615\,031\\
{\fss GGA\_C\_LYP} \cite{Lee1988_PRB_785, Miehlich1989_CPL_200} & \texttt{diatomic}
& 9
& 20
& 222
& 1\\
& \texttt{atomic}
& 0
& 0
& -5
& 0\\
& \opmks{}
& -134
& -125
& -97
& 0\\
\hline
{\fss GGA\_X\_PBE} \cite{Perdew1996_PRL_3865, Perdew1997_PRL_1396} & \xtwodhf{}
& -2.964\,147\,813
& -14.769\,959\,464
& -129.263\,565\,003
& -2\,754.935\,906\\
{\fss LDA\_C\_VWN} \cite{Vosko1980_CJP_1200} & \texttt{diatomic}
& 12
& 609
& 971
& 4\\
& \texttt{atomic}
& 0
& 59
& -66
& 2\\
& \opmks{}
& 149
& 344
& 725
& 4\\
\hline
{\fss GGA\_X\_PBE} \cite{Perdew1996_PRL_3865, Perdew1997_PRL_1396} & \xtwodhf{}
& -2.963\,756\,111
& -14.769\,215\,227
& -129.259\,995\,798
& -2\,754.921\,272\\
{\fss LDA\_C\_PW} \cite{Perdew1992_PRB_13244} & \texttt{diatomic}
& 11
& 610
& 972
& 4\\
& \texttt{atomic}
& -1
& 60
& -66
& 1\\
& \opmks{}
& 324
& 766
& 2778
& 15\\
\hline
{\fss GGA\_X\_PBE} \cite{Perdew1996_PRL_3865, Perdew1997_PRL_1396} & \xtwodhf{}
& -2.892\,934\,867
& -14.629\,947\,789
& -128.866\,427\,587
& -2\,753.416\,110 \\
{\fss GGA\_C\_PBE} \cite{Perdew1996_PRL_3865, Perdew1997_PRL_1396}
& \texttt{diatomic}
& 11
& 1038
& 1379
& 5\\
& \texttt{atomic}
& 0
& 73
& -158
& 2\\
& \opmks{}
& 0
& 73
& -158
& 1\\
\hline
{\fss GGA\_X\_PBE} \cite{Perdew1996_PRL_3865, Perdew1997_PRL_1396} & \xtwodhf{}
& -2.895\,699\,125
& -14.640\,227\,604
& -128.902\,927\,530
& -2\,753.402\,420\\
{\fss GGA\_C\_LYP} \cite{Lee1988_PRB_785, Miehlich1989_CPL_200} & \texttt{diatomic}
& 11
& 550
& 899
& 3\\
& \texttt{atomic}
& 0
& 5
& -66
& 1\\
& \opmks{}
& -142
& -122
& -158
& 3\\
\hline
% {\footnotesize HYB\_GGA\_XC\_B3LYP}
{\fss HYB\_GGA\_XC\_B3LYP} \cite{Stephens1994_JPC_11623} & \xtwodhf{}
& -2.915\,218\,663 % -2.915218663
& -14.673\,328\,221
& -128.980\,973\,214
& -2\,753.851\,519\\ % -2753.8515186462005
& \texttt{diatomic}
& 8
& 151
& 357!!!
& -5\\
& \texttt{atomic}
&  0
& 46
& -24
& -6\\
& \opmks{}
& 20
& 240
& 731!!!
& -5 \\ 
\hline
% {\footnotesize HYB\_LDA\_XC\_LDA0}
{\fss HYB\_LDA\_XC\_LDA0} \cite{Rinke2012_PRL_126404} & \xtwodhf{}
& -2.840\,868\,320 %
& -14.477\,351\,811
& -128.306\,321\,244
& -2\,750.605\,900 \\ % -2.75060589423
& \texttt{diatomic}
& 7
& 21
& -1634
& -11\\
& \texttt{atomic}
& 0
& 12
& -1810
& -11\\
\end{longtable}
}
\end{center}

\subsection{Kinetic potentials \label{sec:kinetic}}

If converged HF or Kohn--Sham orbitals $\phi_i(\gv{r})$ and
eigenvalues $\varepsilon_i$ are available, the total kinetic potential
can be computed as \cite{LevyPS:1984}
\begin{equation}
v_k(\gv{r})=v^P(\gv{r})+v^W(\gv{r})
\end{equation}
where $v^P(\gv{r})$ and $v^W(\gv{r})$ are the  Pauli and
von Weizs\"acker kinetic potentials defined as
\begin{align}
v^P(\gv{r})&=\frac{\tau(\gv{r})-\tau^W(\gv{r})}{\rho(\gv{r})}
+ \sum_i^N(\varepsilon_H-\varepsilon_i)\frac{|\phi_i(\gv{r})|^2}{\rho(\gv{r})}\\
  v^W(\gv{r})&=\frac{|\nabla
\rho(\gv{r})|^2}{8\rho^2(\gv{r})}-\frac{\nabla^2\rho(\gv{r})}{4\rho(\gv{r})}
= \frac{\tau^W(\gv{r})}{\rho(\gv{r})} -\frac{\nabla^2\rho(\gv{r})}{4\rho(\gv{r})}
\end{align}
where $\tau^W(\gv{r})=|\nabla \rho(\gv{r})|^2/8\rho(\gv{r})$,
$\tau(\gv{r})=1/2\sum_i^N |\nabla\phi_i(\gv{r})|^2$ is the
kinetic energy density, $\varepsilon_H$ is the energy of the highest
occupied molecular orbital (HOMO), and the total density is equal to
$\rho(\gv{r})=\sum_i^N q_i |\varphi_i(\gv{r})|^2$.

The FD HF method can be used to compute these properties.  Plots of
the HF total density $\rho_{\rm HF}$, its Laplacian $\nabla^2
\rho_{\rm HF}$, the Pauli kinetic potential $v^P$, the logarithm of
the von Weizs\"acker kinetic potential $\log(v^W)$, and the logarithm
of their sum $\log(v^P+v^W)$ are shown in \cref{fig:fh-hf-kinpot} for
the FH molecule. The values are plotted along the $z$ axis, with the F
atom displaced to $z=0$, and the H atom to the experimental equilibrium distance $z=1.7328a_0$.

\def\scalea{0.48}
\begin{figure}
    \caption{The total kinetic potential for the FH molecule ($R=1.7328a_0$)
    along the internuclear axis and its ingredients computed from a HF wave
    function. The F atom is placed at $z=0$, while the H atom is found at
    $z=R$. The ingredients comprise: the total HF density, $\rho_{\rm HF}$,
    $\nabla^2 \rho_{\rm HF}$, the Pauli kinetic potential, $v^P$, the
    logarithm of the von Weizs\"acker kinetic potential, $\log(v^W)$, and
    the logarithm of their sum, $\log(v^P+v^W)$. }

  \label{fig:fh-hf-kinpot}
\begin{center}
\begin{tabular}{cc}
\includegraphics[width=\scalea\textwidth]{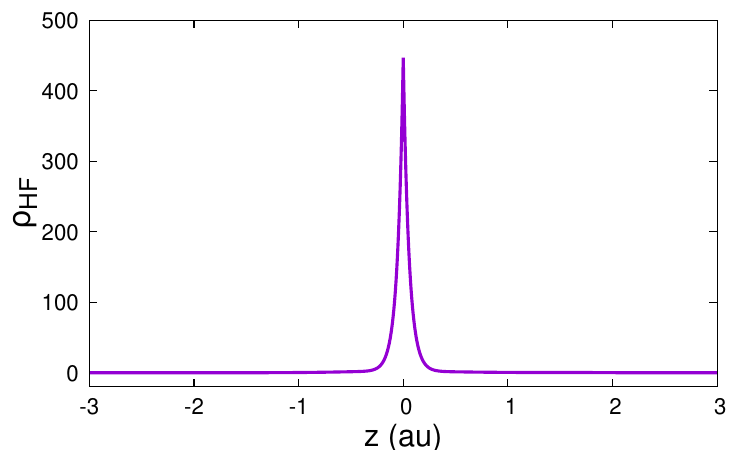}
&\includegraphics[width=\scalea\textwidth]{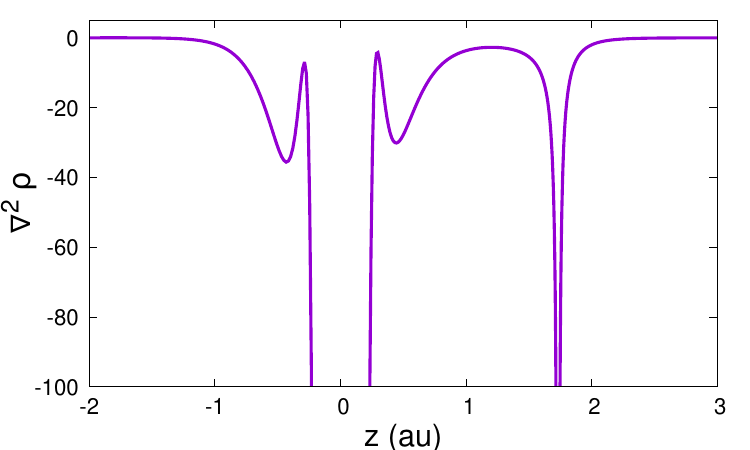}\\
\includegraphics[width=\scalea\textwidth]{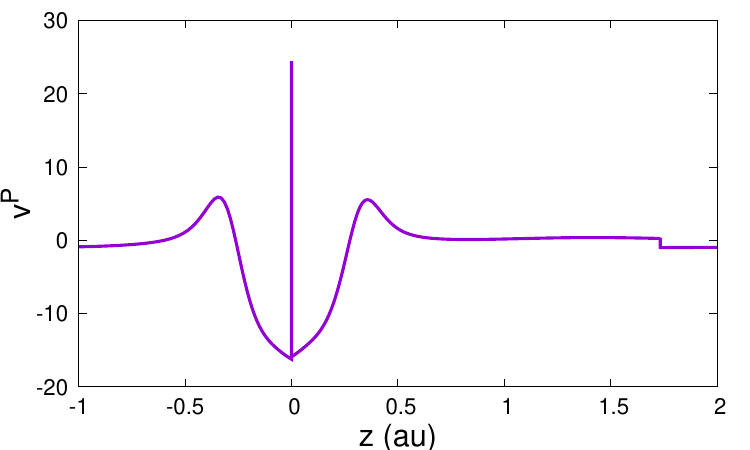}
&\includegraphics[width=\scalea\textwidth]{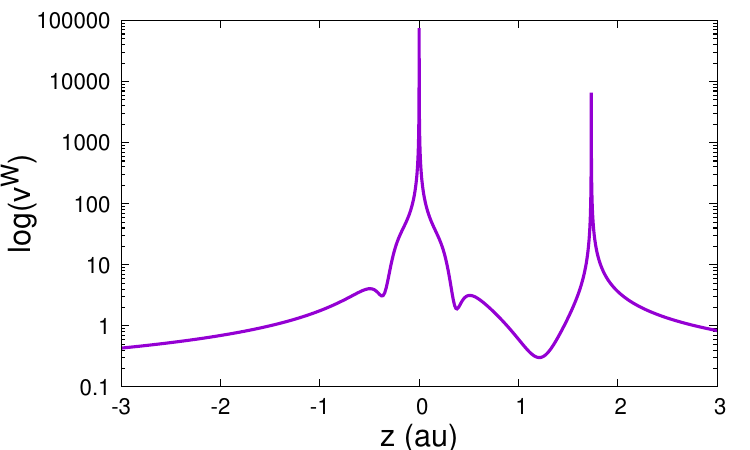}\\
\multicolumn{2}{c}{\includegraphics[width=\scalea\textwidth]{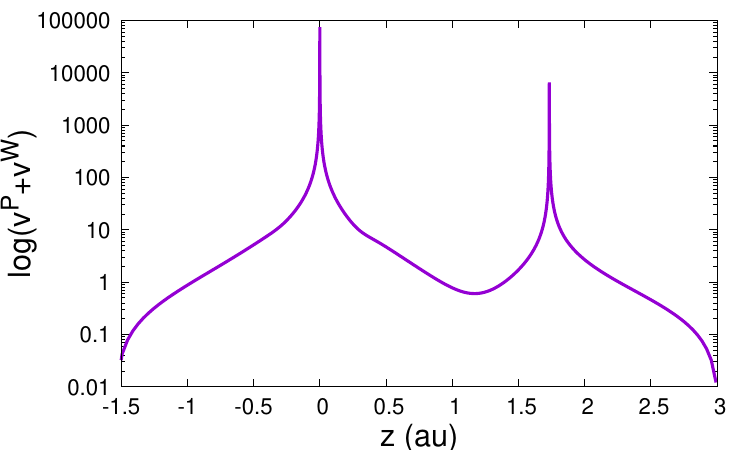}}

\end{tabular}
\end{center}
\end{figure}

\begin{figure}
\caption{ Differences in the DFT total kinetic potential and its
  ingredients compared to HF values shown in \cref{fig:fh-hf-kinpot};
  analogous notation is used.
}
\label{fig:fh-hf-dft-kinpot}
\begin{center}
\begin{tabular}{cc}
\includegraphics[width=\scalea\textwidth]{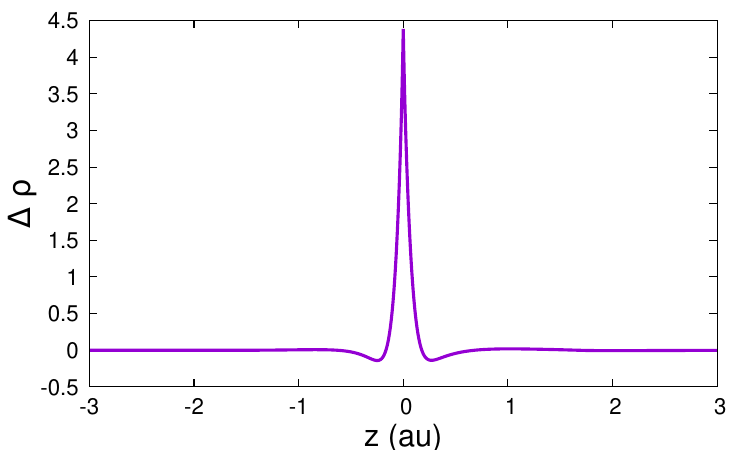}
&\includegraphics[width=\scalea\textwidth]{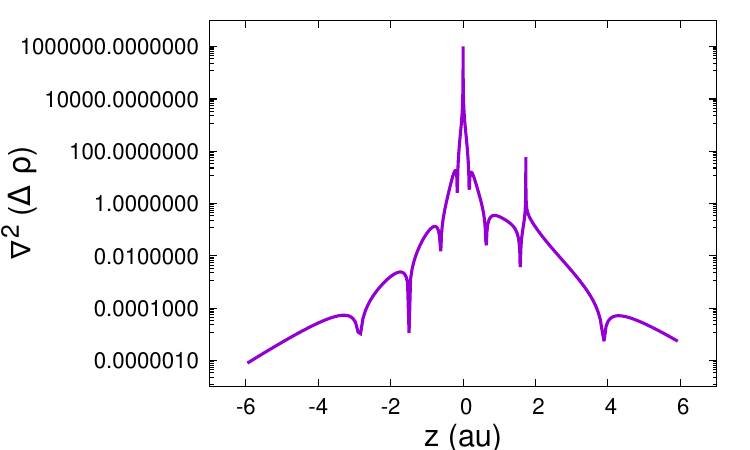}\\
\includegraphics[width=\scalea\textwidth]{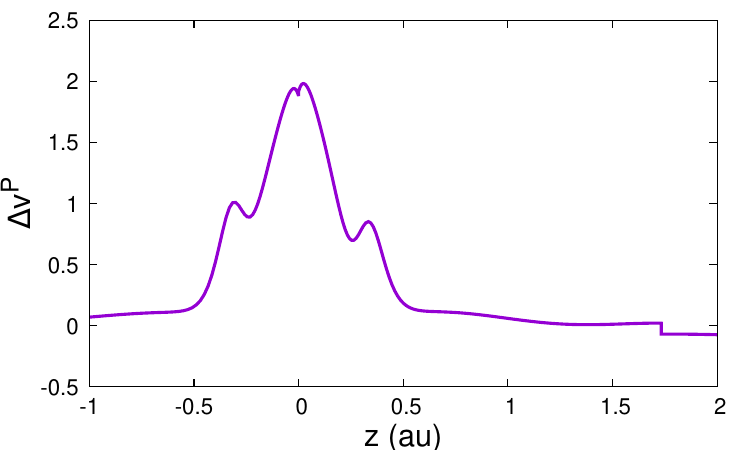}
&\includegraphics[width=\scalea\textwidth]{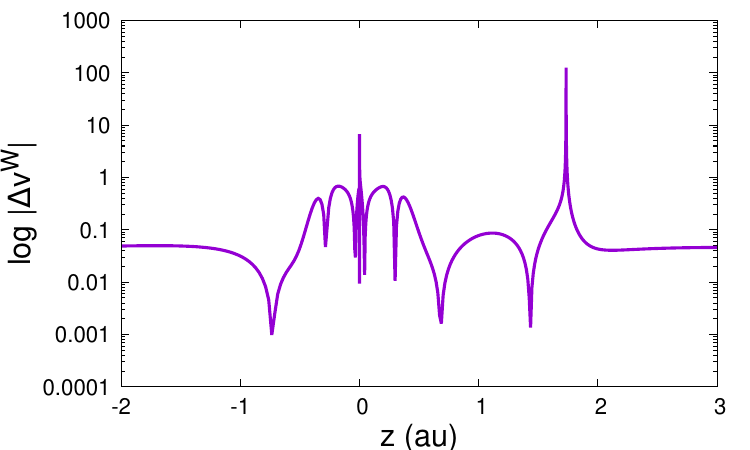}\\
\multicolumn{2}{c}{\includegraphics[width=\scalea\textwidth]{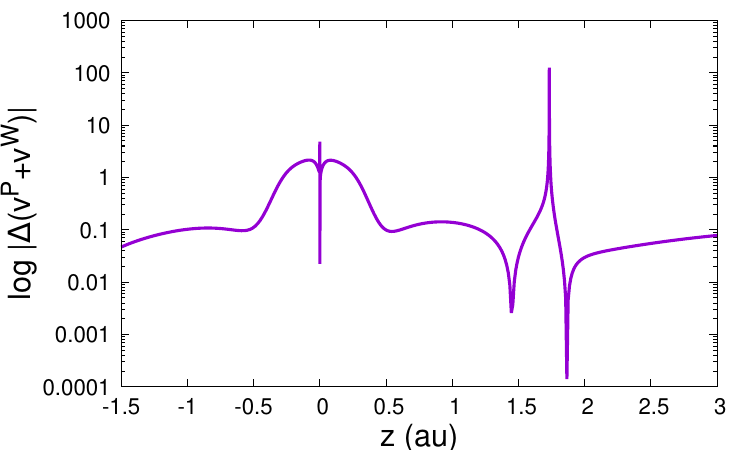}}

\end{tabular}
\end{center}
\end{figure}

For comparison, we also carried out exchange-only LDA
\cite{Bloch1929_ZfuP_545, Dirac1930_MPCPS_376} calculations.  The obtained
differences to the HF data in \cref{fig:fh-hf-kinpot} are shown in
\cref{fig:fh-hf-dft-kinpot}.  To allow displaying the details of the
changes in $\Delta v^W$ and $\Delta (v^P+v^W)$, which can be both
positive and negative, the logarithms of their absolute values are
shown, instead. Some of the plots were also trimmed to focus on the
regions with interesting changes.

Overall, \cref{fig:fh-hf-dft-kinpot} shows that the LDA calculation
reproduces the major features of the HF calculations. The axis scales
in \cref{fig:fh-hf-dft-kinpot} are orders of magnitude smaller than in
\cref{fig:fh-hf-kinpot}; the only difference is the plot of the
density laplacian $\nabla^2 \rho$, which in any case diverges at the
nuclei \cite{Lehtola2023_JCTC_2502}.

\section{Conclusions \label{sec:conclusions}}

We have carried out an extensive review of the restricted open-shell
Hartree--Fock and Kohn--Sham methods for diatomic molecules, as well
as its finite difference discretization. The methods can be also
applied to solving orbitals for atomic model potentials, the
Kramers--Henneberger atom, as well as Hooke's atom (also known as
harmonium).

We have discussed in detail how the discretization takes place, and
how the various boundary conditions for the solution are applied. The
solution of the coupled sets of Poisson equations for the orbitals and
potentials with the successive overrelaxation (SOR) method have also
been presented at depth.

We have described the new, parallelised version 3.0 of \xtwodhf{}, the
finite difference Hartree--Fock program for atoms and diatomic
molecules. The program has been written mostly in Fortran 95, with
some optional C extensions. The program is built with CMake. The
program is hosted openly on GitHub \cite{x2dhf}, and it is provided
with the open source GNU General Public License v2.0, or later
(GPL-2.0-or-later).

Besides extensive internal refactoring, the initialisation process of
\xtwodhf{} has been greatly simplified by including tabulations of
radial LDA and HF atomic orbitals in the program, as well as atomic
LDA Coulomb potentials which afford a better initial guess to the
Poisson problem. As significant new features, \xtwodhf{} can now also
be used to solve Kohn--Sham equations with LDA and GGA functionals
from the Libxc library \cite{Lehtola2018_S_1}, as well as to calculate
molecular orbitals for the superposition of atomic potentials
\cite{Lehtola2019_JCTC_1593, Lehtola2020_JCP_144105}.

Continuing with a discussion of the literature complementing the
recent review article \cite{Lehtola2019_IJQC_25968}, we discussed the
science that has been enabled by earlier versions of \xtwodhf{}, as
well as recent work carried out with alternative approaches.

We exemplified the capabilities of the new version of \xtwodhf{} by
studies of basis set truncation errors in \ce{He2} and other
diatomics, as well as calculations on harmonium and Ar-C at
small internuclear distances. We verified the Libxc interface and the
rectified implementation of GGA functionals by calculations on the He,
Be, Ne, and Kr atoms, and found excellent agreement with reference
calculations carried out with \helfem{} and \opmks{}. Finally, we
plotted kinetic potentials from HF and exchange-only LDA calculations
on the FH molecule.

This article represents the pinnacle of over 40 years of effort on the FD
HF approach on diatomic molecules, much of it by the first author, who is
about to retire. The considerable length of this article is a direct result
of this long and extensive effort. The collaboration of the new and old
generation exhibited in the authorship of this work has aimed to document
in depth some of the finer details of the approach. These details may not
have been explicitly written down in previous work, but they are important
for achieving a thorough understanding of the FD HF approach.

Thanks to the ongoing development of mathematical methods and computer
programming \cite{Lehtola2023_JCP_180901}, alternative ways to perform
fully numerical calculations on diatomic molecules are nowadays
feasible. The use of higher-order numerical methods affords more
accurate results at a similar cost, as was demonstrated on a model
Poisson equation in \cref{sec:discretization-order}. The solution
strategy in \helfem{} \cite{Lehtola2019_IJQC_25944} enables simpler
ways to program as well as set up calculations, saving human time
instead of computer time. We hope to achieve speedups in \helfem{}
calculations based on insight from \xtwodhf{} in future applications.

\subsection*{CRediT authorship contribution statement} {\bf Jacek Kobus}:
Conceptualization, Methodology, Software, Resources, Investigation,
Validation, Formal analysis, Writing – original draft, Visualization,
Project administration.

{\bf Susi Lehtola}: Software, Investigation, Validation, Writing – review \& editing.

\subsection*{Declaration of competing interest}

The authors declare that they have no known competing financial
interests or personal relationships that could have appeared to influence
the work reported in this paper.

\subsection*{Data availability}

Data will be made available on request.

\section*{Acknowledgements}

J.K. thanks Sz. Śmiga for valuable discussions on DFT related problems,
especially for drawing attention to the kinetic potential and its
components, and for the OPMKS results in \cref{tab:lxc1}.  S.L. thanks
the Academy of Finland for financial support under project numbers 350282
and 353749.

\renewcommand{\thesection}{A.\arabic{section}}

\section{Finite nuclear models}\label{sec:nuclear-models}

A detailed discussion of various nuclear
charge distribution models has been published by Andrae
\cite{Andrae:2000}. The \xtwodhf{} program implements finite nuclear
models following Gaussian and Fermi distributions.

\subsection{Gaussian nuclear model}

The Gaussian distribution has the form
\begin{equation}
\rho (r)=
Z\left( \frac \lambda \pi \right) ^{\frac 32}\exp \left(-\lambda r^2\right) \nonumber
\end{equation}
where the exponent of the normalized Gaussian-type function
representing the nuclear distribution is determined by the
root-mean-square radius of the nuclear charge distribution via the
relation $\lambda=3/2 \avg{r^2}$.  A~statistical model gives the
following value for the root-mean-square radius (in fm)
\begin{equation}
  \avg{r^2}^{1/2}=0.836 A^{1/3}+0.570, \qquad A>6 \label{eq:rms}
\end{equation}
where $A$ is the atomic mass of a nucleus. Therefore, when the
distance is expressed in atomic units we have
\begin{equation}
\lambda = \frac{3}{2} 10^{10} \left( \frac{0.529177249}{0.836A^{\frac 13}+0.570}\right) ^2\nonumber
\end{equation}
For any spherically symmetric charge distribution $\rho (r)$ the
potential energy is
\begin{equation}
-rV(r)=4\pi \left(\int_0^r s^2\rho (s){\rm d}s+r\int_r^\infty s\rho (s){\rm d}s \right) \nonumber
\end{equation}
If $\rho (r)=\rho _0\exp(-\lambda r^2)$ then the first integral is equal to
\begin{equation}
  \rho _0 \left(\frac{-r\exp(-\lambda r^2)}{2 \lambda} + \frac{\sqrt{\pi }
      \mbox{erf}(\sqrt{\lambda }r)}{4\lambda^{3/2}} \right) \nonumber
\end{equation}
and the second to $\rho _0 r\exp(-\lambda r^2)/2\lambda.$ Thus
\begin{align*}
-rV(r)&=\rho _0\left(\frac{\pi}{\lambda}\right)^{3/2}\mbox{erf}\left(\sqrt{\lambda }r\right)
         =Z \mbox{erf}\left(\sqrt{\lambda }r\right) \\
      &= \frac{2Z}{\sqrt{\pi}} \left( \gamma\left(\frac{3}{2},\eta r^2\right)
            +\sqrt{\eta} r e^{-\eta r^2}\right)\\
      &= \frac{2Z}{\sqrt{\pi}} \left( \gamma\left(\frac{3}{2},\eta r^2\right)
            +\sqrt{\eta} r \left(1-\gamma(1,\eta r^2)\right)\right)
\end{align*}
since $\rho _0=Z(\lambda /\pi )^{3/2}$ (the last form of the potential can be found in
Parpia's paper \cite{Parpia:1997}).  %(cf. Eq. (6) of \cite{DyallF:93})
In the above equation $\gamma(a,x)$ is the lower incomplete gamma function, namely
\begin{equation*}
  \gamma(a,x)=\int_0^x t^{a-1}e^{-t} {\rm d}t
\end{equation*}
The potential $V(r)$ reduces to the Coulomb potential if
$r\gg \sqrt{\lambda }$.  In case a very tight diatomic system with finite
nuclei is considered the $Z_1 Z_2/r$ contribution to the total energy must
be replaced by
\begin{equation*}
  \frac{Z_1 Z_2}{r}\frac{1}{\sqrt{\pi}}\gamma\left(\frac{1}{2},\lambda_{12} r^2\right)
\end{equation*}
and $1/\lambda_{12}=1/\lambda_1+1/\lambda_2$ \cite{Parpia:1997}.

\subsection{Fermi distribution}

The Fermi distribution is more detailed than the Gaussian model and
has traditionally been employed in fitting nuclear scattering
data. The Fermi distribution is
\begin{equation}
\rho (r)=\frac{\rho _0}{1+e^{(r-c)/a}}  \nonumber
\end{equation}
where $c$ is the \textit{half-density radius} since
$\rho(c)=\rho_0/2$. The parameter $a$ is related to the nuclear
\textit{skin thickness} $t$ through ${t/a}=4\ln 3$. It may be verified
that $\rho(c-t/2)=0.9\rho_0$ and that $\rho(c+t/2)=0.1\rho_0$. The
skin thickness is thus the interval across which the nuclear charge
density drops from $0.9\rho_0$ to $0.1\rho_0$.  It is a standard
practice to take $t=2.30$ fm independent of the atomic mass.  The
parameter $c$ depends on the atomic mass through \cref{eq:rms}
and the following relation
\begin{equation}
 \avg{r^2} \approx \frac{3}{5}c^2+\frac{7}{5}\pi^2 a^2 \nonumber
\end{equation}
The potential of the Fermi distribution reads
\begin{align}
-rV(r)&=\frac{Z}{N}
     \left\{ 6 \left(\frac{a}{c}\right)^3 \left[-S_3 \left(-\frac{c}{a}\right)
                                                 +S_3 \left(\frac{r-c}{a}\right) \right] \right. \nonumber\\
     & \left. +\frac{r}{c}\left[\frac{3}{2} +\frac{1}{2}\pi^2\left(\frac{a}{c}\right)^2
     -3\left(\frac{a}{c}\right)^2 S_2\left(\frac{r-c}{a}\right) - \frac{1}{2} \left(\frac{r}{c}\right)^2\right]\right\} \nonumber
\end{align}
for $r/c < 1$ and
\begin{align}
-rV(r)&= \frac{Z}{N}
     \left\{ N +3 \left(\frac{a}{c}\right)^2
                \left[\frac{r}{c}  S_2\left(\frac{r-c}{a}\right)
              +2\left(\frac{a}{c}\right) S_3\left(\frac{r-c}{a}\right)\right] \right\} \nonumber
 \end{align}
otherwise; $N={1+\pi^2 (a/c)^2 -6(a/c)^3 S_3(-c/a)}$ and $S_k$ is an
infinite series defined as
\begin{equation}
S_k(r)=\sum_{n=1}^{\infty} (-1)^n \frac{e^{n r}}{n^k}  \nonumber
\end{equation}
This is the potential for the Fermi distribution used in the GRASP92
package \cite{Parpia:1996}, and the \xtwodhf{} program follows suit.

\section{Evaluation of one- and two-particle integrals \label{sec:appendix-integrals}}

The volume element in the $(\nu,\mu,\theta)$ coordinates is
\begin{align}
{\rm d} x {\rm d} y {\rm d} z &= {R^3 \over 8} \sinh\mu \;\sin\nu \;(\cosh^2\mu-\cos^2\nu)
\; {\rm d}\nu {\rm d}\mu {\rm d}\theta \label{eq:volxyz}
\end{align}
The expression for the kinetic energy can be calculated
in the ($\nu,\mu,\theta$) coordinates as
\begin{align}
E_T^a&=\int \int \int {\rm d}x {\rm d}y {\rm d}z \; \phi_a^*\left(-{1 \over 2}\nabla^2\right)\phi_a^{} \nonumber \\
&= -\frac{\pi R}{2} \int \int
\sqrt{(\xi^2-1)(1-\eta^2)} f_a(\nu,\mu)T(\nu,\mu)
f_a(\nu,\mu) {\rm d}\nu {\rm d}\mu \label{eq:kinE1}
\end{align}
where
\begin{equation}
T(\nu,\mu)= \frac{\partial^2}{\partial\mu^2}
+\frac{\xi}{\sqrt{\xi^2-1}} \frac{\partial}{\partial \mu}
+\frac{\partial^2}{\partial\nu^2}
+\frac{\eta}{1-\eta^2} \frac{\partial}{\partial \nu}
-m_a^2 \left(\frac{1}{\xi^2-1}+\frac{1}{1-\eta^2}\right) \label{eq:kinO}
\end{equation}
The nuclear potential energy is analogously evaluated as
\begin{equation}
E_n^a=-\frac{\pi R}{2}\int \int \sqrt{(\xi^2-1)(1-\eta^2)}
R\left\{\xi(Z_1+Z_2)+\eta(Z_2-Z_1)\right\} f_a^2 {\rm d}\nu {\rm d}\mu \label{eq:nucE1}
\end{equation}
The two-electron Coulomb and exchange energy contributions
to the total energy are obtained as
\begin{align}
E_C^{ab}&= \int \int \int {\rm d}x{\rm d}y{\rm d}z \;
\phi_a \frac{2}{R\xi}\tilde{V}_C^b \phi_a \nonumber \\
&= \frac{\pi R^2}{2} \int \int \frac{1}{\xi}\sqrt{(\xi^2-1)(1-\eta^2)}(\xi^2-\eta^2)
f_a(\nu,\mu)\tilde{V}_C^b f_a(\nu,\mu) {\rm d}\nu {\rm d}\mu \label{eq:coulE}\nonumber \\
E_x^{ab}&= \int \int \int {\rm d}x{\rm d}y{\rm d}z \; \phi_a \frac{2}{R\xi}
\tilde{V}_x^{ab} \phi_b \nonumber \\
&=\frac{\pi R^2}{2} \int \int \frac{1}{\xi}\sqrt{(\xi^2-1)(1-\eta^2)}(\xi^2-\eta^2)
f_a(\nu,\mu)\tilde{V}_x^{ab} f_b(\nu,\mu) {\rm d}\nu {\rm d}\mu \nonumber
\end{align}
The two-dimensional integration is carried out with \cref{eq:nc-quad}

\section{Interface to Libxc routines \label{sec:appendix-libxc}}

In the following, we use the same notation as Lehtola \etal{}
\cite{Lehtola2018_S_1}, except that orbitals are written as $\phi$ in our
notation.  In the case of the LDA, we have
\begin{equation}
  E_{\rm xc}^{\rm LDA} = \int {\rm d}^3 r n({\v r}) e_{\rm xc}^{\rm LDA}(\{n_{\sigma}({\v r})\}) = \int {\rm d}^3 r \epsilon_{\rm xc}^{\rm LDA}(\{n_{\sigma}({\v r})\}) \label{eq:e-lda}
\end{equation}
where n({\v r}) is the total electron density, $e_{\rm xc}^{\rm
  LDA}(\{n_{\sigma}({\v r})\})$ is the xc energy density per electron
evaluated by Libxc, and $\epsilon_{\rm xc}^{\rm LDA}(\{n_{\sigma}({\v
  r})\})$ is the resulting xc energy density.  The corresponding expression for
the LDA potentials reads
\begin{equation}
  v_{\rm xc}^{\sigma}={\partial \epsilon_{\rm xc} \over \partial n_{\sigma}}, \label{eq:v-lda}
\end{equation}
and like $e_{\rm xc}$, this quantity is readily computed and returned by Libxc.

The functional form of GGAs is often referred to as dependence on the
density gradient
\begin{equation}
E_{\rm xc}^{\rm GGA} = \int {\rm d}^3 r \epsilon_{\rm xc}^{\rm GGA}(\{n_{\sigma}({\v r})\}, \{\nabla n_{\sigma}({\v r}))\}). \label{eq:e-gga-formal}
\end{equation}
However, the gradient dependence in GGA functionals can actually be written in the form
\begin{equation}
E_{\rm xc}^{\rm GGA} = \int {\rm d}^3 r \epsilon_{\rm xc}^{\rm GGA}(\{n_{\sigma}({\v r})\}, \{\gamma_{\sigma \sigma'}({\v r}))\}), \label{eq:e-gga}
\end{equation}
where the reduced gradient is
\begin{equation}
\gamma_{\sigma\sigma'}({\v r})=\nabla n_{\sigma} \cdot \nabla n_{\sigma'}, \label{eq:red-grad}
\end{equation}
and the input and output of Libxc is formulated with respect to this
variable \cite{Lehtola2018_S_1}.
Due to the new ingredient, the GGA expression for the local potential
to be employed in the orbital optimization is somewhat more involved
\begin{equation}
v_{\rm xc}^{\sigma}={\partial \epsilon_{\rm xc}^{\rm GGA} \over \partial n_{\sigma}}
-\nabla \cdot \left( 2
{\partial \epsilon_{\rm xc}^{\rm GGA} \over \partial
\gamma_{\sigma\sigma'}} \nabla n_{\sigma}
+{\partial \epsilon_{\rm xc}^{\rm GGA} \over \partial
\gamma_{\sigma\sigma'}} \nabla n_{\sigma'}\right). \label{eq:v-gga}
\end{equation}
The GGA potential \cref{eq:v-gga} can be computed in two ways: by computing
the divergence term by finite differences, or by evaluating the divergence
analytically, which generates complicated expressions involving higher
derivatives of the density functional and Hessians of the density, for
example. We adopted the former way to evaluate \cref{eq:v-gga}, which
appears to be the standard way the problem is approached also in the plane
wave community, for example.

Since the total density is defined as
\begin{equation*}
  n = \sum_a q_a \varphi^*_a \varphi_a,
\end{equation*}
 where $q_a$ are again orbital occupation numbers, $\nabla^2 n$ can be
 evaluated as
\begin{equation*}
  \nabla^2 n = 2 \sum_a q_a \varphi^*_a \nabla^2 \varphi_a + 2 \sum_a
  q_a \nabla \varphi^*_a \nabla \varphi_a
\end{equation*}
When perfoming DFT calculations using generalized gradient approximation
one thus needs to evaluate $\nabla^2 (f^*f)$ and $\nabla f^* \nabla g$ in
the (transformed) prolate spheroidal coordinates where $f(x,y,z)$ and
$g(x,y,z)$ are orbitals or their densities. These densities are real
functions, since they do not contain any $\exp(im_f)$ or $\exp(im_g)$
terms; this also means that the densities have $\sigma$-type
symmetry. Several relevant formulae for the evaluation of various DFT
potentials have been collected below to help the reader to examine, check
or modify the code in \xtwodhf{}. Employing the definition of prolate
spheroidal coordinates (\cref{eq:nu,eq:mu,eq:xi,eq:eta}) and the general
form of the functions (\cref{eq:ffunct}) one can write
\begin{equation}
  {\partial f  \over \partial x}={\partial \xi  \over \partial x}
  {\partial \mu  \over \partial \xi} {\partial f \over \partial \mu}+
  {\partial \eta  \over \partial x} {\partial \nu  \over \partial \eta}
  {\partial f \over \partial \nu} + i m_f {\partial \theta  \over \partial x} \nonumber
\end{equation}
\begin{equation}
  {\partial f  \over \partial y}={\partial \xi  \over \partial y}
  {\partial \mu  \over \partial \xi} {\partial f \over \partial \mu}+
  {\partial \eta  \over \partial y} {\partial \nu  \over \partial \eta}
  {\partial f \over \partial \nu} + i m_f f {\partial \theta  \over \partial y} \nonumber
\end{equation}
\begin{equation}
  {\partial f  \over \partial z}={\partial \xi  \over \partial z}
  {\partial \mu  \over \partial \xi} {\partial f \over \partial \mu}+
  {\partial \eta  \over \partial z} {\partial \nu  \over \partial \eta}
  {\partial f \over \partial \nu} \phantom{ + {\partial \theta  \over \partial y}
    {\partial f \over \partial \theta} } \nonumber
\end{equation}
since $\partial f /\partial \theta = im_f f$, $\partial \theta/
\partial z =0$ (now $f\equiv f(\nu,\mu)$).  We also have
\begin{align}
  x^2+y^2 & = \frac{R^2}{4}(\xi^2-1)(1-\eta^2) \nonumber\\
  r_1 & = \sqrt {x^2 + y^2 + (z+R/2)^2} \nonumber\\
  r_2 & = \sqrt {x^2 + y^2 + (z-R/2)^2} \nonumber\\
  r_1 r_2 & = \frac{R^2}{4}(\xi^2-\eta^2)\nonumber \\
  x & = \frac{R}{2}\sqrt{\xi^2-1}\sqrt{1-\eta^2}\cos(\theta)\nonumber\\
  y & = \frac{R}{2}\sqrt{\xi^2-1}\sqrt{1-\eta^2}\sin(\theta)\nonumber\\
  z & = \frac{R}{2}\xi\eta \nonumber
\end{align}
and therefore
\begin{equation}
{\partial \mu \over \partial x} = {1 \over \sqrt{\xi^2-1}} {\partial \xi \over \partial x}
\qquad
{\partial \mu \over \partial y} = {1 \over \sqrt{\xi^2-1}} {\partial \xi \over \partial y}
\qquad
{\partial \mu \over \partial z} = {1 \over \sqrt{\xi^2-1}} {\partial \xi \over \partial z} \nonumber
\end{equation}
\begin{equation}
{\partial \nu \over \partial x} = {-1 \over \sqrt{1-\eta^2}} {\partial \eta \over \partial x}
\qquad
{\partial \nu \over \partial y} = {-1 \over \sqrt{1-\eta^2}} {\partial \eta \over \partial y}
\qquad
{\partial \nu \over \partial z} = {-1 \over \sqrt{1-\eta^2}} {\partial \eta \over \partial
  z} \nonumber
\qquad
\end{equation}
\begin{equation}
{\partial \xi  \over \partial x} = {4 \over R^2(\xi^2-\eta^2)}\xi x
\qquad
{\partial \xi  \over \partial y} = {4 \over R^2(\xi^2-\eta^2)}\xi y
\qquad
{\partial \xi  \over \partial z} = {4 \over R^2(\xi^2-\eta^2)}(\xi z-\frac{1}{2}R\eta) \nonumber
\end{equation}
 \begin{equation}
 {\partial \eta  \over \partial x} = {-4 \over R^2(\xi^2-\eta^2)}\eta x
\qquad
 {\partial \eta  \over \partial y} = {-4 \over R^2(\xi^2-\eta^2)}\eta y
\qquad
 {\partial \eta  \over \partial z} =  {-4 \over R^2(\xi^2-\eta^2)}(\eta z-\frac{1}{2}R\xi) \nonumber
 \end{equation}
\begin{equation}
  {\partial \theta \over \partial x}  = {\partial \over \partial x}
  \cos^{-1} \left( {2x \over R \sqrt{\xi^2-1}\sqrt{1-\eta^2}} \right) =
  {2\over R} {\partial \over \partial x} \left({x \over \sqrt{\xi^2-1}\sqrt{1-\eta^2}} \right)
  {1 \over \sqrt{1-\tilde{x}^2}} \nonumber
\end{equation}
\begin{align}
  {\partial \theta \over \partial x}
  &= {\partial \over \partial x}\cos^{-1} \left( {2x \over R \sqrt{\xi^2-1}\sqrt{1-\eta^2}} \right)\nonumber \\
  &= {2\over R} {\partial \over \partial x} \left({x \over \sqrt{\xi^2-1}\sqrt{1-\eta^2}} \right)
                                          {1 \over \sqrt{1-\tilde{x}^2}}\nonumber\\
  &= {2\over R\sin(\theta)}{1 \over \sqrt{\xi^2-1}\sqrt{1-\eta^2}}
      \left(1 - {4x^2 \over {R^2(\xi^2-\eta^2)}}\right) \nonumber
\end{align}
where $\tilde{x} = 2x/(R\sqrt{\xi^2-1}\sqrt{1-\eta^2})=\cos(\theta)$.
Likewise
\begin{align}
  {\partial \theta \over \partial y}
  &= {\partial \over \partial y}\cos^{-1} \left( {2y \over R \sqrt{\xi^2-1}\sqrt{1-\eta^2}} \right)\nonumber\\
  &= {2\over R} {\partial \over \partial y} \left({y \over \sqrt{\xi^2-1}\sqrt{1-\eta^2}} \right)
      {1 \over \sqrt{1-\tilde{y}^2}} \nonumber\\
  &= {2\over R\cos(\theta)}{1 \over \sqrt{\xi^2-1}\sqrt{1-\eta^2}}
      \left(1 - {4y^2 \over {R^2(\xi^2-\eta^2)}}\right) \nonumber
\end{align}
since $\tilde{y} = 2y/(R\sqrt{\xi^2-1}\sqrt{1-\eta^2})=\sin(\theta)$.

In case the $f$ and $g$ functions are of $\sigma$-type symmetry ($m_f=m_g=0$) we have
\begin{align}
  {\partial f \over \partial x}
  &= {4 \over R^2(\xi^2-\eta^2)} \left\{
      \frac{\xi x}{\sqrt{\xi^2-1}}{\partial f \over \partial \mu} +
      \frac{\eta x}{\sqrt{1-\eta^2}}{\partial f \over \partial \nu}\right\} \label{eq:partial-fx}\\
  {\partial f \over \partial y}
  &= {4 \over R^2(\xi^2-\eta^2)} \left\{
      \frac{\xi y}{\sqrt{\xi^2-1}}{\partial f \over \partial \mu} +
      \frac{\eta y}{\sqrt{1-\eta^2}}{\partial f \over \partial \nu}\right\} \label{eq:partial-fy}\\
  {\partial f \over \partial z}
  &= {4 \over R^2(\xi^2-\eta^2)} \left\{
      \frac{(\xi z -\frac{1}{2}R\eta)}{\sqrt{\xi^2-1}}{\partial f \over \partial \mu} +
      \frac{(\eta z- \frac{1}{2}R\xi)}{\sqrt{1-\eta^2}}{\partial f \over \partial \nu}\right\}\nonumber
\end{align}
\begin{align}
  {\partial f \over \partial x}{\partial g \over \partial x}
  &=
      \left[{4 \over R^2(\xi^2-\eta^2)}\right]^2 \left\{
      \frac{\xi^2 }{(\xi^2-1)} {\partial f \over \partial \mu}{\partial g \over \partial \mu} +
      \frac{\eta^2}{(1-\eta^2)}{\partial f \over \partial \nu}{\partial g \over \partial
      \nu}\right.\nonumber \\
  &\phantom{=}
                \phantom{\left[{4 \over R^2(\xi^2-\eta^2)}\right]^2}
                \left.+\frac{\xi \eta}{\sqrt{\xi^2-1}\sqrt{1-\eta^2}}
                \left({\partial f \over \partial \mu} {\partial g \over \partial \nu} +
                {\partial f \over \partial \nu}{\partial g \over \partial \mu} \right)\right\}x^2\nonumber
\end{align}
\begin{align}
  {\partial f \over \partial y}{\partial g \over \partial y}
  &= \left[{4 \over R^2(\xi^2-\eta^2)}\right]^2 \left\{
        \frac{\xi^2 }{(\xi^2-1)} {\partial f \over \partial \mu}{\partial g \over \partial \mu} +
        \frac{\eta^2}{(1-\eta^2)}{\partial f \over \partial \nu}{\partial g \over \partial \nu}\right. \nonumber\\
  &\phantom{=} \phantom{\left[{4 \over R^2(\xi^2-\eta^2)}\right]^2}
                \left.+\frac{\xi \eta}{\sqrt{\xi^2-1}\sqrt{1-\eta^2}}
                \left({\partial f \over \partial \mu} {\partial g \over \partial \nu} +
                {\partial f \over \partial \nu}{\partial g \over \partial \mu} \right)\right\}y^2\nonumber
\end{align}
\begin{align}
  {\partial f \over \partial z}{\partial g \over \partial z}
  &=
      \left[{4 \over R^2(\xi^2-\eta^2)}\right]^2 \left\{
      \frac{(\xi z-\frac{R}{2} \eta)^2}{(\xi^2-1)}
      {\partial f \over \partial \mu}{\partial g \over \partial \mu} +
      \frac{(\eta z -\frac{R}{2}\xi)^2}{(1-\eta^2)}
      {\partial f \over \partial \nu}{\partial g \over \partial \nu}\right.\nonumber \\
  &\phantom{=}
                \phantom{\left[{4 \over R^2(\xi^2-\eta^2)}\right]^2 }
                \left.+\frac{(\xi z-\frac{R}{2}\eta) (\eta z-\frac{R}{2}\xi)}
                {\sqrt{\xi^2-1}\sqrt{1-\eta^2}}
                \left({\partial f \over \partial \mu} {\partial g \over \partial \nu} +
                {\partial f \over \partial \nu}{\partial g \over \partial \mu} \right)\right\}\nonumber
\end{align}
\begin{align}
  \nabla f \nabla g
  &= \left[{4 \over R^2(\xi^2-\eta^2)}\right]^2 \nonumber \\
  &\phantom{=}\left\{
                \left(\frac{R^2}{4}\xi^2 {(1-\eta^2)}+\frac{(\xi
                z-\frac{R}{2} \eta)^2}{(\xi^2-1)} \right) {\partial f \over \partial
                \mu}{\partial g \over \partial \mu} + \right. \nonumber \\
 &\phantom{=} \phantom{\left\{ \right. }\left(\frac{R^2}{4}\eta^2 {(1-\xi^2)}+\frac{(\eta
	  z-\frac{R}{2} \xi)^2}{(1-\eta^2)} \right)
           {\partial f \over \partial \nu}{\partial g \over \partial \nu} + \nonumber\\
 &\phantom{=} \phantom{\left\{ \right. }\left.\left( \frac{R^2}{4}\xi \eta\sqrt{\xi^2-1}\sqrt{1-\eta^2} +
	      \frac{(\xi z-\frac{R}{2}\eta) (\eta z-\frac{R}{2}\xi)}
               {\sqrt{\xi^2-1}\sqrt{1-\eta^2}}\right)
               \left({\partial f \over \partial \mu} {\partial g \over \partial \nu} +
			    {\partial f \over \partial \nu}{\partial g \over \partial \mu}
                            \right)\right\}\nonumber \\
\nabla^2 f &= {4 \over R^2(\xi^2-\eta^2)}
     \left( {\partial^2 f\over \partial\mu^2}
     +{\xi \over \sqrt{\xi^2-1}} {\partial f \over \partial \mu}
     +{\partial^2 f\over \partial\nu^2}
    +{\eta \over \sqrt{1-\eta^2}} {\partial f \over \partial \nu} \right)  \nonumber
\end{align}

In case $m_f\ne 0$\ie{} $f=e^{im_f\theta}f(\nu,\mu)$, the partial
derivatives in \cref{eq:partial-fx,eq:partial-fy} contain additional terms
$i m_f f\partial \theta /\partial x$ and
$i m_f f\partial \theta /\partial y$ and therefore now
$\nabla f^{*} \nabla g$ and the Laplacian $\nabla^2 f$ formulae take the
form
\begin{align}
  \nabla f \nabla g
  &= \left[{4 \over R^2(\xi^2-\eta^2)}\right]^2 \nonumber \\
  &\phantom{=} \left\{
                \left(\frac{R^2}{4}\xi^2 {(1-\eta^2)}+\frac{(\xi
                z-\frac{R}{2} \eta)^2}{(\xi^2-1)} \right) {\partial f \over \partial
                \mu}{\partial g \over \partial \mu} + \right. \nonumber \\
  &\phantom{=}  \phantom{\left\{ \right. }\left(\frac{R^2}{4}\eta^2 {(1-\xi^2)}+\frac{(\eta
                z-\frac{R}{2} \xi)^2}{(1-\eta^2)} \right)
                {\partial f \over \partial \nu}{\partial g \over \partial \nu} + \nonumber\\
  &\phantom{=}  \phantom{\left\{ \right. }\left.\left( \frac{R^2}{4}\xi \eta\sqrt{\xi^2-1}\sqrt{1-\eta^2} +
	      \frac{(\xi z-\frac{R}{2}\eta) (\eta z-\frac{R}{2}\xi)}
               {\sqrt{\xi^2-1}\sqrt{1-\eta^2}}\right)
               \left({\partial f \over \partial \mu} {\partial g \over \partial \nu} +
			    {\partial f \over \partial \nu}{\partial g \over \partial \mu}
               \right)\right\}\nonumber \\
  & +  m_f m_g f g  {4 \over R^4  \sqrt{(\xi^2-1)(1-\eta^2)}} \nonumber\\
  &   \left\{ {1 \over \sin^2(\theta)} \left( 1- {4x^2 \over R^2 (\xi^2-1) (1-\eta^2)} \right)^2
     +  {1 \over \cos^2(\theta)}  \left( 1 - {4y^2 \over R^2 (\xi^2-1) (1-\eta^2)} \right)^2
     \right\} \nonumber
\end{align}
\begin{align}
\nabla^2 f &= {4 \over R^2(\xi^2-\eta^2)}
               \left( {\partial^2 f\over \partial\mu^2}
               +{\xi \over \sqrt{\xi^2-1}} {\partial f \over \partial \mu}
               +{\partial^2 f\over \partial\nu^2}
               +{\eta \over \sqrt{1-\eta^2}} {\partial f \over \partial \nu} \right) \nonumber \\
           & + {4 m_f^2 f\over R^2(\xi^2-1)(1-\eta^2)}   \nonumber \\
           &= {4 \over R^2(\xi^2-\eta^2)}
               \left\{ {\partial^2 f\over \partial\mu^2}
               +{\xi \over \sqrt{\xi^2-1}} {\partial f \over \partial \mu}
               +{\partial^2 f\over \partial\nu^2}
               +{\eta \over \sqrt{1-\eta^2}} {\partial f \over \partial \nu} \right. \nonumber \\
           & \left. + m_f^2 \left( {1 \over \xi^2-1} + {1 \over 1-\eta^2}\right)f \right\}\nonumber
\end{align}
When $\nabla^2 (f^{*} f)$ is needed there is no $\theta$-dependency and the
density must be treated as a $\sigma$-type orbital.

The extension of this approach to meta-GGA functionals is non-trivial,
as~discussed by Zahariev \etal{} \cite{Zahariev2013_JCP_244108}, for
instance. We note here for completeness that the need to compute
expressions similar to \cref{eq:v-gga} can be avoided in a basis set
expansion by the use of integrations by parts
\cite{Lehtola2020_M_1218}, following the work of Pople \etal{} in
\cite{Pople1992_CPL_557}. The integration by parts can be used to move
the differentiation from the xc potential onto the basis functions,
which typically have analytical derivatives, and most atomic-orbital
implementations follow this approach that is easily extended to
meta-GGAs, as well \cite{Neumann1996_MP_1, Lehtola2020_M_1218}. The
finite element implementation of GGAs and meta-GGAs in \helfem{} also
employs this technique \cite{Lehtola2019_IJQC_25944,
  Lehtola2019_IJQC_25945, Lehtola2020_PRA_12516,
  Lehtola2023_JCTC_2502}.

\section{Boundary conditions for potentials at $r_\infty$ \label{sec:potential-boundary}}
The boundary conditions for the potentials $\tilde{V}^{ab}$ at the
practical infinity are obtained from the multipole expansion. In
particular, we have
\begin{align}
\tilde{V}^{a}_C&=\frac{R\xi}{2}\sum_{l=0}^{l_{\rm max}}
\frac{P_{l,0}(\cos\theta)}{r^{l+1}} Q_{l,0}^{aa} \label{eq:pot-multipole-C} \\
\tilde{V}^{ab}_x&=\frac{R\xi}{2} \sum_{l=0}^{l_{\rm max}}(-1)^{|\Delta m|}
\frac{(l-|\Delta m|)!}{(l+|\Delta m|)!} \frac{P_{l|\Delta m|}(\cos\theta)}{r^{l+1}}
{Q_{l,|\Delta m|}^{ab}} \label{eq:pot-multipole-x}
\end{align}
where $Q_{l,|m|}^{ab} = \langle \phi_a | r^l P_{l,|m|}(\cos\theta) |
\phi_b \rangle$ are the multipole moments and $P_{l,|\Delta m|}$ are the
associated Legendre functions. In the present version of the program,
the multipole moments are computed in parallel if OpenMP support is
available.

Note that the Coulomb potential in \cref{eq:pot-multipole-C} only contains
the $\Delta m=0$ term, because orbital densities $|\phi_a|^2$ have
cylindrical symmetry ($|\exp(i m \theta)|^2 = 1$), while exchange
potentials arise from orbital density products $\phi^*_a \phi_b$ that are
$\theta$ dependent via $\Delta m = m_a - m_b$.  \xtwodhf{} employs the
default value $l_{\rm max}=4$ unless orbitals of $\phi$ symmetry are
detected, in which case the default value is increased to $l_{\rm max}=8$.

\section{Boundary conditions for orbitals at $r_\infty$ \label{sec:orbital-boundary}}

The asymptotic behaviour of the orbitals is used to estimate their
values in the last few grid points in the $\mu$ direction, close to
$r_\infty$. The asymptotic behaviour of the orbitals of diatomic
molecules can be modeled as that of the orbitals of the corresponding
united atom. Therefore, one can consider the second-order differential
equation \cite{Fischer:1977}
\begin{equation}
\frac{{\rm d}^2y_a}{{\rm d} r^2}=\left( \varepsilon_a - \frac{g_1(r)}{r}+
\frac{g_2(r)}{r^2}\right)y_a =F_a(r)y_a \label{eq:orb-boundary-eqn}
\end{equation}
with $y(0)=0$ and $y(r)\rightarrow0$ as $r\rightarrow\infty$, where
$\varepsilon_a$ is the orbital energy, and $g_1(r)$, $g_2(r)$, and
$F_a(r)$ are functions whose exact definitions are not essential here,
as the details of the approach can be found in the work of
Froese-Fischer \cite{Fischer:1977}. The solution $y_a(r)$ to this
equation exhibits exponential asymptotic decay as
\begin{equation}
y_a(r) \propto F_a(r)^{-1/4}\;
\exp\left(-\int_{r_{0}}^rF_a(r^\prime)^{1/2} {\rm d}r^\prime\right) \label{eq:orb-asympt-form}
\end{equation}
By discretizing and approximating the integral by a rectangular rule, the above equation
yields the appropriate expression of the boundary condition for the orbitals at the
practical infinity in the form
\begin{equation}
y_a(r_{i+1})\approx y_a(r_i)\;
\left(\frac{F_a(r_i)}{F_a(r_{i+1})}\right)^{1/4}
\exp\left(-\sqrt{F_a(r_i)}(r_{i+1}-r_i)\right) \label{eq:eq22}
\end{equation}

\section{Interpolation of boundary values \label{sec:appendix-interpolation}}

The properties discussed in \cref{eq:nu-symmetry,eq:mu-symmetry,eq:pi-nu-symmetry} can be used to provide the additional values along
the remaining sides of the rectangular region
$[\nu_{1},\nu_{N_\nu}] \times [\mu_{1},\mu_{N_\mu}]$.
If the solution $f$ is an odd
function, it should vanish along the $(0,\mu)$, $(\pi,\mu)$ and $(\nu,0)$ lines, and the
corresponding values are thereby set to~zero.%
If the solution $f$ is an even function, suitable values are calculated
using the Lagrange $9$-point interpolation formula for an equally spaced abscissa
\cite{AbramowitzS:77}

\begin{equation}
f(x_0+ph)\approx \sum_k A_k^n(p)u_k \nonumber
\end{equation}
where $A_k^n$ is the interpolation constant given for even and odd values of $n$ by the
following formulae
\begin{align*}
A_k^n(p) & =  \frac{(-1)^{\frac{1}{2}n+k}}{\left( \frac{n-2}{2}+k\right)!
\left( \frac{1}{2}n-k\right)! \left( p-k \right)}
\prod_{t=1}^n \left( p+\frac{1}{2}n-t\right), \\
& \phantom{xxxxxxxxxxxxxxxxxxxxxxxxxxx} -\frac{1}{2}(n-2)\le k \le \frac{1}{2}n \nonumber\\
A_k^n(p) & =  \frac{(-1)^{\frac{1}{2}(n-1)+k}}{\left( \frac{n-1}{2}+k\right)!
\left( \frac{n-1}{2}-k\right)! \left( p-k \right)}
\prod_{t=0}^{n-1} \left( p+\frac{n-1}{2}-t\right),\\
& \phantom{xxxxxxxxxxxxxxxxxxxxxxxxxxx} -\frac{1}{2}(n-1)\le k \le \frac{1}{2}(n-1) \nonumber\\
\end{align*}
Assuming that $f(-x_i)=f(x_i)$ and $f_k=f(x_0+kh)$, solving $f_0$ from
the equation for $f_5$ yields
\begin{equation}
f_0={1 \over 126}(210f_1-120f_2+45f_3-10f_4+f_5). \nonumber
\end{equation}

\section{Assignments of the 2D arrays in the \xtwodhf{} code \label{sec:appendix-initArrays}}

\def\es{\\[10pt]}

\begin{tabular}{cccc}\label{initArrays}
  \texttt{VXI}$=\xi=\cosh\mu$  & \texttt{VXISQ}$=\xi^2$ & \texttt{VXI1}$=\sqrt{\xi^2-1}$
  & \texttt{VXI2}$=\frac{\xi}{\sqrt{\xi^2-1}}$\es
  \texttt{VETA}$=\eta=\cos\nu$ & \texttt{VETASQ}$=\eta^2$ & \texttt{VETA1}$=\sqrt{1-\eta^2}$
 &\texttt{VETA2}$=\frac{\eta}{\sqrt{1-\eta^2}}$\es
   \multicolumn{2}{c}{\texttt{BORB}$=\frac{\cosh\mu}{\sinh\mu} = \frac{\xi}{\sqrt{\xi^2-1}}$}
   &\multicolumn{2}{c}{\texttt{BPOT}$=\frac{1}{\sqrt{\xi^2-1}}-\frac{2\sqrt{\xi^2-1}}{\xi}$}\es
   \multicolumn{2}{c}{\texttt{D}$=\frac{\cos\nu}{\sin\nu}=\frac{\eta}{\sqrt{1-\eta^2}}$}
   &\multicolumn{2}{c}{\texttt{E}$=-\left(\frac{1}{\xi^2-1}+\frac{1}{1-\eta^2}\right)$}\es
  \multicolumn{2}{c}{\texttt{G}$=-\frac{\pi R^3\xi}{2} (\xi^2-\eta^2)$}
  & \multicolumn{2}{c}{\texttt{F0}$=R(Z_a+Z_b)\xi+R(Z_b-Z_a)\eta$}\es
  \multicolumn{1}{c}{\texttt{F1}$=\frac{R^2}{2}(\xi^2-\eta^2)$}
& \multicolumn{1}{c}{\texttt{F2}$=-\frac{R}{\xi}(\xi^2-\eta^2)$}
& \multicolumn{1}{c}{\texttt{F3}$=-\frac{2}{\xi^2}$}
& \multicolumn{1}{c}{\texttt{F4}$=\frac{R\xi}{2}$}\es

\multicolumn{2}{c}{\texttt{WGT1}$=\frac{\pi R}{2} \sqrt{\xi^2-1} \sqrt{1-\eta^2}$}
&\multicolumn{2}{c}{\texttt{WGT2}$=\frac{\pi R^2}{2}\sqrt{\xi^2-1}\sqrt{1-\eta^2} (\xi^2 - \eta^2)/\xi$}\es

\multicolumn{4}{c}{\texttt{WGT2*F4}$=\frac{\pi R^3}{4} \sqrt{\xi^2-1}\sqrt{1-\eta^2}(\xi^2-\eta^2)
=\frac{\pi R^3}{4} \sinh \mu \sin \nu (\cosh^2 \mu -\cos^2\nu)$}

\end{tabular}

\bibliography{books-articles-jk,susi}

\end{document}